\documentclass[a4paper]{article}

\usepackage{CJK}
\usepackage[paperwidth=185mm,paperheight=230mm,textheight=190mm,textwidth=145mm,left=20mm,right=20mm, top=25mm, bottom=20mm]{geometry}
\usepackage{amsmath,amssymb,graphicx, setspace, subfigure, url, multirow, verbatim, bm,threeparttable,algorithm,color,amsthm, rotating, float}
\usepackage{amsthm}
\usepackage{calc}
\usepackage{supertabular}
\usepackage{longtable}
\usepackage{color}
\usepackage{enumerate}
\usepackage{colortbl,booktabs}
\usepackage[noend]{algpseudocode}
\newtheorem{theorem}{Theorem}

\begin{document}

\title{Projective Resampling Imputation Mean Estimation Method for Missing Covariates Problem}

\author{Zishu Zhan$^{1,2}$, Xiangjie Li$^{3,*}$, Jingxiao Zhang$^{1,2,*}$}

\maketitle

\begin{center}
	
	{\it $^1$Center for Applied Statistics, Renmin University of China,\\
		$^2$ School of Statistics, Remin University of China,\\
		$^3$ State Key Laboratory of Cardiovascular Disease,
		Fuwai Hospital, National Center for Cardiovascular Diseases, Chinese Academy of Medical Sciences and Peking Union Medical College
	}
\end{center}

\begin{abstract}
     Missing data is a common problem in clinical data collection, which causes difficulty in the statistical analysis of such data. To overcome problems caused by incomplete data, we propose a new imputation method called projective resampling imputation mean estimation (PRIME), which can also address ``the curse of dimensionality" problem in imputation with less information loss. We use various sample sizes, missing-data rates, covariate correlations, and noise levels in simulation studies, and all results show that PRIME outperformes other methods such as iterative least-squares estimation (ILSE), maximum likelihood (ML), and complete-case analysis (CC). Moreover, we conduct a study of influential factors in cardiac surgery-associated acute kidney injury (CSA-AKI), which show that our method performs better than the other models. Finally, we prove that PRIME has a consistent property under some regular conditions. 

\textbf{ Keywords }: projective resampling, missing covariates problem, imputation method, linear regression
\end{abstract}

\section{Introduction}
\label{section1}
In medical research, an investigator's ultimate interest may be in inferring prognostic markers, given the patients' genetic, cytokine, and/or environmental backgrounds \cite{Wu2018, Chen2020}. However, in practical applications, data are often missing. The most common approach to address missing-data problems is complete-case analysis (CC), which is simple but inefficient. CC can also lead to biased estimates when the data are not missing completely at random. The maximum likelihood method (ML \cite{Dempster1977, Jiang2020}) and the inverse probability weighting method (IPW \cite{Seaman2013,Sun2018}) are also widely used approaches to address missing data. However, likelihood-based methods are sensitive to model assumptions, and re-weighting methods do not always make full use of the available data. Alternatively, imputation \cite{Conn1989, Yin2016} is a more flexible approach to couple with missing data. 

However, a preliminary analysis of cardiac surgery-associated acute kidney injury (CSA-AKI) data used in Chen et al. \cite{Chen2020} indicates that the missing-data patterns vary across individuals. Accordingly, new and more capable quantitative methods are needed for this individual-specific missing data.
Furthermore, it is common for only a small fraction of records to have complete information across all sources. Existing methods do not work well when the percentage of unavailable data is high. To estimate the coefficients (rather than predicting them), Lin et al. \cite{Lin2019} proposed the iterative least-squares estimation (ILSE) method to deal with individual-specific missing-data patterns using the classical regression framework, but it needs a complete set of observations to obtain the initial values, and its results might not converge when based on bad initial values. Furthermore, Lin et al. \cite{Lin2019} may have difficulty accommodating data missing from both important and unimportant variables.

In this study, based on the idea of projection resampling/random projection, we propose Projection Resampling Imputation Mean Estimation (PRIME), a method that tackles the aforementioned drawbacks of existing methods. The key idea behind projection resampling/random projection was given in the Johnson-Lindenstrauss lemma \cite{Johnson1984}, which preserves pairwise distances after projecting a set of points to a randomly chosen low-dimensional subspace. There are several previous studies on projection resampling/random projection for dimension reduction, including Schulman \cite{Schulman2000} for clustering, Donoho \cite{Donoho2006} for signal processing, Shi et al. \cite{Shi2010} for classification, Maillard and Munos \cite{Maillard2012} for linear regression, and Le et al. \cite{Le2013} for kernel approximation. Specifically, the idea of PRIME is to project the covariates along randomly sampled directions to obtain samples of scalar-valued predictors and kernels (dimension reduction). Next, a simple geometric average is taken on the scalar-predictor-based kernel to impute the missing parts (using all-sided information). Our method has several advantages, including the following. First, PRIME can deal with a high degree of missing data, even data containing no complete observations, while most existing methods require at least a fraction of the subjects to have fully complete observations. Second, we can average the imputed estimates from multiple projection directions and fully utilize the available information to reach a more reliable and useful result. Third, to reduce the undesirable influence of unimportant variables, PRIME can be easily extended to sparse PRIME (denoted as SPRIME), which has a profound impact in practical applications.

The remainder of the paper is organized as follows. Section \ref{section2} introduces the basic setup of PRIME and SPRIME. Theoretical properties are discussed in Section \ref{section3}. Sections \ref{section4} and \ref{section5} present the numerical results using simulated and real data examples, respectively. Section \ref{section6} presents some concluding remarks. In addition, our proposed method is implemented using R and the scripts to reproduce our results are available at \url{https://github.com/eleozzr/PRIME}. The proof of the theorem is available the Appendix. 

\section{Projective resampling Imputation mean estimation (PRIME)}
\label{section2}
\subsection{Model and estimation by PRIME}
In this paper, let ${\bm Y}=(Y_1,Y_2,\cdots,Y_n)^{\top}$ be the response variable of interest and ${\bm X}=\{X_{ij}: i=1,2,\cdots,n, j=1,2,\cdots,p\}$ be the covariate matrix. We assume that $n>p$. We consider the presence of missing covariates by $r_{ij}$, which denotes the missing-data indicator for $X_{ij}$, where $r_{ij}$ is 1 if $X_{ij}$ is missing and is 0 otherwise. For each unit $i$, $A_i=\{j:r_{ij}=0, j=1,2,\cdots,p\}$ denotes the available covariates set, e.g., $A_i=\{1,2,\cdots,p\}$ for the complete case. 

In this study, we focus on a linear regression model. Assume that the random sample $\{(Y_i, {\bm X}_i):i=1,2,\cdots,n, \ j=1,2,\cdots,p\}$ is generated by:
\begin{eqnarray}\label{gen}
Y_i=\sum_{j=1}^{p}\beta_jX_{ij}+\varepsilon_i={\bm X}_i^{\top}{\bm \beta}+\varepsilon_i,
\end{eqnarray}
where ${\bm \beta}=(\beta_1,\beta_2,\cdots,\beta_p)^{\top}$ is the coefficient vector for the covariates and the $\varepsilon_i$'s are independently identically distributed random errors. We assume 
$E(\varepsilon_i|{\bm X}_i)=0$ and $E(\varepsilon_i^2|{\bm X}_i)=\sigma^2$.

\begin{figure}[t]
	\begin{center}
		\includegraphics[width=10cm,height=8cm]{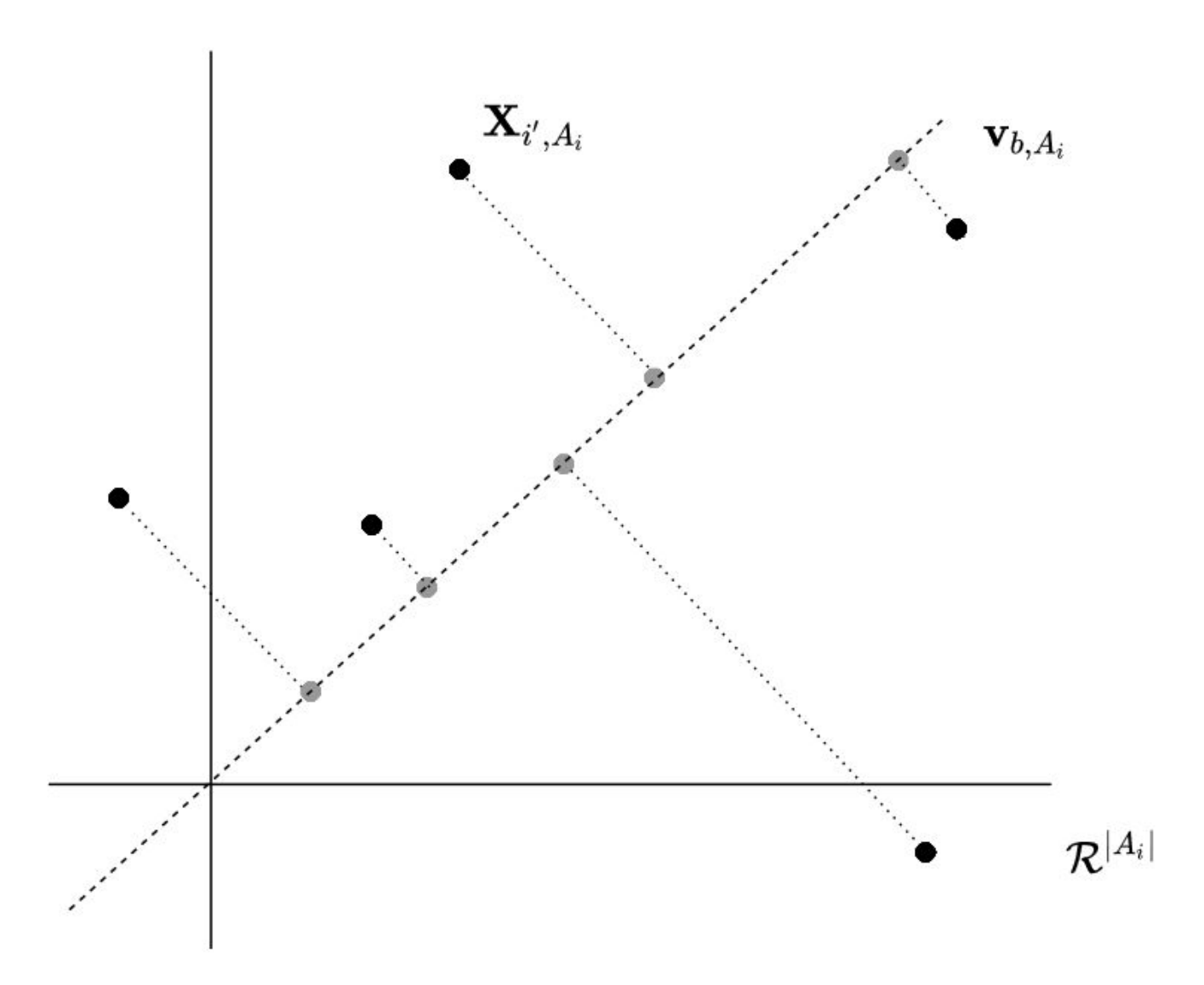}
	\end{center}
	\caption{Random projection}
	\label{fig:idea}
\end{figure}

Covariates can be divided into two parts based on $A_i$: ${\bm X}_{i,A_i}=(X_{ij}: j\in A_i)^{\top}$ for observed covariates and ${\bm X}_{i,\bar{A}_i}=(X_{ij}: j\notin A_i)^{\top}$ for missing covariates. Thus, equation (\ref{gen}) can be expressed as 
\begin{eqnarray*}\label{pred}
	Y_i={\bm X}_{i,A_i}^{\top}{\bm \beta}_{A_i}+{\bm X}_{i,\bar{A}_i}^{\top}{\bm \beta}_{\bar{A}_i}+\varepsilon_i,
\end{eqnarray*} 
where ${\bm \beta}_{A_i}$ and ${\bm \beta}_{\bar{A}_i}$ denote the regression coefficients for the complete and incomplete covariates, respectively. For ${\bm X}_{i,\bar{A}_i}$, which is unobserved, refer to Lin et al. \cite{Lin2019}, we take the expectation of $Y_i$ given the observed covariates. Thus, we obtain the following equation:
\begin{equation}\label{cm}
E(Y_i|{\bm X}_{i,A_i})={\bm X}_{i,A_i}^{\top}{\bm \beta}_{A_i}+E({\bm X}_{i,\bar{A}_i}^{\top}{\bm \beta}_{\bar{A}_i}|{\bm X}_{i,A_i}).
\end{equation}
Hence, by equation (\ref{cm}), we can impute the incomplete part using the information on ${\bm X}_{i,A_i}$ to obtain an estimator for ${\bm \beta}$. We use the following estimator to estimate the missing components of the covariates for unit $i$:
\begin{eqnarray}\label{est}
\tilde{X}_{ij}=\frac{\sum_{i'=1}^{n}I(A_{i'}\supset A_i\cup j)X_{i'j}K_{h}({\bm X}_{i',A_i}-{\bm X}_{i,A_i})}{\sum_{i'=1}^{n}I(A_{i'}\supset A_i\cup j)K_{h}({\bm X}_{i'A_i}-{\bm X}_{i,A_i})}, \quad \left(j \notin A_i\right),
\end{eqnarray} 
In equation (\ref{est}), $K_h(\cdot)=K(\cdot/h)/h$, where $K(\cdot)$ is a kernel function and $h$ is a bandwidth. In this way, we can make use of the information as fully as possible. To tackle the problem of ``the curse of dimensionality'', we further transform the estimator in equation (\ref{est}) using the projective resampling method. As shown in Figure \ref{fig:idea}, for subject $i'$, we project ${\bm X}_{i',A_i}$ onto random directions $\{{\bm v}_{b,A_i} \in \mathcal{R}^{1\times |A_i|}, b=1,2,\cdots,B\}$, where $|A|$ denotes the cardinality of a set A, then obtain $B$ kernel values using the resulting scalars $\{{\bm X}_{i',A_i}^{\top}{\bm v}_{b,A_i}, b=1,2,\cdots,B\}$, and finally integrate the $B$ kernels through the geometric mean.

\begin{equation}\label{est2}
\hat{X}_{ij}=\frac{\sum_{i'=1}^{n} I\left(A_{i'} \supset A_i \cup j\right) X_{i^{\prime} j} \left[\prod_{b=1}^{B}K_{h}\left(\mathbf{X}_{i',A_i}^{\top} {\bm v}_{b,A_i}-{\bm X}^{\top}_{i,A_i}{\bm v}_{b,A_i}\right)\right]^{\frac{1}{B}}}{\sum_{i'=1}^{n} I\left(A_{i'} \supset A_i \cup j\right) \left[\prod_{b=1}^{B}K_{h}\left(\mathbf{X}_{i',A_i}^{T}{\bm v}_{b,A_i}-{\bm X}^{\top}_{i,A_i}{\bm v}_{b,A_i}\right)\right]^{\frac{1}{B}}}, \quad \left(j \notin A_i\right),
\end{equation}
where ${\bm v}_{b,A_i}$ is a random vector, with each entry $v_{b,A_i,j}$ chosen independently from a distribution $\mathcal{D}$ that is symmetric about the origin with $E(v_{b,A_i,j}^2)=1$. In practice, we usually generate $v_{b,A_i,j}$ from $N(0,1)$ or $U(-1,1)$, but random vectors from other possible distributions can also be used.

Applying the above imputation strategy, we can obtain ${\bm Z}_i$ by using observed data for part $A_i$ and imputed data for part $\bar{A}_i$.
\begin{eqnarray}\label{imput}
{\bm Z}_i=(X_{ij}I(j\in A_i)+\hat{X}_{ij}I(j\notin A_i): j=1,2,\cdots,p)^{\top}.
\end{eqnarray} 
Therefore, we propose the following estimation equation:
\begin{equation}\label{ee}
U({\bm \beta})=\frac{1}{n}\sum_{i=1}^n{\bm Z}_i\{Y_i-{\bm Z}_i^{\top}{\bm \beta}\}.
\end{equation}
Thus, the estimator of the regression coefficient can be solved by
\begin{equation*}
\hat{\bm \beta}=\left\{\sum_{i=1}^{n}{\bm Z}_i{\bm Z}_i^{\top}\right\}^{-1}\sum_{i=1}^{n}{\bm Z}_iY_i.
\end{equation*}
Specifically, we propose the following algorithm:
\begin{algorithm}[t]
	\caption{algorithm for PRIME} 
	\hspace*{0.02in} {\bf Input:} 
	$\{(Y_i,{\bm X}_i,A_i): i=1,2,\cdots,n, j=1,2,\cdots,p\}$, \\
	\hspace*{0.02in} {\bf Output:} 
	$\hat{\bm \beta}$
	\begin{algorithmic}[1]
		\State Sample entries of ${\bm v}_{b,A_i} \ (b=1,2,\cdots,B)
		\in \mathcal{R}^{1\times |A_i|}$ i.i.d. from $N(0,1)$ or $U(-1,1)$
		\For{$1\le i\le n$ and $1 \le j \le p$} 
		\State 1) Apply the equation (\ref{est2}) to obtain $\hat{X}_{ij}, (j\notin A_i)$
		\State 2) Apply the equation(\ref{imput}) to obtain the imputed data $Z_i$ based on ${\bm v}_{b,A_i} (b=1,2,\cdots,B)$
		\EndFor 
		\State Solve the closed-form equation to get $\hat{\bm \beta}$.\\
		\Return $\hat{\bm \beta}$
	\end{algorithmic}
\end{algorithm}

\subsection{Simultaneous fitting and selection by sparse PRIME}
Because the number of disease-associated biomarkers is not expected to be large, it is of great importance to take the sparse assumption into account when not all variables contribute to outcome variables. Hence, we assume the linear regression model in equation (\ref{gen}) is sparse and define the index set of the active and inactive predictors by $\mathcal{I}_1=\{j:\beta_j\neq0\}$ and $\mathcal{I}_2=\{j:\beta_j=0 \}$, respectively. Our practical goal is to identify which biomarkers in the CSA-AKI datasets are disease-related as well as to estimate the corresponding coefficients. The main idea of sparse PRIME is to replace the estimation equation \ref{ee} with the penalized estimation equations as follows:
\begin{equation}\label{pee}
\left\{
\begin{aligned}
\sum_{i=1}^n{Z}_{i1}\{Y_i-{\bm Z}_i^{\top}{\bm \beta}\} &+ \lambda_n \gamma |\beta_1|^{\gamma-1}sign(\beta_1) = 0,\\
&\vdots&\\
\sum_{i=1}^n{Z}_{ip}\{Y_i-{\bm Z}_i^{\top}{\bm \beta}\} &+\lambda_n \gamma |\beta_p|^{\gamma-1}sign(\beta_p) =  0.
\end{aligned}
\right.
\end{equation}
where $\lambda \gamma |\beta_j|^{\gamma-1}sign(\beta_j)$ is the partial derivative for the penalty function with respect to $\beta_j$. The least absolute shrinkage and selection operator (LASSO) estimator is defined to satisfy $\gamma=1$. Optimizing the functions in (\ref{pee}) with $\gamma=1$ is computationally cumbersome because the functions are non-differentiable. Fortunately, the shooting algorithm proposed in Fu \cite{Fu1998} can be used to compute the LASSO estimator. Moreover, Fu \cite{Fu1998, Fu2003} proved that the unique estimator of (\ref{pee}) is equivalent to the solution of the penalized objective function as follows:
\begin{equation*}\label{eql}
\min_{\bm \beta} \frac{1}{2n}\sum_{i=1}^{n}\left\{Y_i-{\bm Z}_i^{\top}{\bm \beta}\right\}^2+\frac{\lambda_n}{n}\sum_{j=1}^{p}|\beta_j|^{\gamma}.
\end{equation*}
We can solve this penalized regression-form problem and get a sparse estimator $\hat{\bm \beta}_s$ using the {\it glmnet} package in R.

To decrease the processing time, we can use the following sparse random projection with i.i.d. entries:
\begin{equation*}\label{vdist}
v_{b,A_i,j}=\sqrt{s_i} \left\{
\begin{aligned}
1, &\ \ \text{with probability}\  \frac{1}{2s_i}\\
0, &\ \ \text{with probability}\  1-\frac{1}{s_i}\\
-1, &\ \ \text{with probability}\  \frac{1}{2s_i}.
\end{aligned}
\right.
\end{equation*}
where Achlioptas \cite{Achlioptas2003} used $s_i=1$ or $s_i=3$ and Li et al. \cite{Li2006} showed that one can use $s_i=\sqrt{|A_i|}$ or even $s_i=\frac{|A_i|}{\log |A_i|}$ to significantly reduce the computing time with little loss of accuracy.

\section{Theoretical properties} 
\label{section3}
We study the consistency of the projection resampling least estimator $\hat{\bm \beta}$. Denote the true value of ${\bm \beta}$ by ${\bm \beta}_0$. We make the following assumptions:

\begin{description}
	\item[(A1)] $h=O(n^{-l})$ with $1/4<l<1/2$;
	\item[(A2)] $A_i \bot {\bm X}_i$;
	\item[(A3)] The kernel function $K(\cdot)$ is a symmetric density function with compact support $[0,1]$ and a bounded derivative;
	\item[(A4)] ${\bm \beta}_0 \in \mathcal{B}$, where $\mathcal{B}$ is a bounded set;
	\item[(A5)] For a missing-data pattern $A$, let $e_{j, A}({\bm w}_A)= E\left(X_{i j}|\mathbf{X}_{i, A}={\bm w}_A\right)$ be the conditional expectation of $X_{i j}$ and $f_A({\bm w}_A)$ be the density of $\mathbf{X}_{i, A}.$ Assume $e_{j, A}({\bm w}_A)$ and $f_{A}({\bm w}_A)$ have continuous second derivatives with respect to ${\bm w}_A$ on the corresponding support;
	\item[(A6)] $E(v^{4}_{bj})<\infty \ (1\le j\le p)$;
	\item[(A7)] ${\bm X}_i$ is bounded for all $i\ge 1$, and the limit $\lim_{n\to \infty}\sum_{i=1}^{n}{\bm X}_i/n={\bm X}_0$ exists;
	\item[(A8)] $D_n=\frac{1}{n}\sum_{i=1}^n{\bm X}_i{\bm X}_i^{\top}\to D$ and $\frac{1}{n}\max_{1\le i\le n}{\bm X}_i^{\top}{\bm X}_i\to 0$, where $D$ is a finite and positive definite matrix;
	\item[(A9)] There exists a constant $C_0>0$, $\inf_{{\bm w}_A}P_r(A_{i^{\prime}}\supset A\cup j)f_A({\bm w}_A)>C_0$.
\end{description}

Assumptions (A1)--(A5) are the same as the conditions in Lin et al. \cite{Lin2019}. Specifically, Assumption (A1) requires under-smoothing to obtain a root$-n$ consistent estimator, which is a commonly used regularity assumption in semiparametric regression. Assumption (A2) addresses the missing data mechanism and ensures the PRIME estimator is consistent. As mentioned in Lin et al. \cite{Lin2019}, Assumption (A2) is weaker than assuming the data are missing completely at random. 
Assumptions (A3)--(A5) are standard in nonparametric regression. Assumption (A3) is achieved when the kernel function is the Gaussian kernel, but it is more general. Assumption (A6) is a moment bound required in Arriaga and Vempala \cite{Arriaga2006} and Li et al. \cite{Li2006}, a necessary technical condition. Assumptions (A7) and (A8) are commonly used in penalized estimation problems \cite{Fu1998, Fu2003}. We assume $\inf_{{\bm w}_A}P_r(A_{i^{\prime}}\supset A\cup j)f_A({\bm w}_A)>C_0$ to make sure that there are enough samples being used to estimate $e_{j, A}({\bm w}_A)$.

\begin{theorem}\label{th1}
	Suppose Assumptions (A1)-(A6) hold and $j \in \mathcal{I}_1=\{j:\beta_j\neq0\}, j=1,2,\cdots,p$, as $n \to \infty$, then $\hat{\bm \beta}\to {\bm \beta}_0$ in probability.
\end{theorem}

\begin{theorem}\label{th2}
	Suppose Assumptions (A1)-(A8) hold, $j \in \mathcal{I}_1=\{j:\beta_j\neq0\}, j=1,2,\cdots,q$, $j \in \mathcal{I}_2=\{j:\beta_j =0\}, j=q+1,\cdots,p$ and tuning parameter $\lambda_n=o(\sqrt{n})$ as $n \to \infty$, then $\hat{\bm \beta}_s\to {\bm \beta}_0$ in probability.
\end{theorem}

\section{Simulation}
\label{section4}
In this section, we consider several simulated scenarios to highlight the properties of PRIME in contrast to some other methods. We experimentally investigate the performance of the following methods:
\begin{description}
	\item[Full:] the least-squares estimator based on the full data as a benchmark;
	\item[PRIME:] the proposed method; 
	\item[ILSE:] the iterative least-square method in Lin et al. \cite{Lin2019};
	\item[ML:] the maximum likelihood method proposed in Jiang et al. \cite{Jiang2020};
	\item[CC:] the complete-case analysis method.
\end{description}

For each model setting with a specific choice of parameters, we repeat the simulation 100 times and evaluate the performance of models using the normalized absolute distance (NAD) and the mean squared error (MSE), defined as follows:
\[
{\rm NAD}_j=\frac{1}{N}\sum_{i=1}^{N}\frac{|\hat{\beta}_j-\beta_{0j}|}{\beta_{0j}}, \ j=1,2\cdots,p,
\]

\[
{\rm MSE}=\frac{1}{N}\sum_{i=1}^{N}\frac{1}{p}\sum_{j=1}^{p}(\hat{\beta}_j-\beta_{0j})^2.
\]
In addition, we calculate the optimal MSE rate, defined as the proportion of times each method (except Full) produced the smallest MSE in repetitions. The MSE based on N repetitions is partitioned into MSE=Variance+Bias$^2$, as follows:
\[
\frac{1}{N}\sum_{i=1}^{N}\frac{1}{p}\sum_{j=1}^{p}(\hat{\beta}_j^{(i)}-\beta_{0j})^2=\frac{1}{p}\sum_{j=1}^{p}\frac{1}{N}\sum_{i=1}^{N}(\hat{\beta}_j^{(i)}-\bar{\beta}_{j})^2+\frac{1}{p}\sum_{j=1}^{p}(\bar{\beta}_j-\beta_{0j})^2,
\]
where $\bar{\beta}_j=\frac{1}{N}\sum_{i=1}^{N}\hat{\beta}_j^{(i)}$ and $N=100$ in this simulation study. For simplicity, we set bandwidth $h = n^{-1/3}$ for PRIME, SPRIME and ILSE. In the following sections, we compare the methods using various settings for sample size, missing data rates, noise levels, and feature correlations. 

\subsection{Scenario 1: Different noise levels}
The data generation model has the linear expression
\begin{eqnarray*}
	Y_i=\sum_{j=1}^p\beta_jX_{ij}+\varepsilon_i,\ i=1,2,\cdots,n,
\end{eqnarray*}
where $n=100,200$, $p=12$, ${\bm \beta}=(1, -0.6, 1.5, 1, 1.2,0.4, -1, -0.7, 1.3, 0.5, 1.1, -1.4, 0.9)^{\top}$. 

We generate $(X_{i1},\cdots,X_{ip})$ from the multivariate normal distribution $N_{p}(0,\Sigma)$. We set the non-diagonal elements $\rho_{ij}$ of $\Sigma$ equal to $0.5$ and the diagonal elements of $\Sigma$ equal to $1$. For $\varepsilon_i$, we use the error distribution
$N(0,\sigma^2)$, where $\sigma^2$ changes with $R^2=\text{Var}({\bm X}_i^{\top}{\bm \beta})/\{\text{Var}({\bm X}_i^{\top}{\bm \beta})+\sigma^2\}$. We consider the cases in which $R^2=0.1,0.2,\cdots,0.9$. 

Missing data are divided into three generative scenarios or assumptions. Missing completely at random (MCAR) means the missingness is independent of the values of the data. Missing at random (MAR) means the propensity of data to be missing depends on the observed values, whereas missing not at random (MNAR) covers the remaining scenario that the mechanism depends on the unobserved values (the variables that are missing). In our study, we consider situations that differ from the classical MCAR, MAR, and MNAR mechanism. 

For each sample, variables $X_{10}, X_{11}, X_{12}$ are always available. There are twelve ``typical" missing patterns considered, the details are shown in Table \ref{tab:pattern}. We divide the 12 patterns into two groups. Specifically, the first group ${\bm A}^{(1)}$ consists of $(A_1,A_2,\cdots,A_6)$, and the second group ${\bm A}^{(2)}$ consists of the rest missing patterns. We randomly assign the missing patterns in ${\bm A}^{(1)}$ to the sample with missing probability $P=a$. Furthermore, we set the missing probability of $i$th unit for the patterns in ${\bm A}^{(2)}$ as $P=\{1+\exp(b\varepsilon_i+c)\}^{-1}$. Then we randomly assign the patterns in ${\bm A}^{(2)}$ to the missing samples. The settings in Table \ref{tab:mis} are used for the missing rate (MR).

\begin{table}[t]
	\centering
	\caption{Missing pattern for all simulation examples.}
	\label{tab:pattern}
	\begin{tabular}{cccccccccccccc}
		\hline 
		\textbf{Group} &\textbf{Pattern} & \multicolumn{12}{c}{\textbf{Variable}}  \\ 
		\hline 
		&Full & $X_1$ & $X_2$ & $X_3$ & $X_4$ & $X_5$ & $X_6$ & $X_7$ & $X_8$ & $X_9$ & $X_{10}$ & $X_{11}$ & $X_{12}$ \\ 
		\multirow{6}{*}{${\bm A}^{(1)}$}
		&$A_1$ & \checkmark   &  \checkmark   &  &  \checkmark   &  \checkmark   &  \checkmark   &  \checkmark   &  \checkmark   &  \checkmark   &  \checkmark   &  \checkmark   &  \checkmark   \\ 
		
		&$A_2$ &  \checkmark   &  \checkmark   &  \checkmark   &  \checkmark   &  \checkmark   &  &  \checkmark   &  \checkmark   &  \checkmark   &  \checkmark   &  \checkmark   &  \checkmark   \\ 
		
		&$A_3$ &  \checkmark   &  \checkmark   &  \checkmark   &  \checkmark   &  \checkmark   &  \checkmark   &  \checkmark   &  \checkmark   & &  \checkmark   &  \checkmark   &  \checkmark   \\ 
		
		&$A_4$ &  &  \checkmark   &  \checkmark   &  &  \checkmark   &  \checkmark   &  \checkmark   &  \checkmark   &  \checkmark   &  \checkmark   &  \checkmark   &  \checkmark   \\ 
		
		&$A_5$ &  &  \checkmark   &  \checkmark   &  \checkmark   &  \checkmark   &  \checkmark   &  &  \checkmark   &  \checkmark   &  \checkmark   &  \checkmark   &  \checkmark   \\ 
		
		&$A_6$ &  \checkmark   &  \checkmark   &  \checkmark   &  &  \checkmark   &  \checkmark   &  &  \checkmark   &  \checkmark   &  \checkmark   &  \checkmark   &  \checkmark   \\ 
		\hline
		\multirow{6}{*}{${\bm A}^{(2)}$}
		&$A_7$ &  & & &  \checkmark   &  \checkmark   &  \checkmark   &  \checkmark   &  \checkmark   &  \checkmark   &  \checkmark   &  \checkmark   &  \checkmark   \\ 
		
		&$A_8$ &  \checkmark   &  \checkmark   &  \checkmark   & & & &  \checkmark   &  \checkmark   &  \checkmark   &  \checkmark   &  \checkmark   &  \checkmark   \\ 
		
		&$A_9$ &  \checkmark   &  \checkmark   &  \checkmark   &  \checkmark   &  \checkmark   &  \checkmark   & & & &  \checkmark   &  \checkmark   &  \checkmark   \\ 
		
		&$A_{10}$ & & &  & &  &  &  \checkmark   &  \checkmark   &  \checkmark   &  \checkmark   &  \checkmark   &  \checkmark   \\ 
		
		&$A_{11}$ &  &  &  &  \checkmark   &  \checkmark   &  \checkmark   &  &  &  &  \checkmark   &  &  \checkmark   \\ 
		
		&$A_{12}$ &  \checkmark   &  \checkmark   &  \checkmark   &  &  &  &  &  &  &  \checkmark   &  \checkmark   &  \checkmark   \\ 
		\hline 
	\end{tabular} 
\end{table}

\begin{table}[t]
	\centering
	\caption{Missing rating setting for all simulation examples.}
	\label{tab:mis}
	\begin{tabular}{ccccccccccc}
		\hline
		\multirow{1}{*}{\textbf{Missing rate}}& &\multicolumn{9}{c}{${\bm R}^2$}\\
		\cline{3-11}
		& &0.1&0.2&0.3&0.4&0.5&0.6&0.7&0.8&0.9\\
		\cline{1-0}
		\hline
		\multirow{3}{*}{$60\%$}
		&$a$ & 0.1  & 0.1 & 0.1 & 0.1 & 0.1 & 0.1& 0.1  & 0.1 & 0.1 \\
		&$b$ & -1.5 &-1.5 & -1.5 & -1.5  & -1.5  & -1.5 & -2 & -2 & -4  \\
		&$c$ & -4  & -2 & -2 & -1.5 & -1.5 & -1 & -1 & -1 & -1  \\
		\hline
		\multirow{3}{*}{$90\%$}
		&$a$ & 0.75  & 0.75 & 0.75  & 0.75 & 0.75 & 0.75 & 0.75 & 0.75 & 0.65 \\
		&$b$ & -1.5 & -1.5  & -2 & -2 & -3 & -3 & -3.5 & -3.5 & -4  \\
		&$c$ & -4   & -4 & -4  & -4 & -4 & -4 &-4 & -4 & -4 \\
		\hline
	\end{tabular}
\end{table} 

\begin{figure}[H]
	\centering
	\subfigure[]{
		\begin{minipage}[t]{0.33\linewidth}
			\centering
			\includegraphics[width=1.9in]{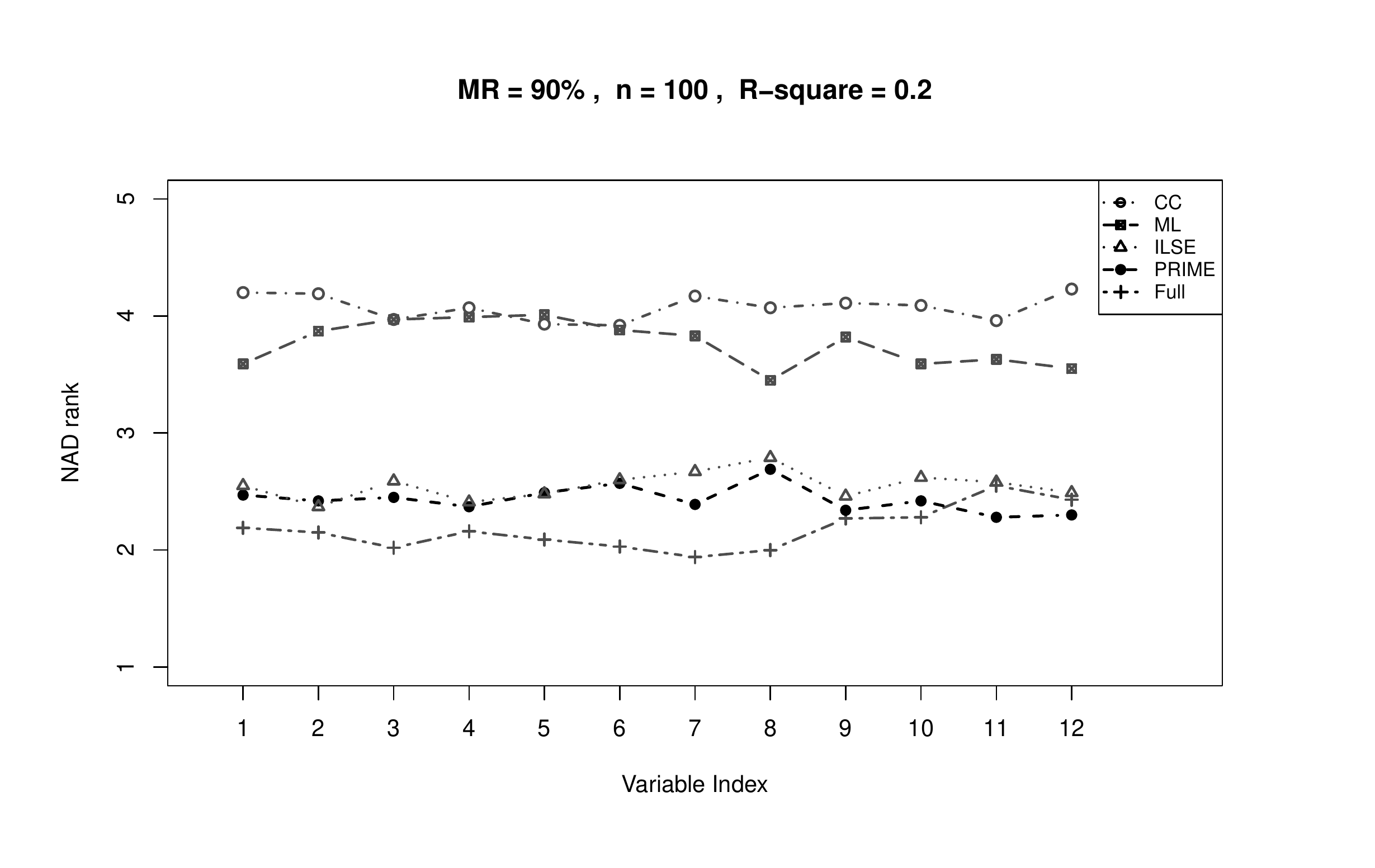}
		\end{minipage}%
	}%
	\subfigure[]{
		\begin{minipage}[t]{0.33\linewidth}
			\centering
			\includegraphics[width=1.9in]{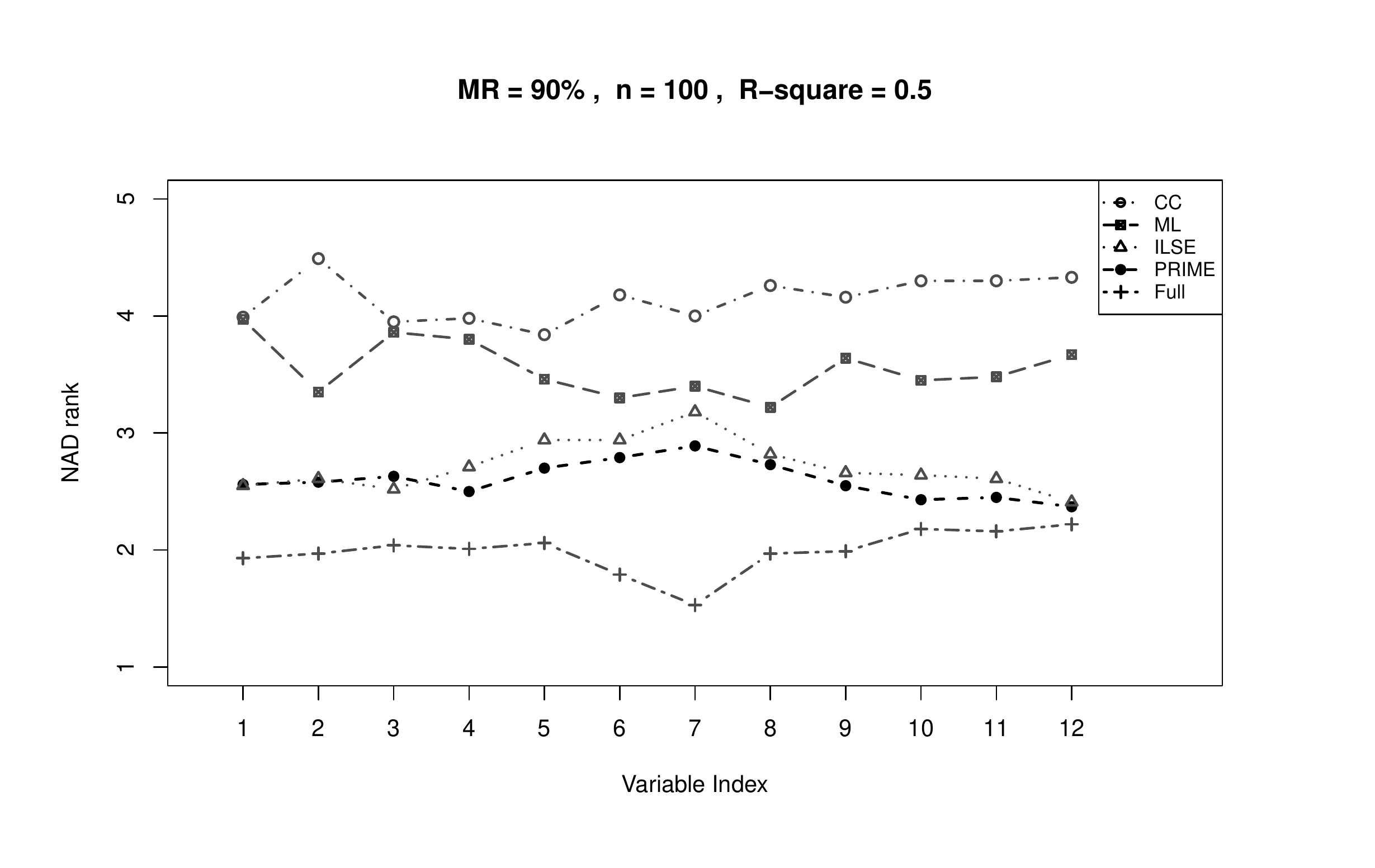}
		\end{minipage}
	}%
	\subfigure[]{
		\begin{minipage}[t]{0.33\linewidth}
			\centering
			\includegraphics[width=1.9in]{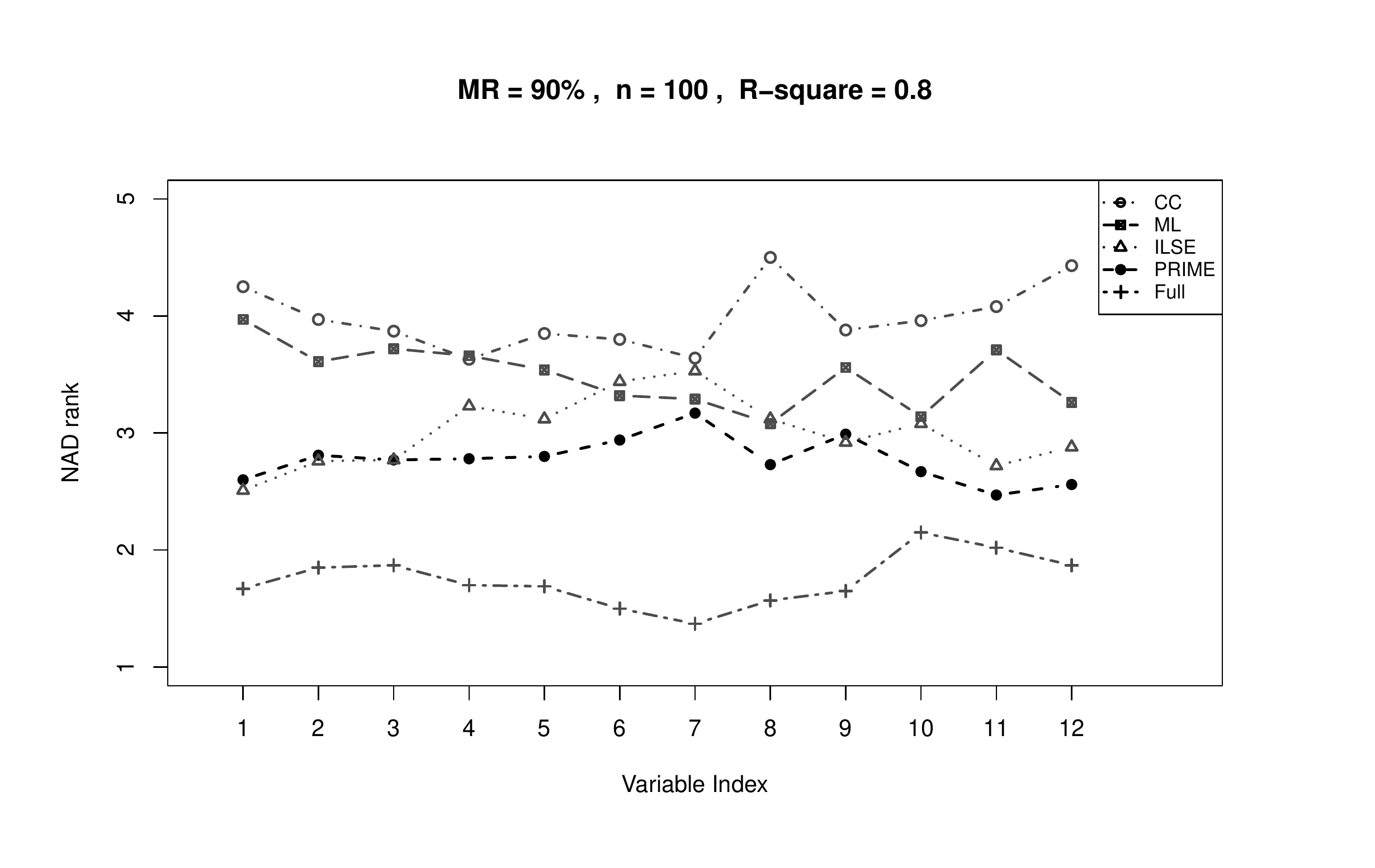}
		\end{minipage}
	}%
	\centering
	\caption{NAD with $n=100$ and 90\% missing data for different methods.}
	\label{fig:RNAD91}
\end{figure}

\begin{figure}[H]
	\centering
	\subfigure[]{
		\begin{minipage}[t]{0.33\linewidth}
			\centering
			\includegraphics[width=1.9in]{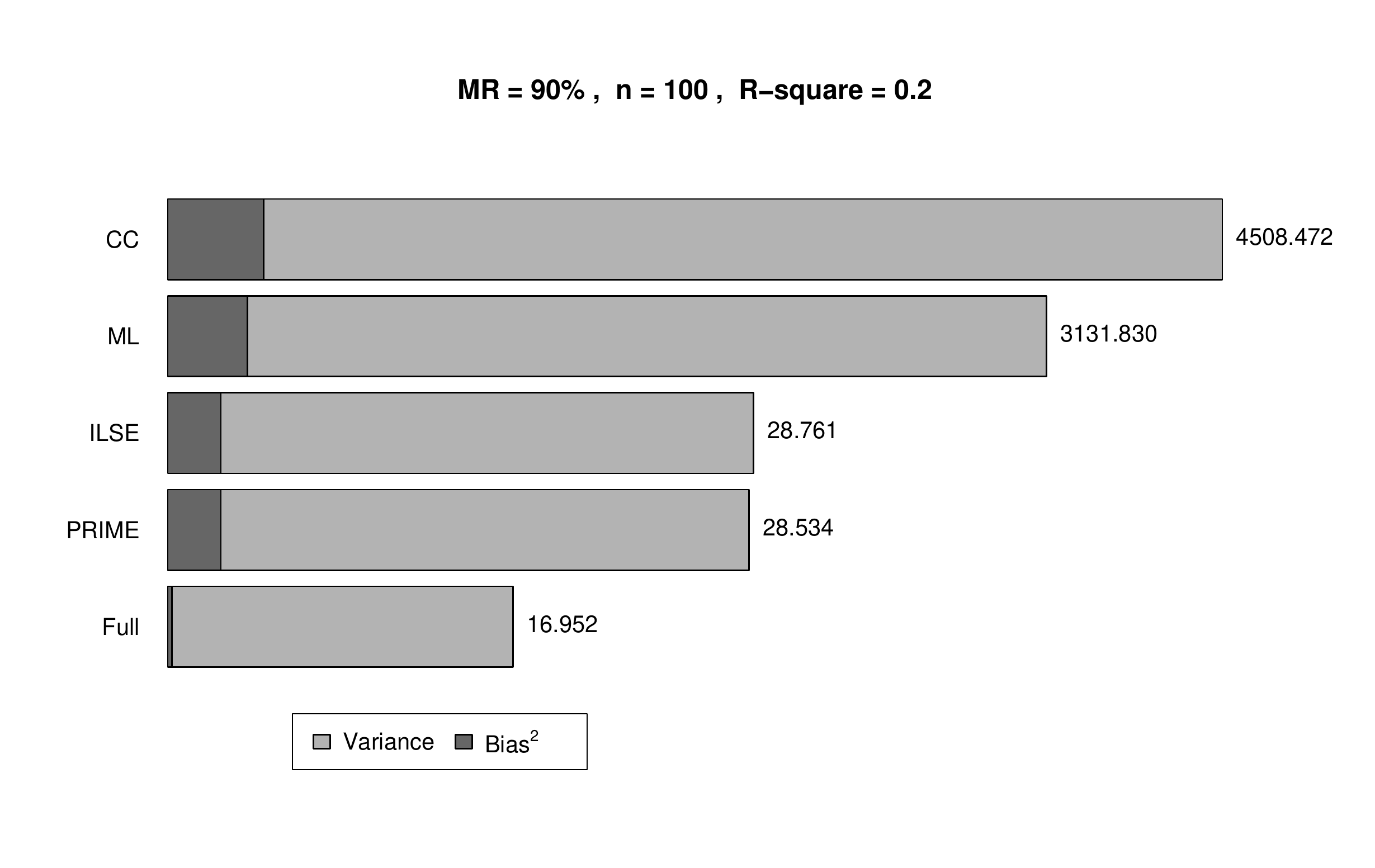}
		\end{minipage}%
	}%
	\subfigure[]{
		\begin{minipage}[t]{0.33\linewidth}
			\centering
			\includegraphics[width=1.9in]{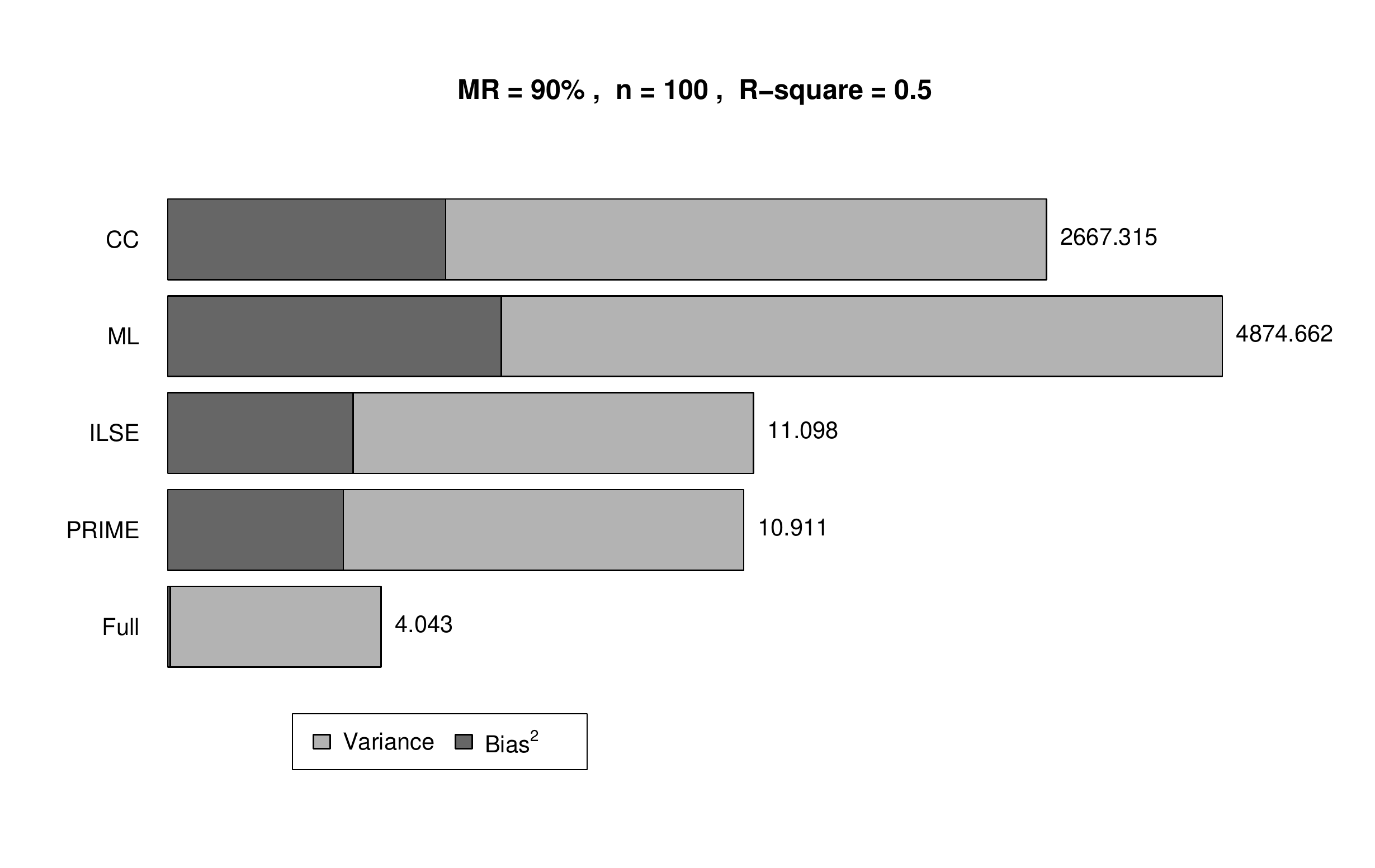}
		\end{minipage}
	}%
	\subfigure[]{
		\begin{minipage}[t]{0.33\linewidth}
			\centering
			\includegraphics[width=1.9in]{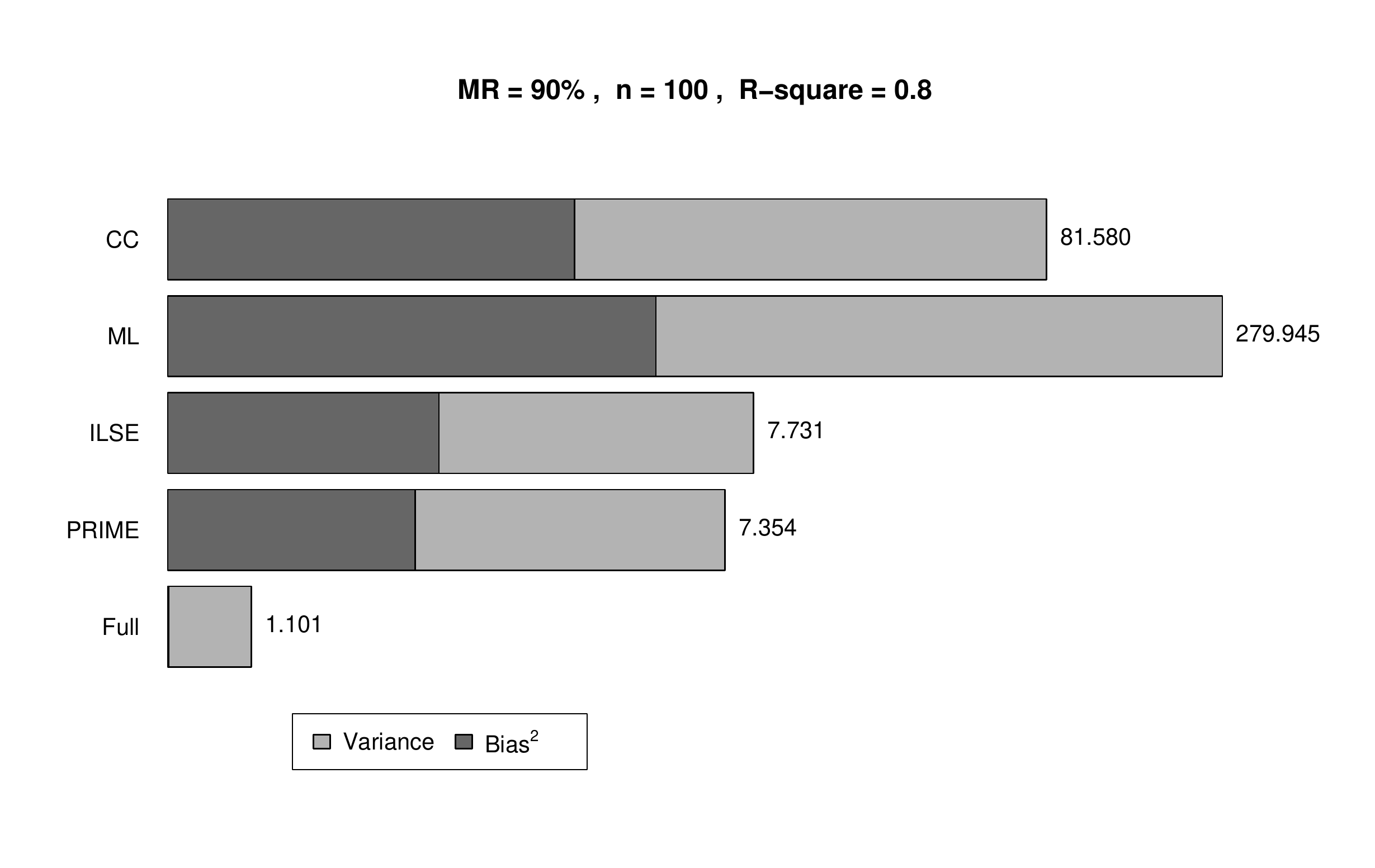}
		\end{minipage}
	}%
	\centering
	\caption{MSE with $n=100$ and 90\% missing data for different methods.}
	\label{fig:RMSE91}
\end{figure}

\begin{figure}[H]
	\begin{center}
		\includegraphics[width=12cm,height=7cm]{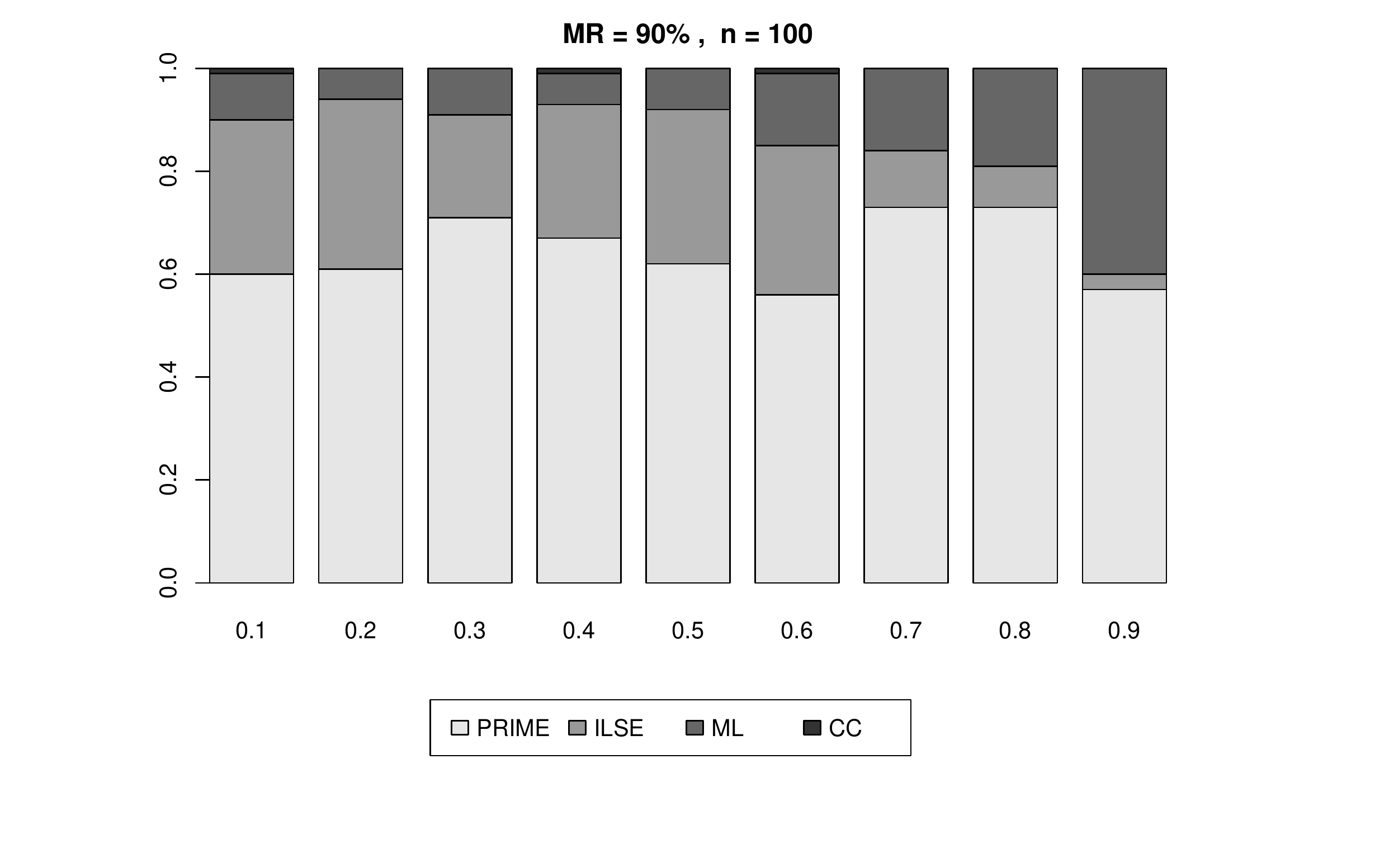}
	\end{center}
	\caption{Optimal rate of MSE with $n=100$ and 90\% missing data for different methods.}
	\label{fig:ORR91}
\end{figure}

\begin{figure}[H]
	\centering
	\subfigure[]{
		\begin{minipage}[t]{0.33\linewidth}
			\centering
			\includegraphics[width=1.9in]{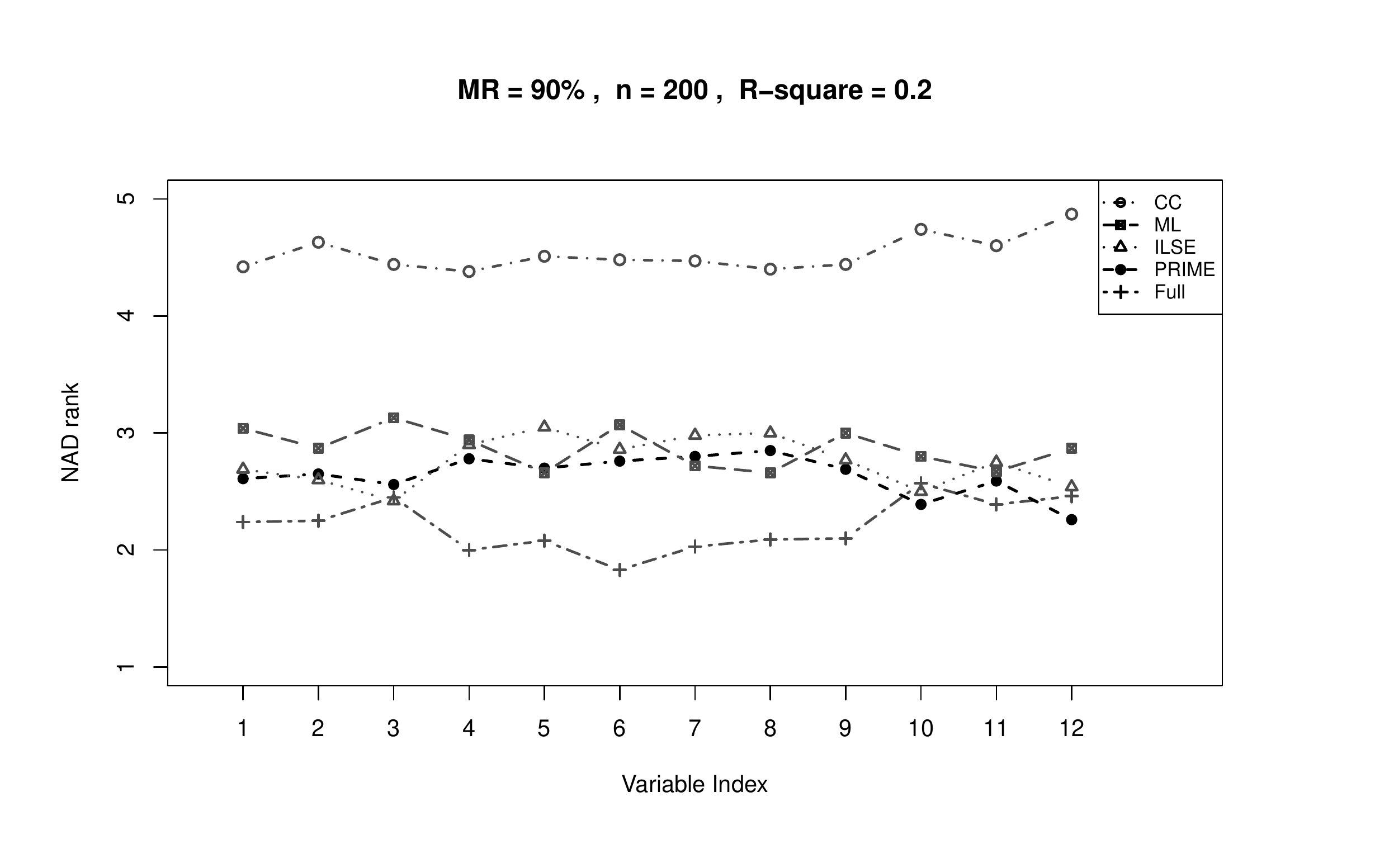}
		\end{minipage}%
	}%
	\subfigure[]{
		\begin{minipage}[t]{0.33\linewidth}
			\centering
			\includegraphics[width=1.9in]{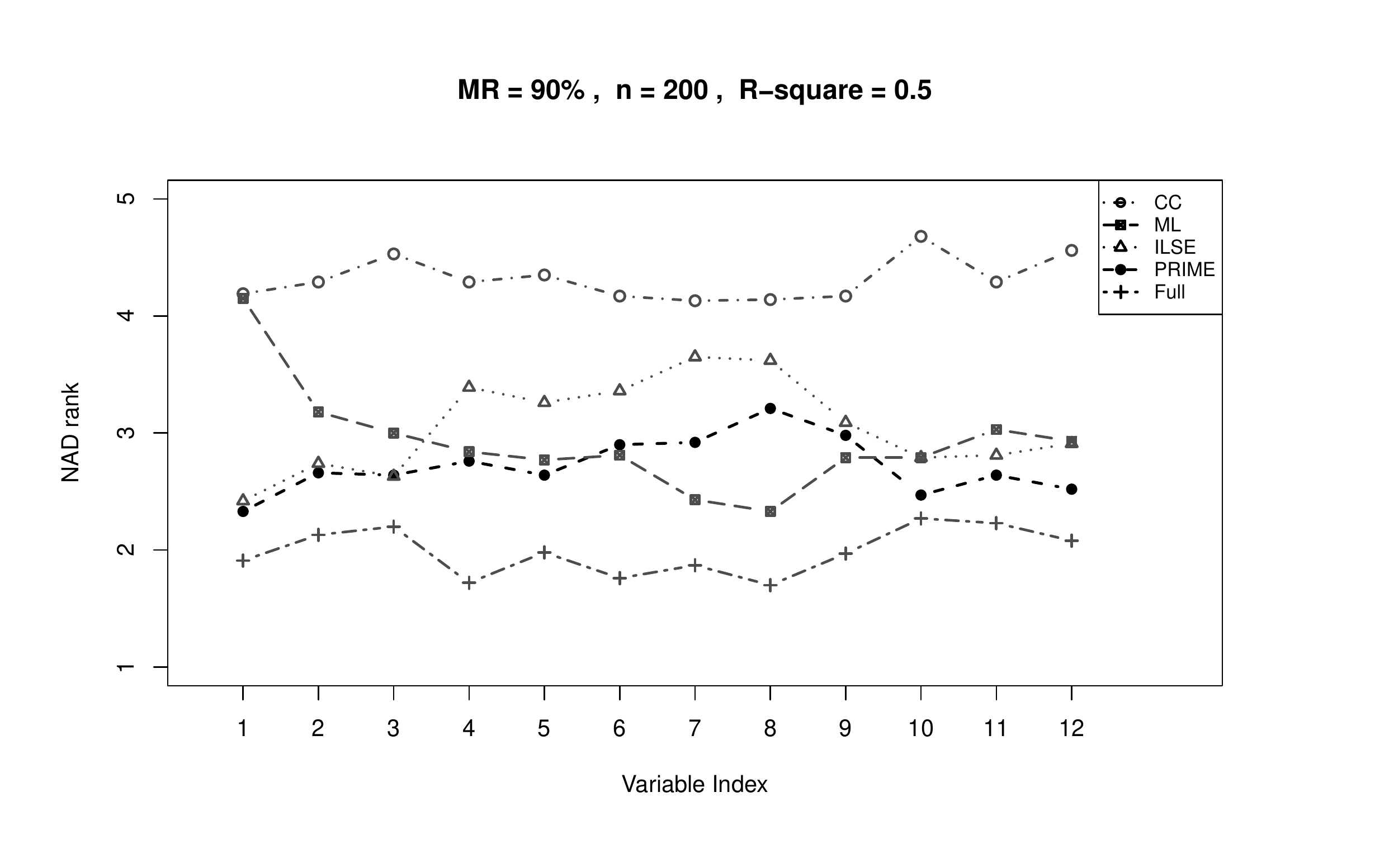}
		\end{minipage}
	}%
	\subfigure[]{
		\begin{minipage}[t]{0.33\linewidth}
			\centering
			\includegraphics[width=1.9in]{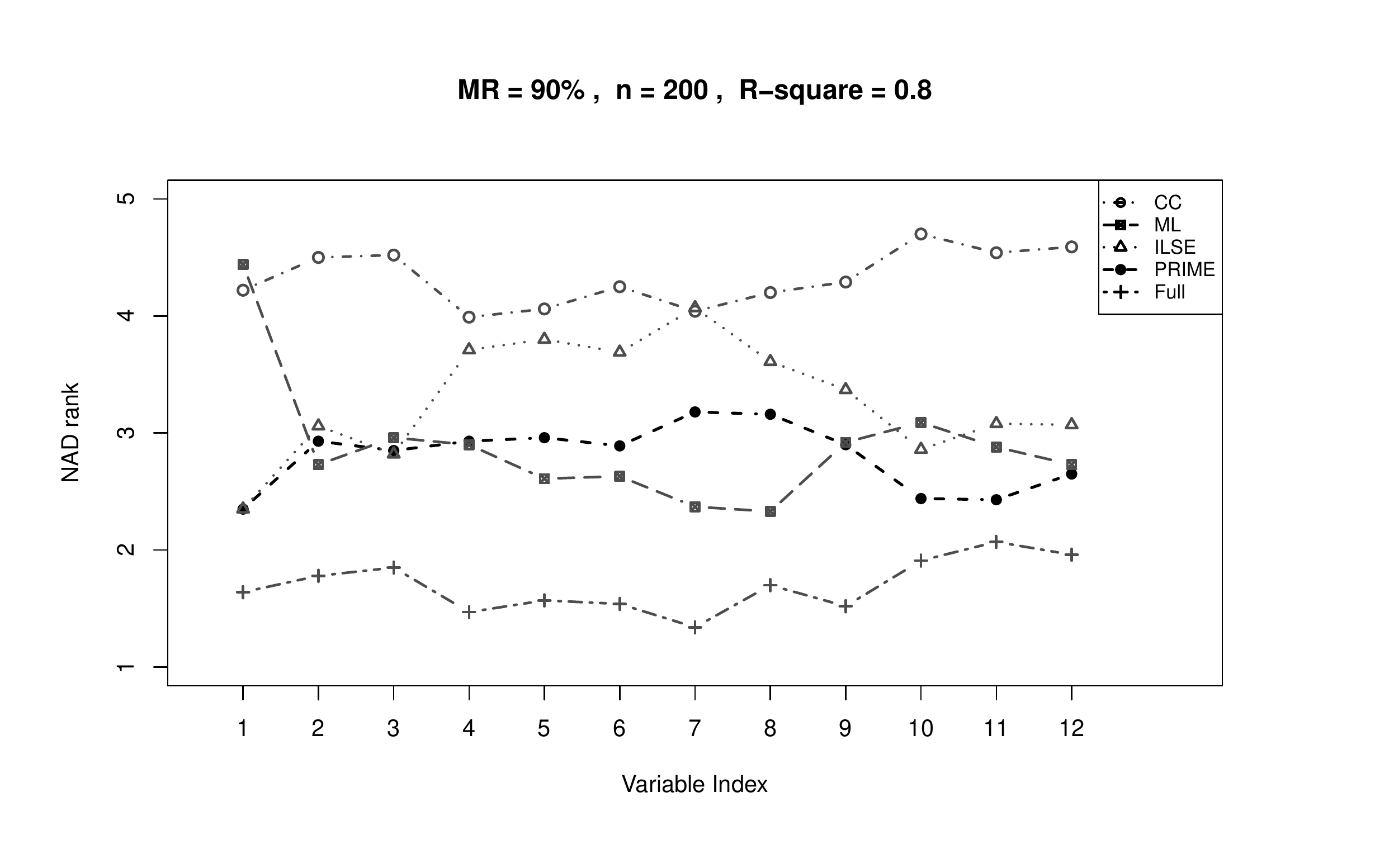}
		\end{minipage}
	}%
	\centering
	\caption{NAD with $n=200$ and 90\% missing data for different methods.}
	\label{fig:RNAD92}
\end{figure}

\begin{figure}[H]
	\centering
	\subfigure[]{
		\begin{minipage}[t]{0.33\linewidth}
			\centering
			\includegraphics[width=1.9in]{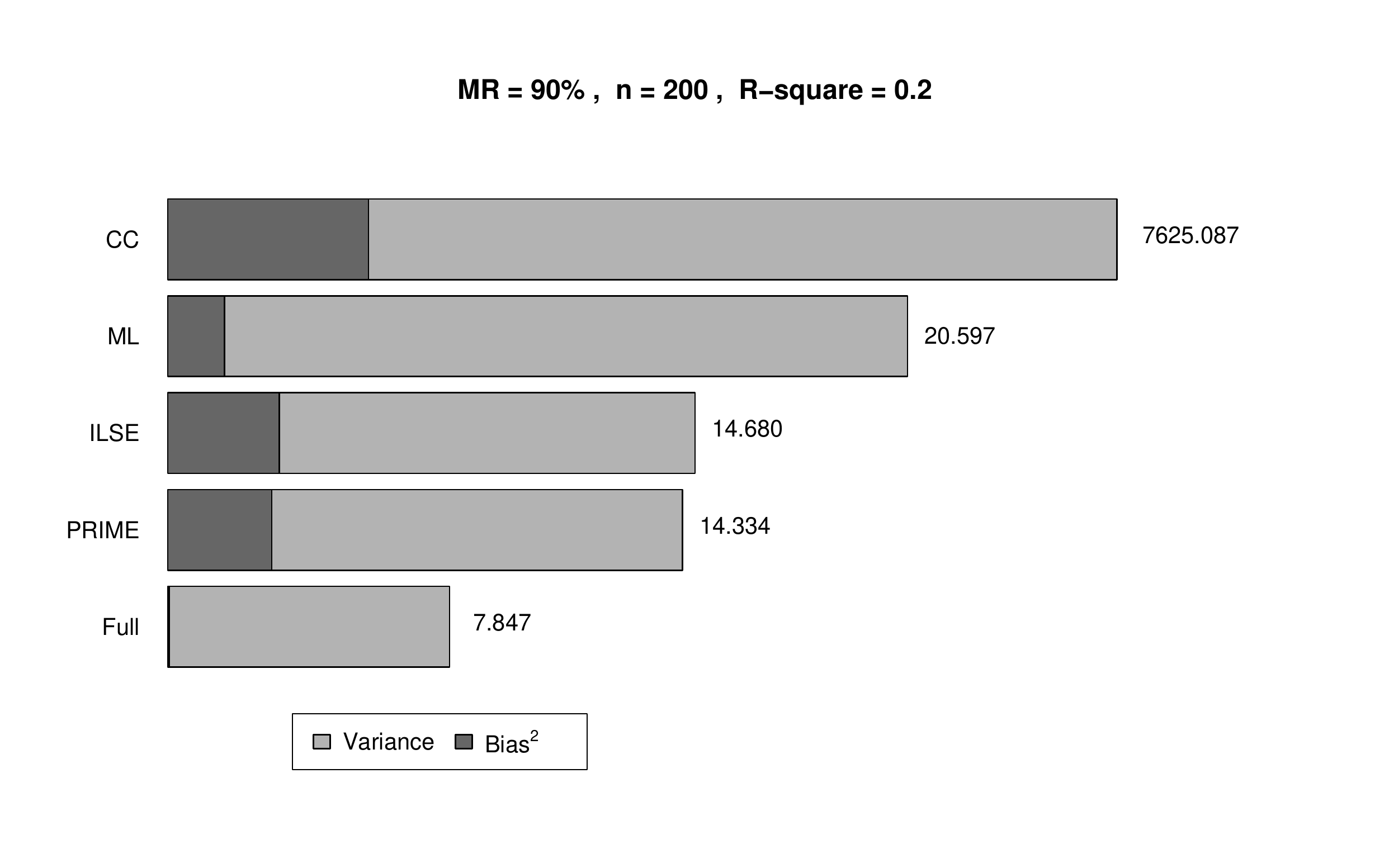}
		\end{minipage}%
	}%
	\subfigure[]{
		\begin{minipage}[t]{0.33\linewidth}
			\centering
			\includegraphics[width=1.9in]{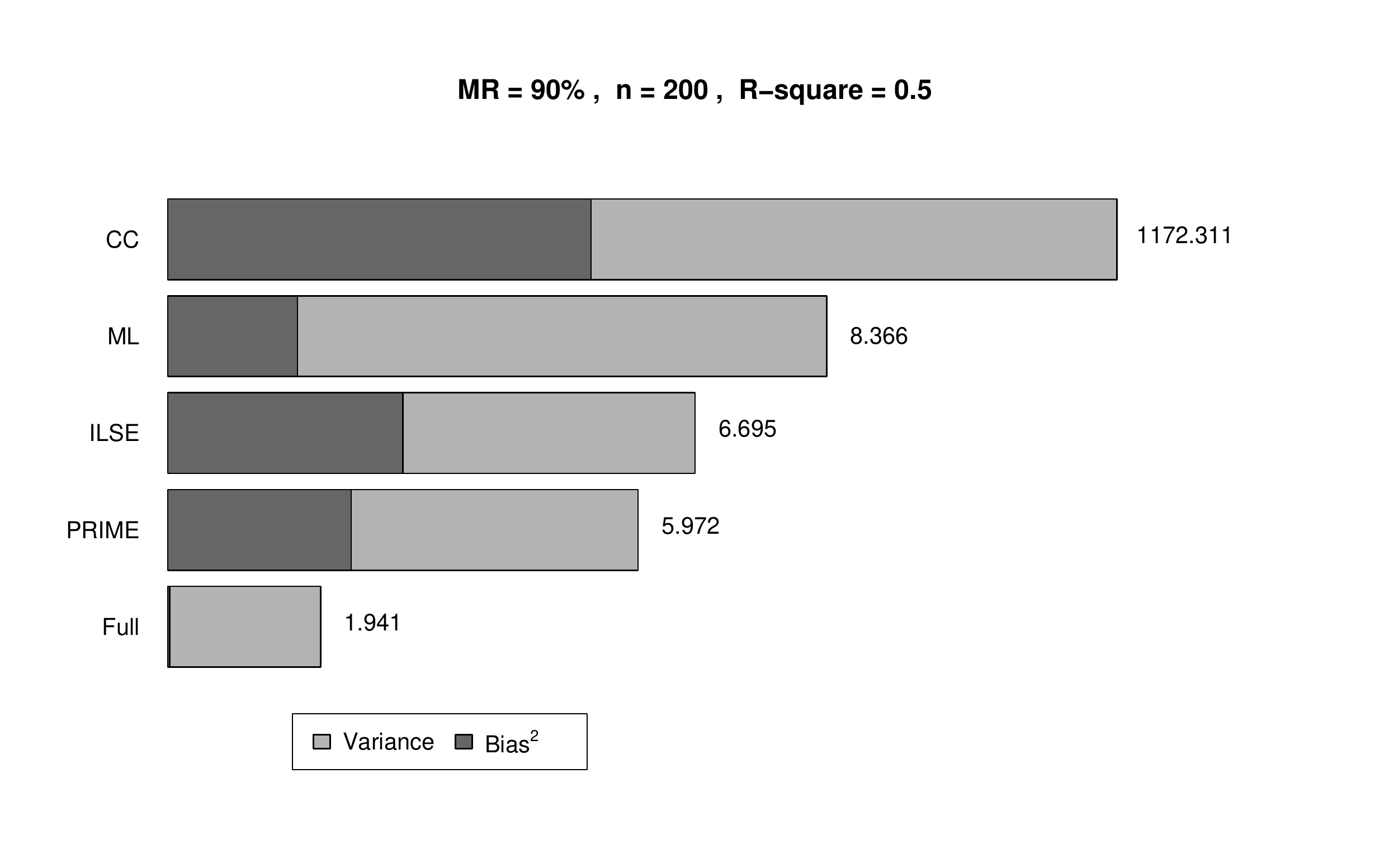}
		\end{minipage}
	}%
	\subfigure[]{
		\begin{minipage}[t]{0.33\linewidth}
			\centering
			\includegraphics[width=1.9in]{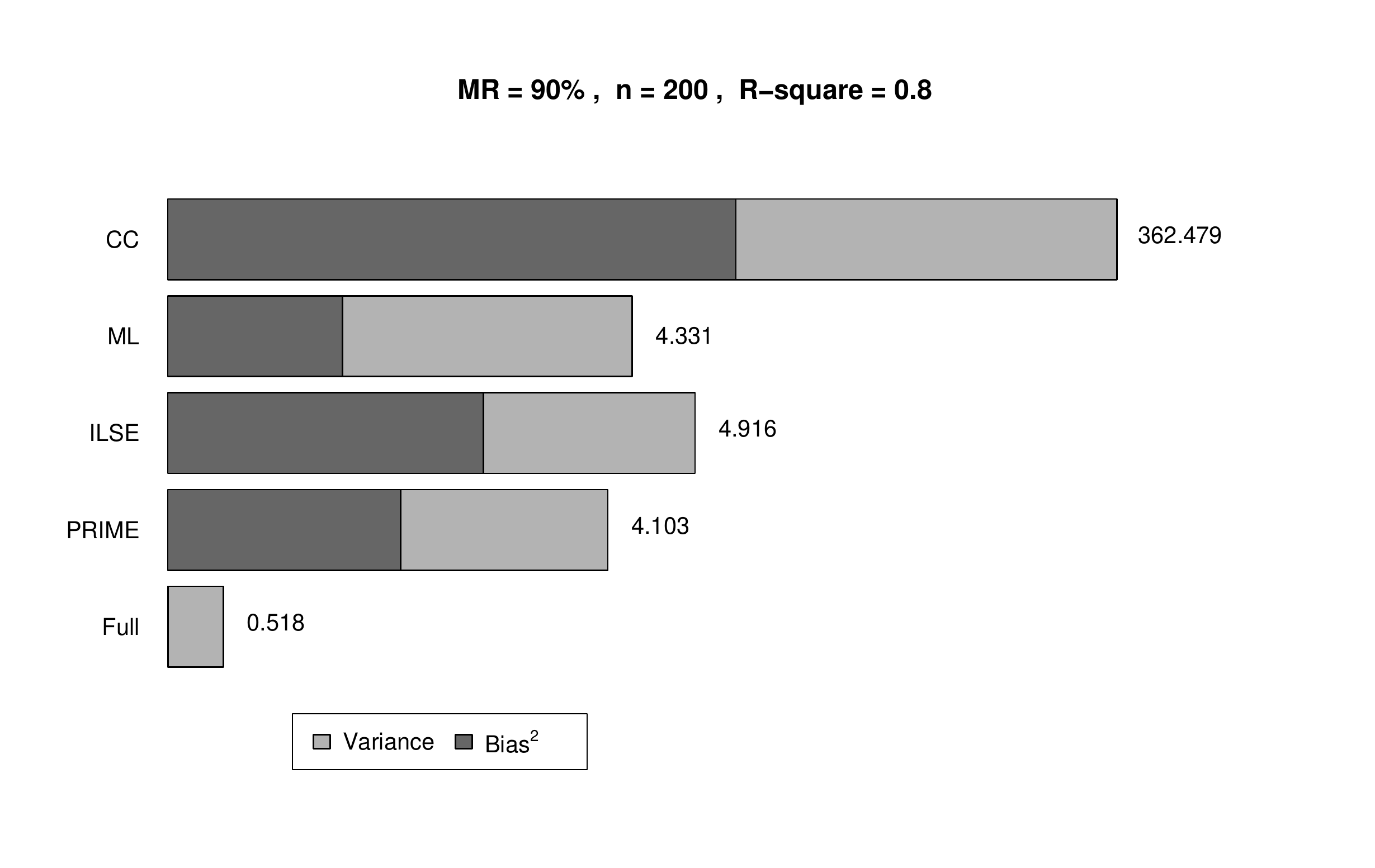}
		\end{minipage}
	}%
	\centering
	\caption{MSE with $n=200$ and 90\% missing data for different methods.}
	\label{fig:RMSE92}
\end{figure}

\begin{figure}[H]
	\begin{center}
		\includegraphics[width=12cm,height=7cm]{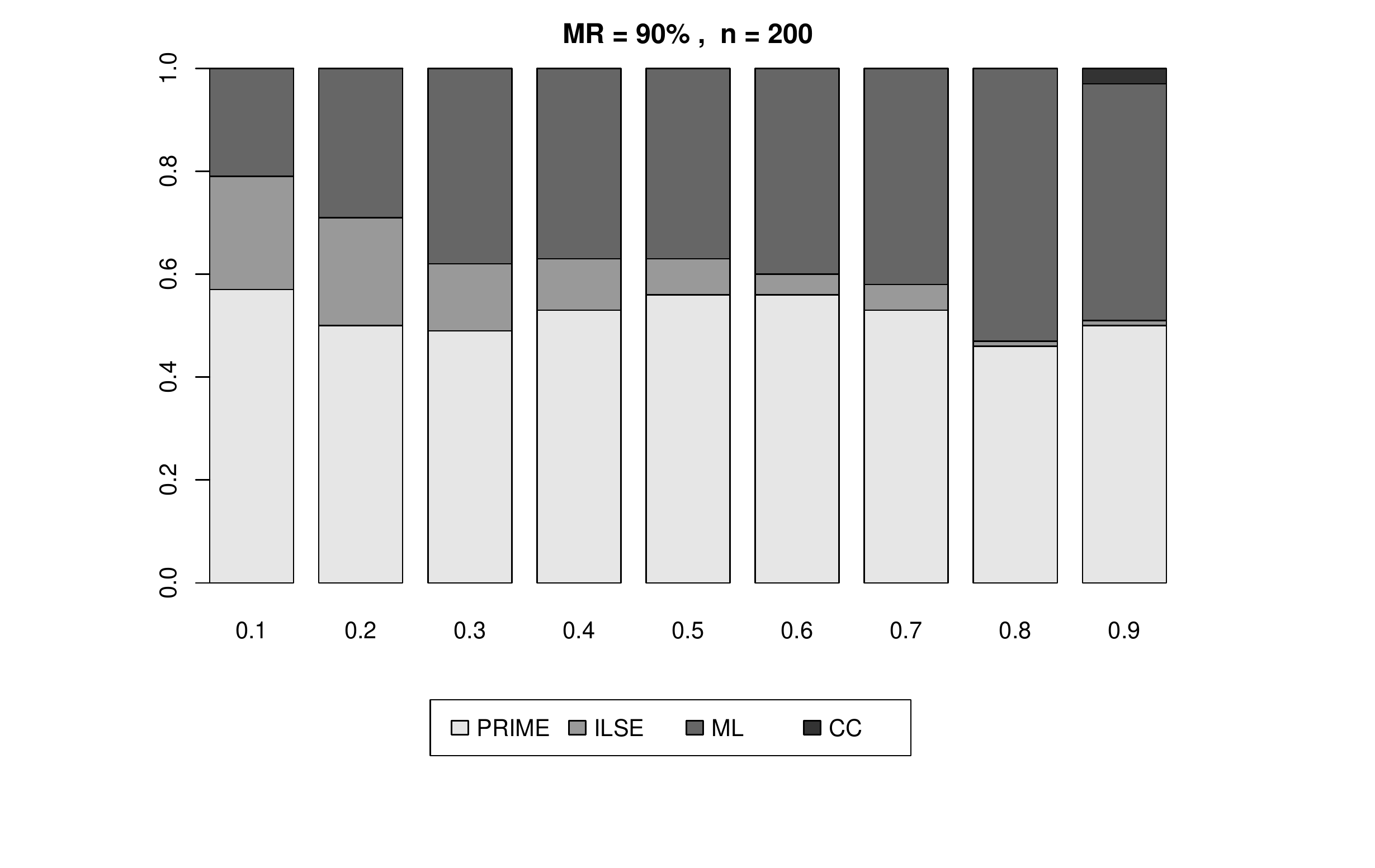}
	\end{center}
	\caption{Optimal rate of MSE with $n=200$ and 90\% missing data for different methods.}
	\label{fig:ORR92}
\end{figure}

\begin{figure}[H]
	\centering
	\subfigure[]{
		\begin{minipage}[t]{0.33\linewidth}
			\centering
			\includegraphics[width=1.9in]{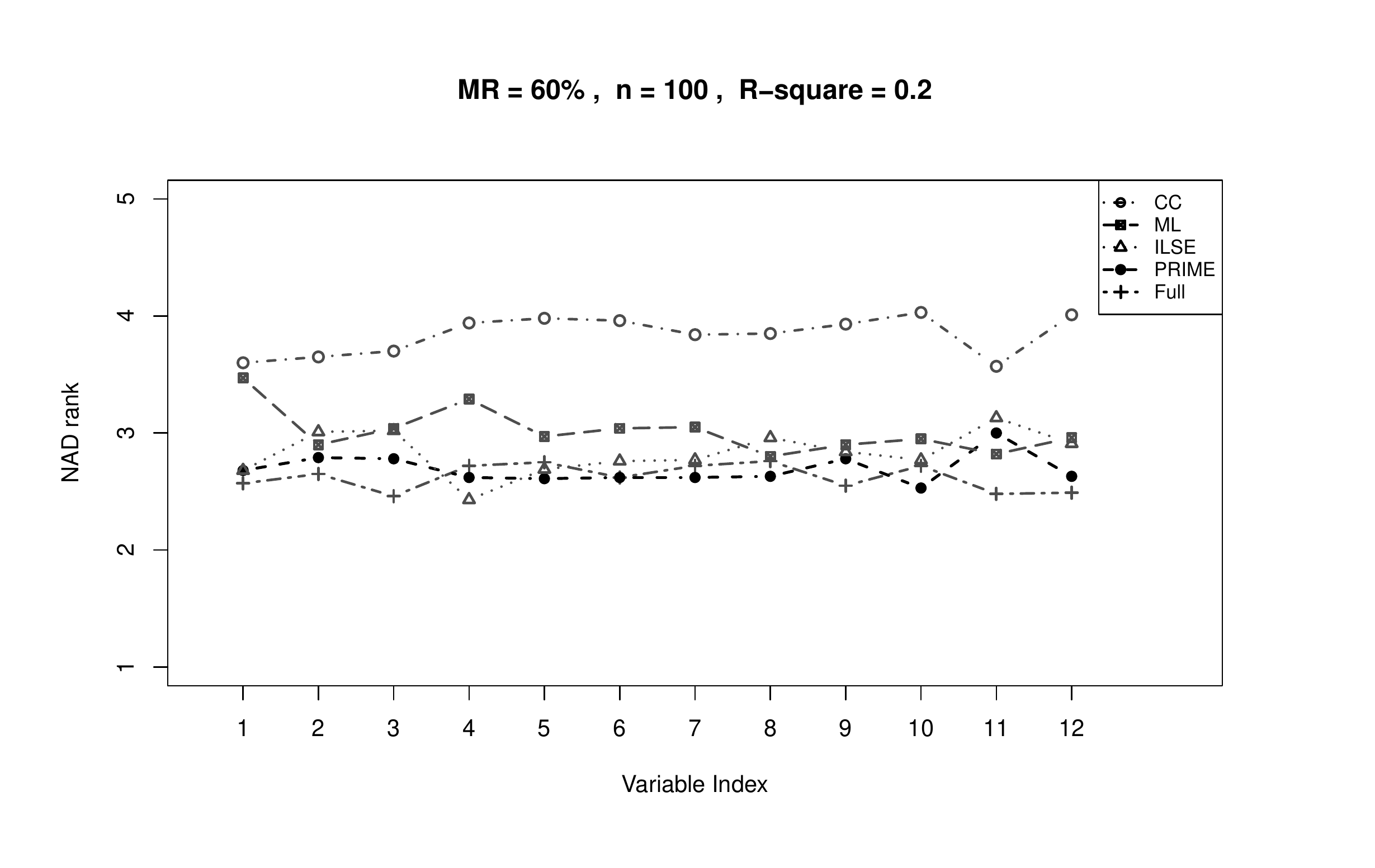}
		\end{minipage}%
	}%
	\subfigure[]{
		\begin{minipage}[t]{0.33\linewidth}
			\centering
			\includegraphics[width=1.9in]{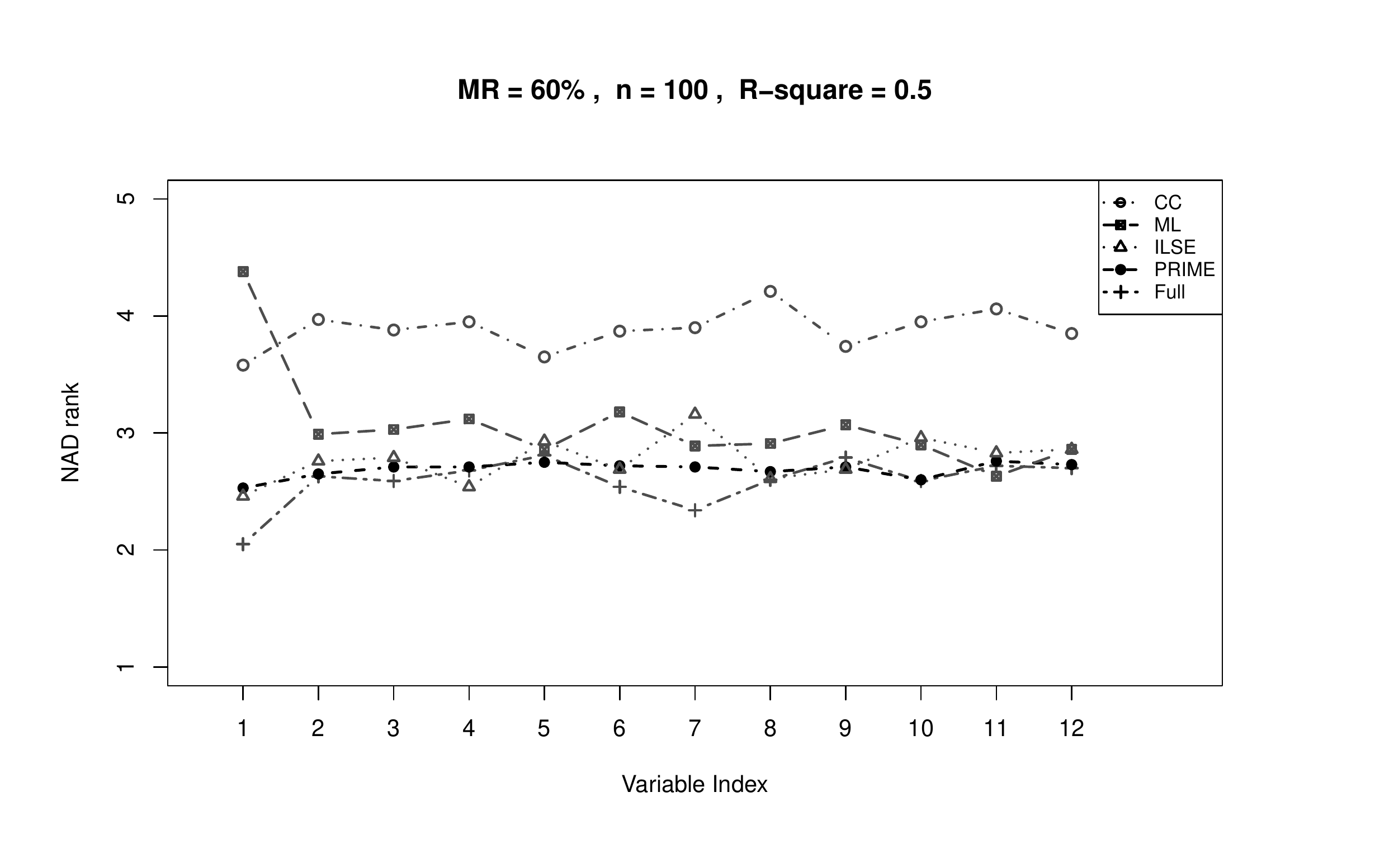}
		\end{minipage}
	}%
	\subfigure[]{
		\begin{minipage}[t]{0.33\linewidth}
			\centering
			\includegraphics[width=1.9in]{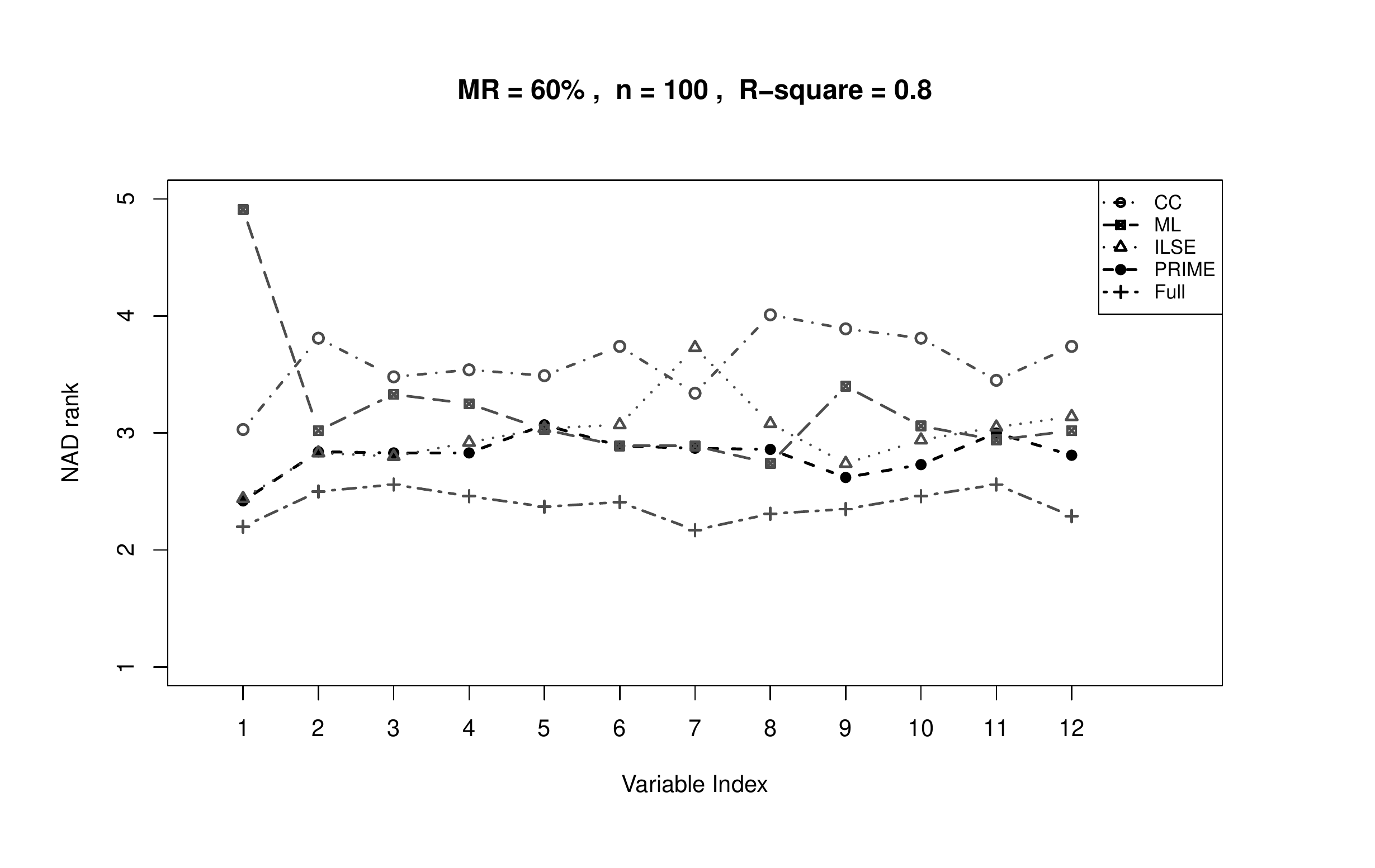}
		\end{minipage}
	}%
	\centering
	\caption{NAD with $n=100$ and 60\% missing data for different methods.}
	\label{fig:RNAD61}
\end{figure}

\begin{figure}[H]
	\centering
	\subfigure[]{
		\begin{minipage}[t]{0.33\linewidth}
			\centering
			\includegraphics[width=1.9in]{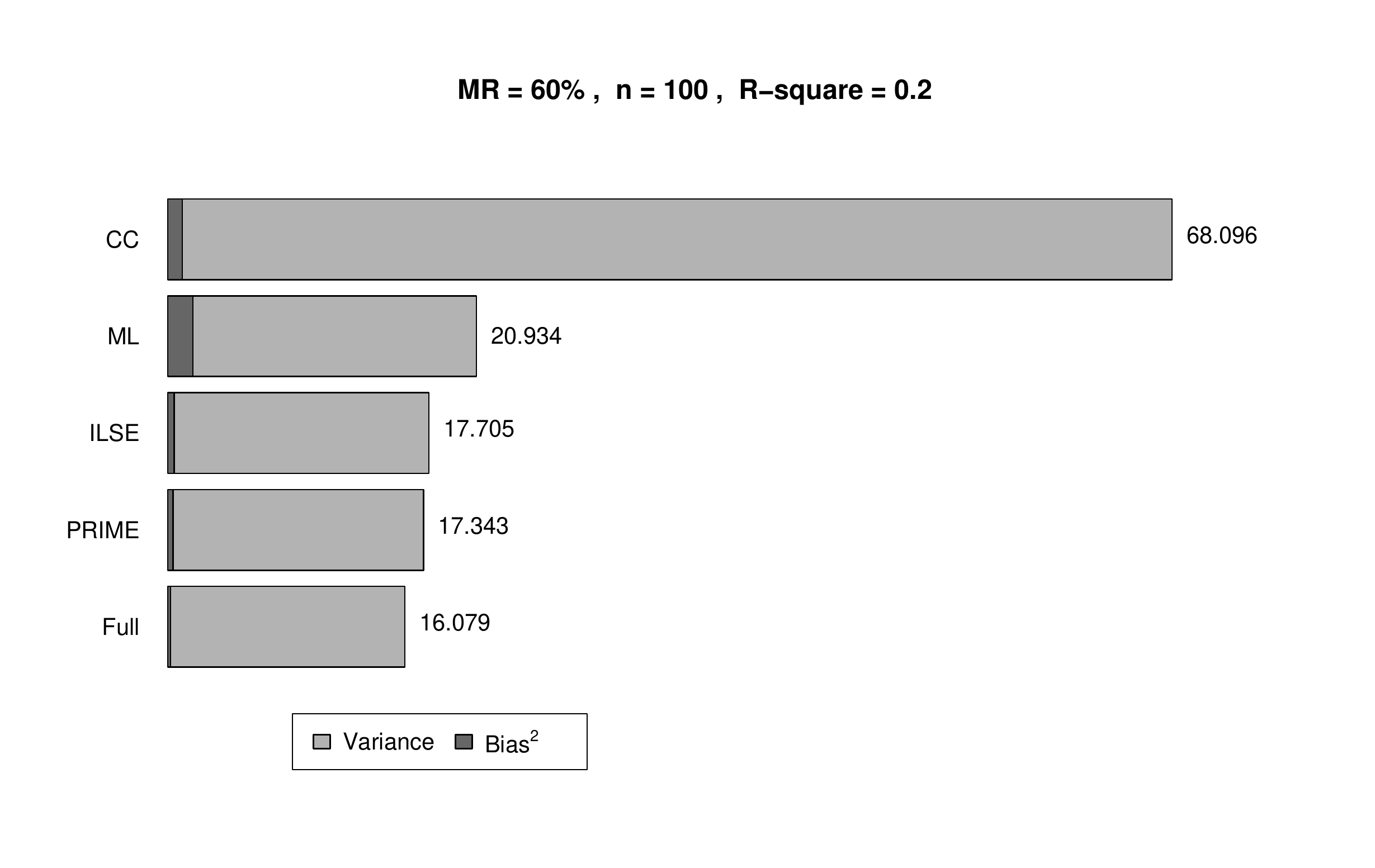}
		\end{minipage}%
	}%
	\subfigure[]{
		\begin{minipage}[t]{0.33\linewidth}
			\centering
			\includegraphics[width=1.9in]{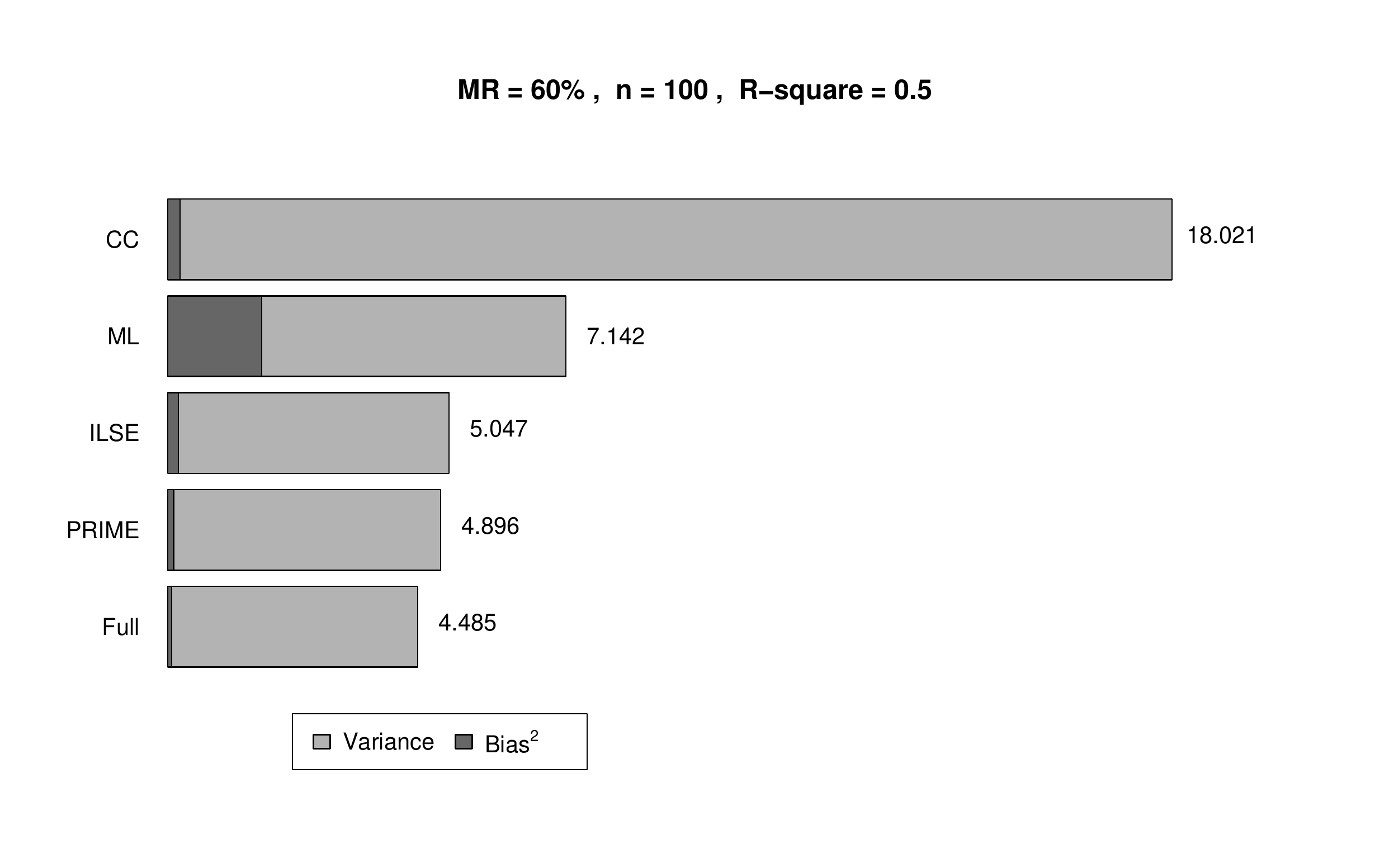}
		\end{minipage}
	}%
	\subfigure[]{
		\begin{minipage}[t]{0.33\linewidth}
			\centering
			\includegraphics[width=1.9in]{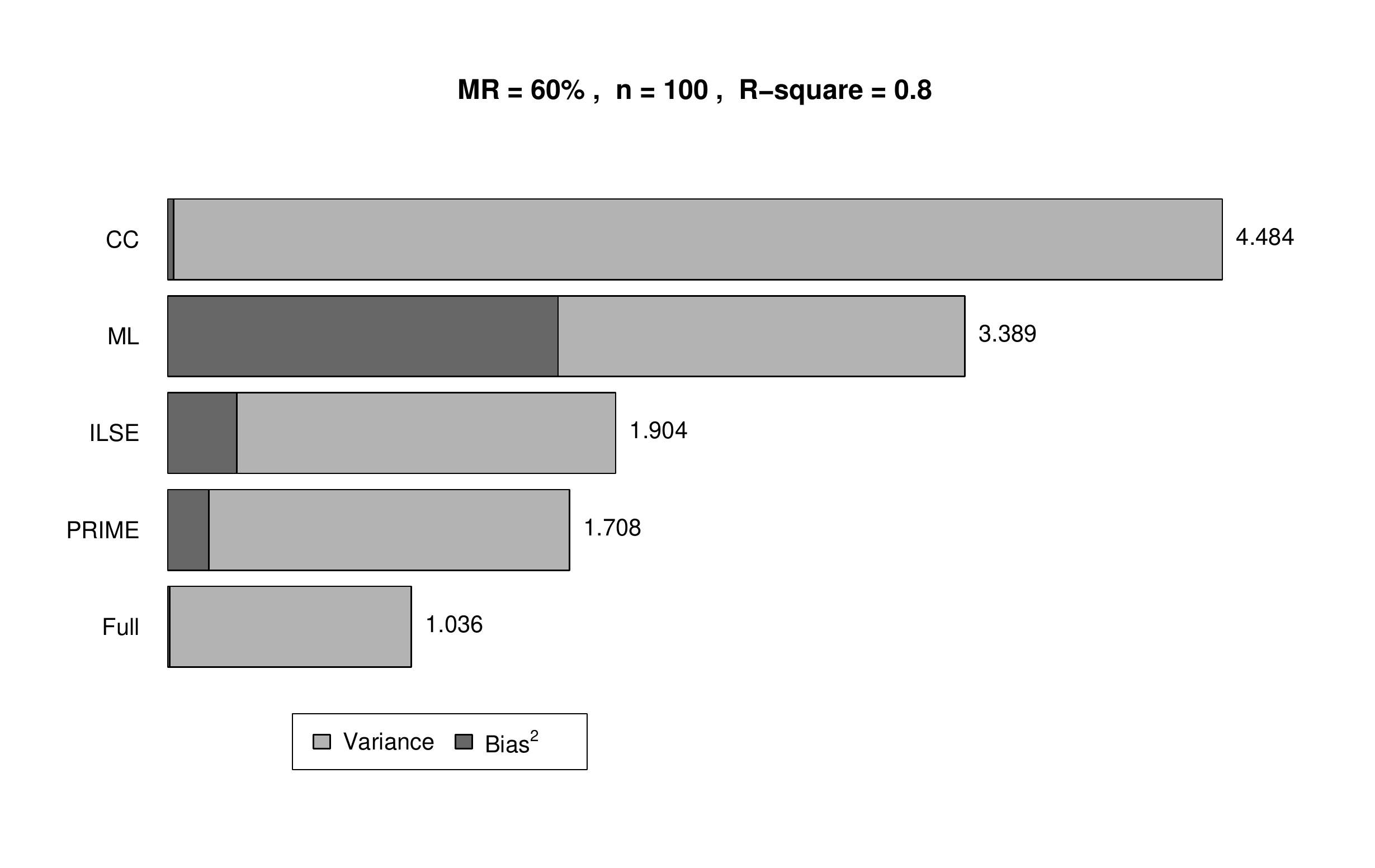}
		\end{minipage}
	}%
	\centering
	\caption{MSE with $n=100$ and 60\% missing data for different methods.}
	\label{fig:RMSE61}
\end{figure}

\begin{figure}[H]
	\begin{center}
		\includegraphics[width=12cm,height=7cm]{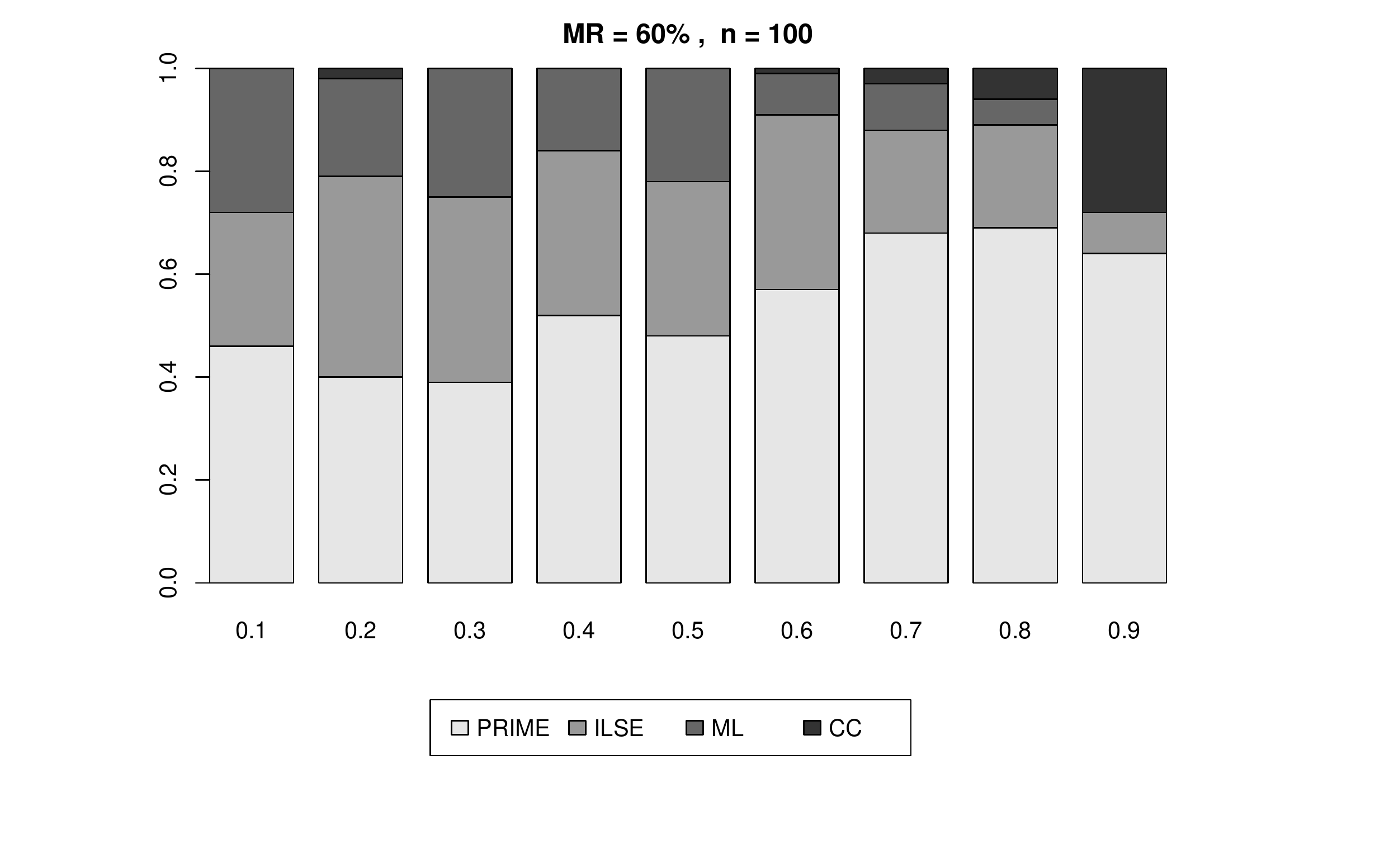}
	\end{center}
	\caption{Optimal rate of MSE with $n=100$ and 60\% missing data for different methods.}
	\label{fig:ORR61}
\end{figure}

\begin{figure}[H]
	\centering
	\subfigure[]{
		\begin{minipage}[t]{0.33\linewidth}
			\centering
			\includegraphics[width=1.9in]{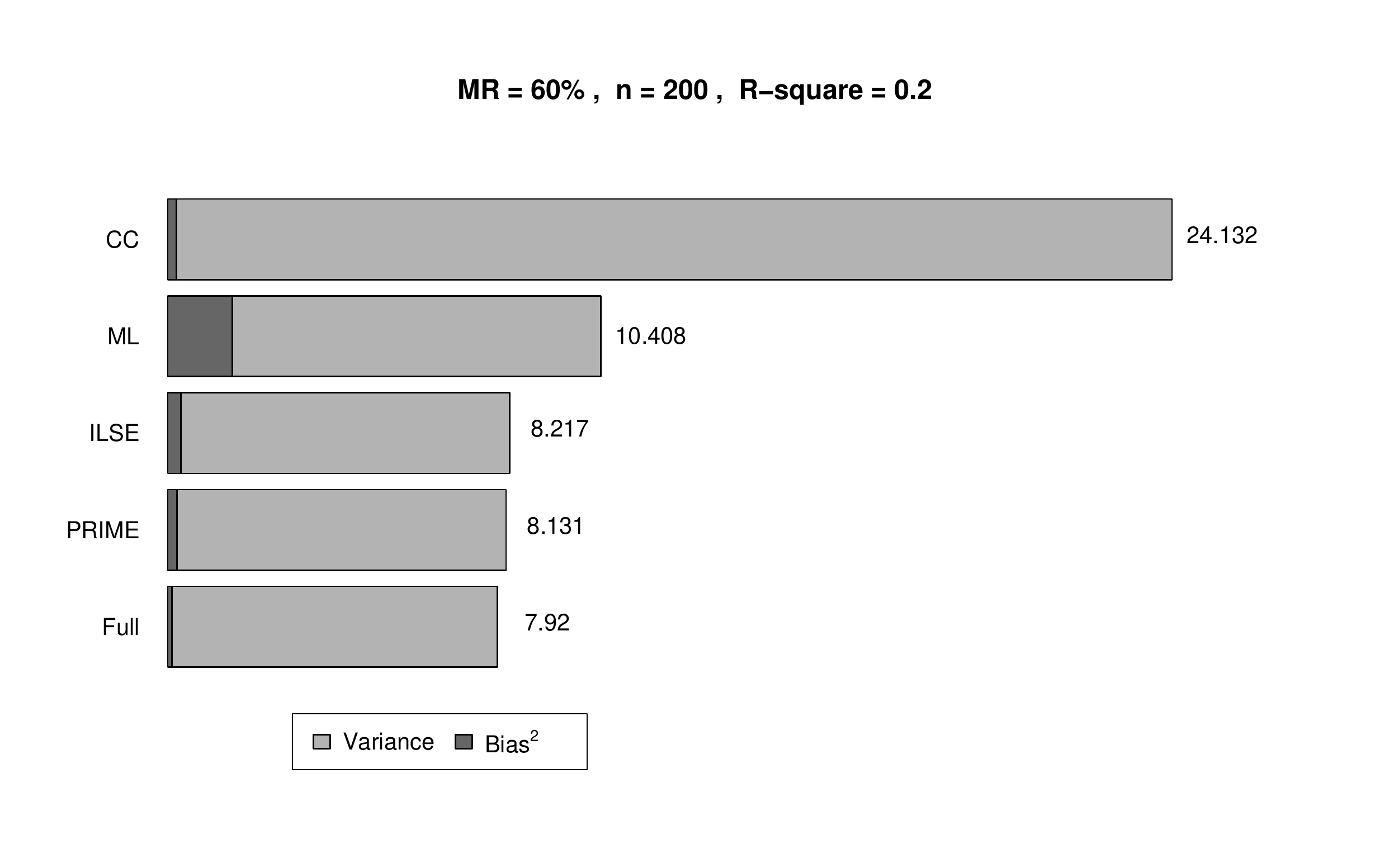}
		\end{minipage}%
	}%
	\subfigure[]{
		\begin{minipage}[t]{0.33\linewidth}
			\centering
			\includegraphics[width=1.9in]{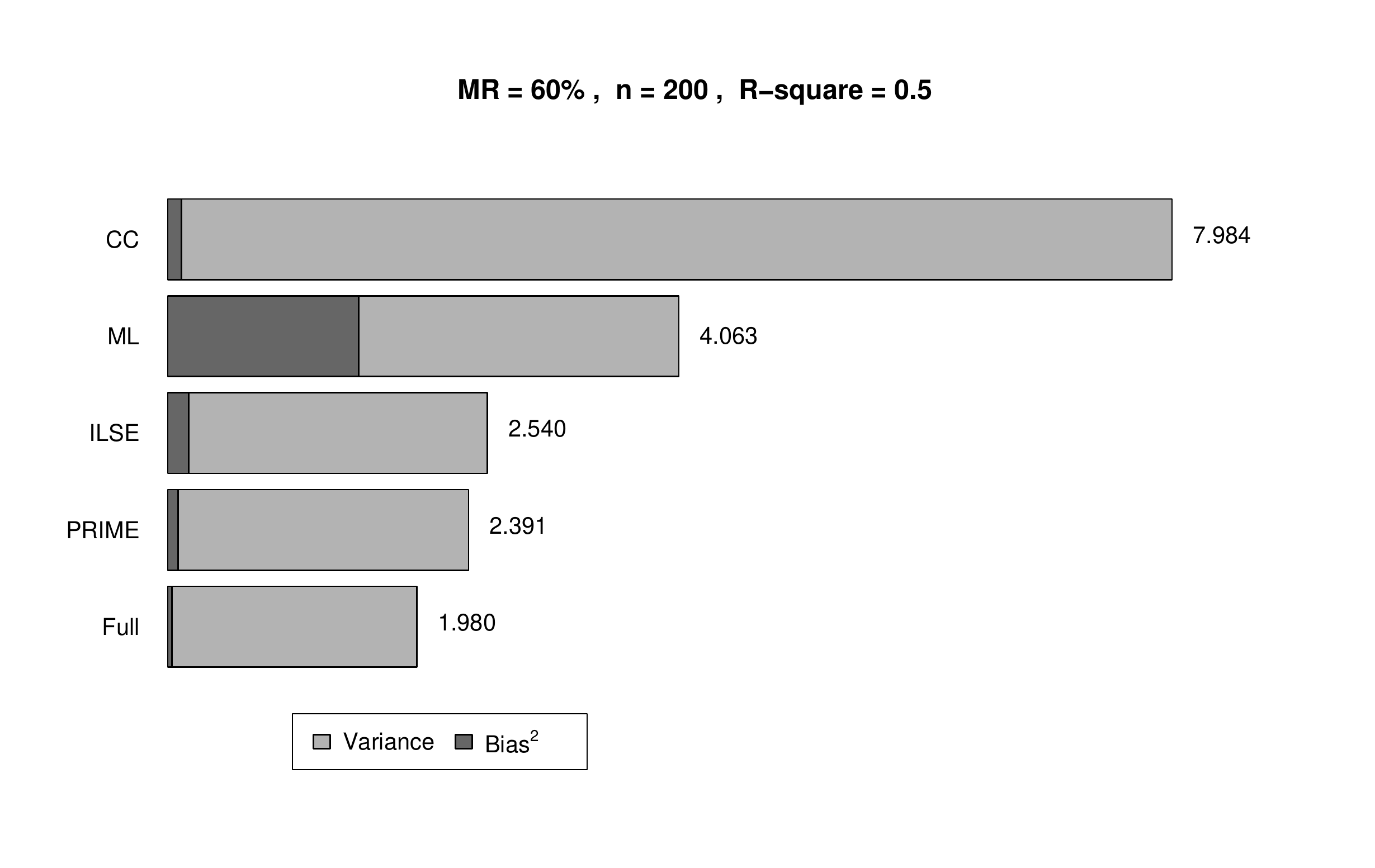}
		\end{minipage}
	}%
	\subfigure[]{
		\begin{minipage}[t]{0.33\linewidth}
			\centering
			\includegraphics[width=1.9in]{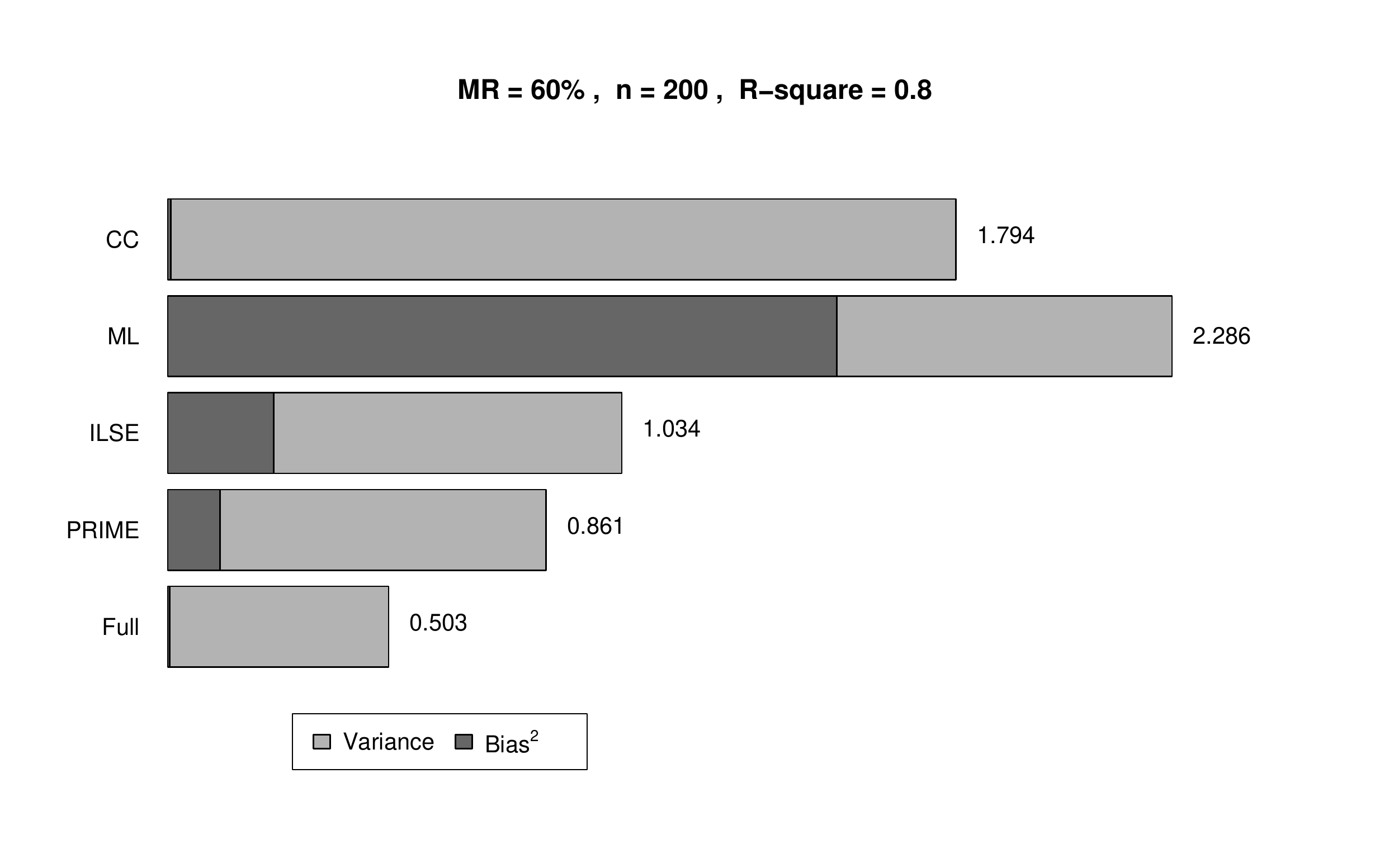}
		\end{minipage}
	}%
	\centering
	\caption{MSE with $n=200$ and 60\% missing data for different methods.}
	\label{fig:RMSE62}
\end{figure}

\begin{figure}[H]
	\begin{center}
		\includegraphics[width=12cm,height=7cm]{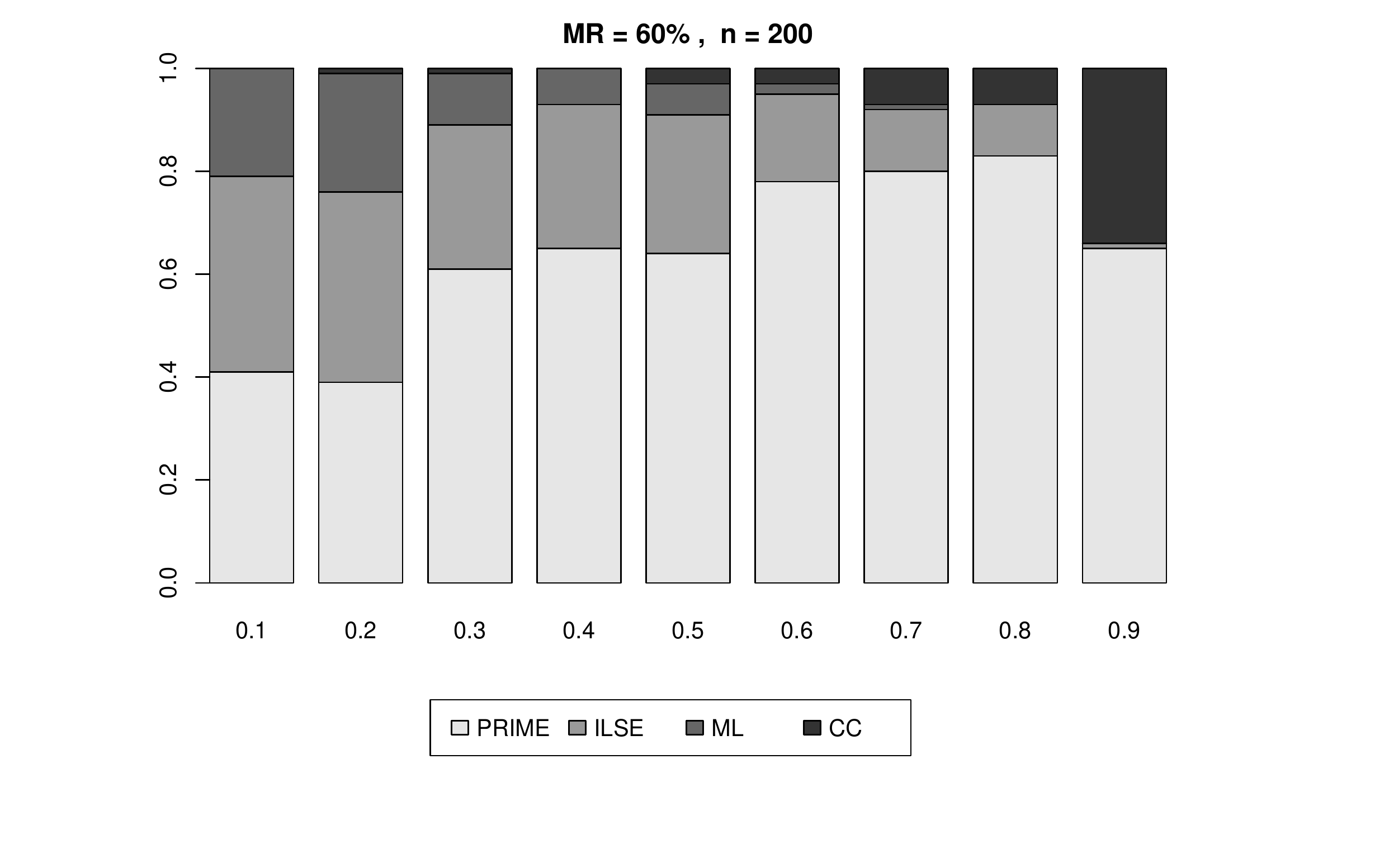}
	\end{center}
	\caption{Optimal rate of MSE with $n=200$ and 60\% missing data for different methods.}
	\label{fig:ORR62}
\end{figure}

The MSE results are shown in Figures \ref{fig:RMSE91}, \ref{fig:RMSE92}, \ref{fig:RMSE61}, and \ref{fig:RMSE62} for $R^2=0.2,0.5,0.8$. The results of optimal rate of MSE are displayed in Figures \ref{fig:ORR91}, \ref{fig:ORR92}, \ref{fig:ORR61}, and \ref{fig:ORR62}. To determine the relative performance, we rank the NADs of the five methods at each repetition. The mean NAD ranks are displayed in Figures \ref{fig:RNAD91}, \ref{fig:RNAD92}, \ref{fig:RNAD61}, and \ref{fig:RNAD62} for $R^2=0.2,0.5,0.8$. The results of the cases not shown here are available in the supplementary materials. In general, they exhibit patterns which is similar to those shown here.

To make the comparison of Full, PRIME, and ILSE easier, the MSE bar plots for CC and ML are manually scaled because these two methods lead to much higher MSEs than the other methods. The main conclusions are as follows:

\begin{enumerate}[1.]
	\item When $n$ and $R^2$ increase, the MSEs of Full, ILSE, and PRIME generally decrease, as expected. However, the relative performance of these three methods does not change.
	
	\item Generally, PRIME outperformes ILSE and ML in terms of NAD, and all three significantly outperforme CC. Surprisingly, PRIME was close or even superior to the Full method in estimating the coefficients of $X_{11}$ and $X_{12}$ when $\text{MR}=90\%, n=100,$ and $R^2=0.2$ and in estimating the coefficients of $X_{10}$ and $X_{12}$ when $\text{MR}=90\%, n=200,$ and $R^2=0.2$. These results confirm the superiority of the PRIME method.
	
	\item The comparison of results shows that the proposed PRIME method performed better than the other three methods (ILSE, CC, and ML). The bias and variance decomposition figures show that ILSE produces more biased estimates than PRIME. The CC method?s estimation has extremely high variance in almost all ranges of $R^2$, and CC?s approximation error was larger when $R^2$ was not very high. Furthermore, the CC method produces biased estimates, as expected, because of the missing-data mechanism. The ML performance is not stable: ML's performance is close to our proposed method?s when $n=200, \text{MR}=90\%,$ and $R^2=0.8$, but ML has the poorest performance when $n=100, \text{MR}=90\%,$ and $R^2=0.8$ because ML estimators can be highly biased when the MAR assumption does not hold.
	
	\item The optimal rates of MSE show a proportion over $40\%$ for PRIME, and this indicates that PRIME yields the smallest MSE of all the competitors in more than 40\% of the trials in Scenario 1. When PRIME yields the optimal rate, ILSE or ML most often yields the second-smallest MSE. When PRIME does not yield the lowest MSE, ILSE and ML most often do. 
	Not surprisingly, CC rarely produces the smallest MSE except when $\text{MR}=60\%$ and $R^2=0.9$.
	
\end{enumerate}

\begin{figure}[H]
	\centering
	\subfigure[]{
		\begin{minipage}[t]{0.33\linewidth}
			\centering
			\includegraphics[width=1.9in]{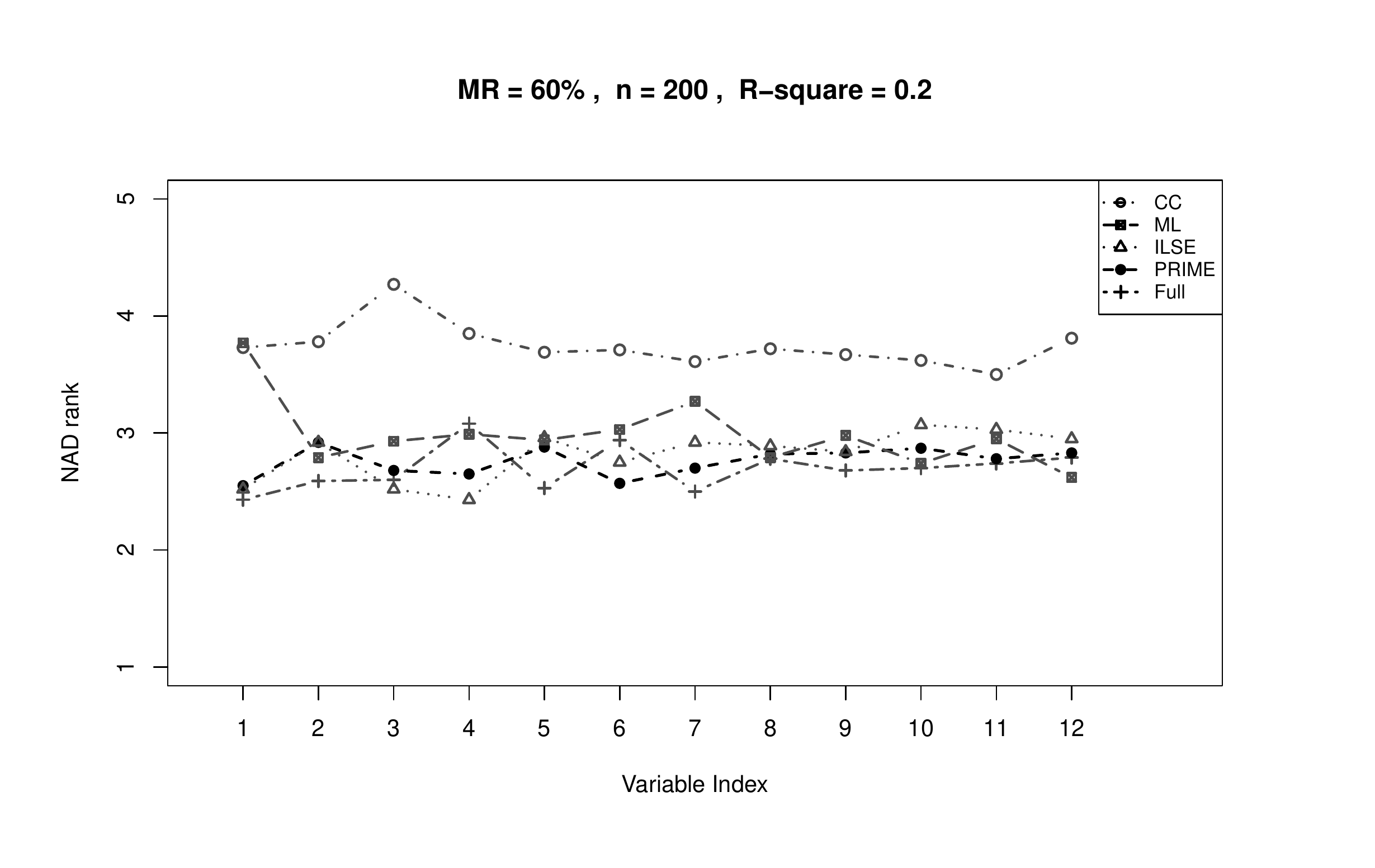}
		\end{minipage}%
	}%
	\subfigure[]{
		\begin{minipage}[t]{0.33\linewidth}
			\centering
			\includegraphics[width=1.9in]{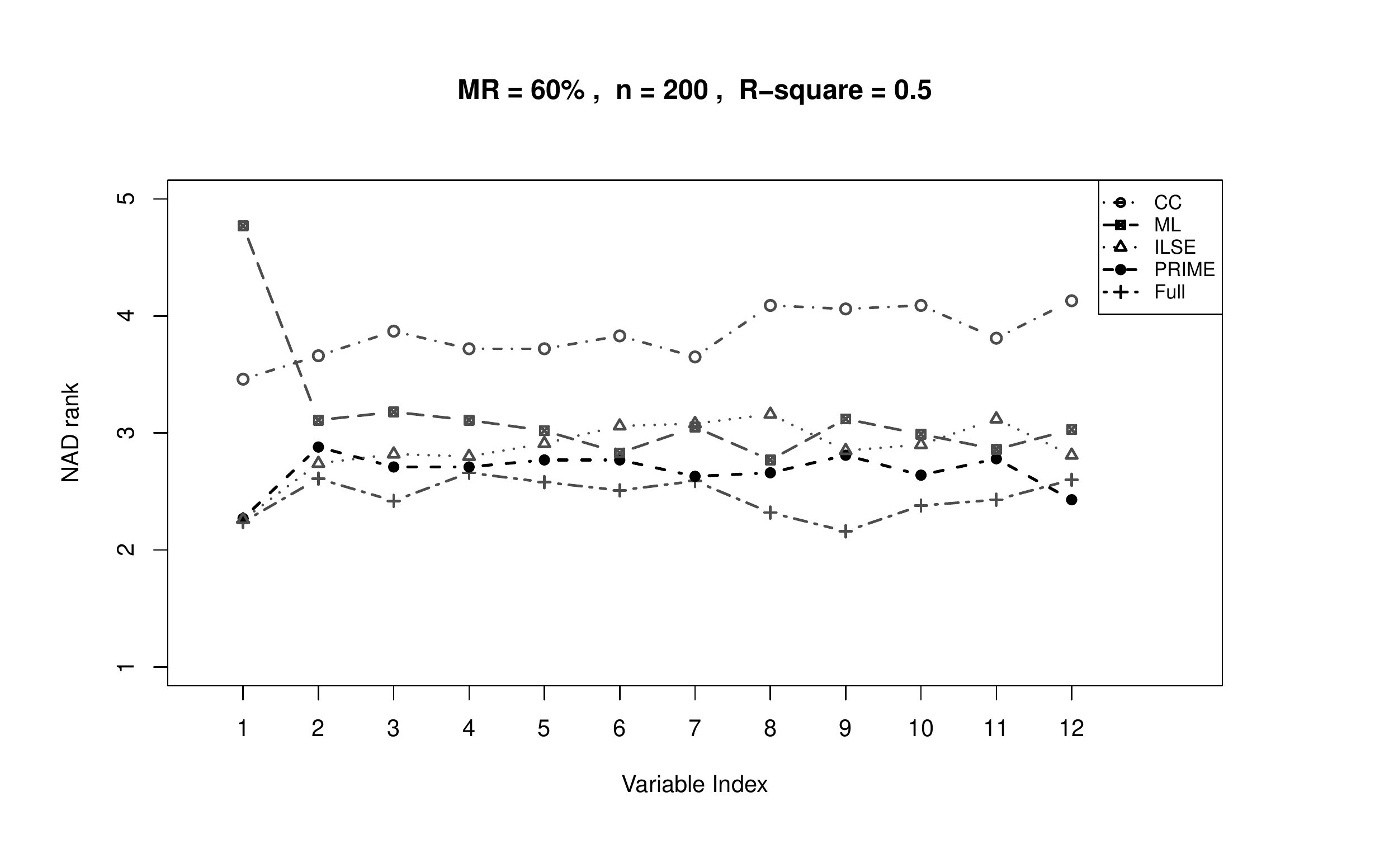}
		\end{minipage}
	}%
	\subfigure[]{
		\begin{minipage}[t]{0.33\linewidth}
			\centering
			\includegraphics[width=1.9in]{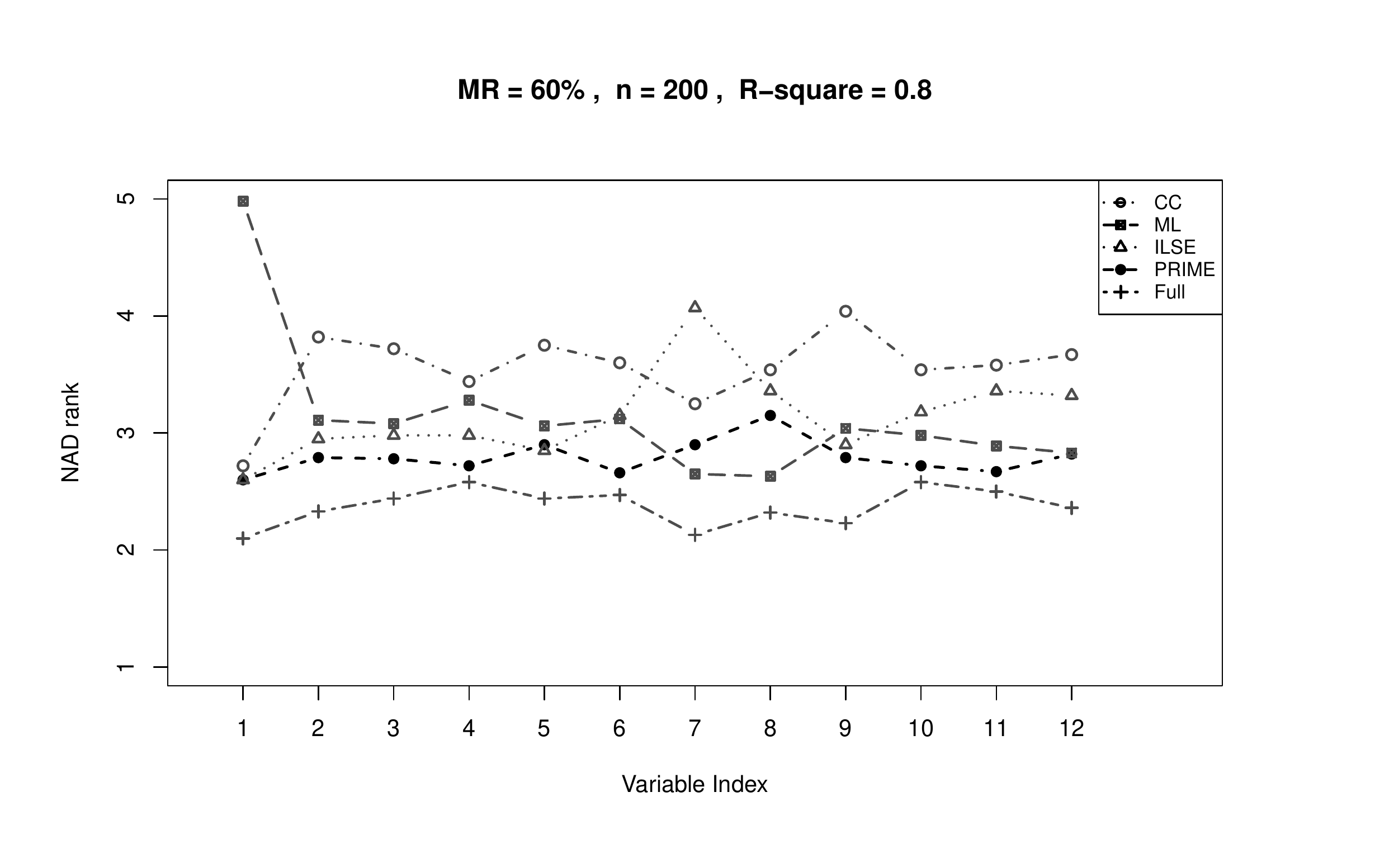}
		\end{minipage}
	}%
	\centering
	\caption{NAD with $n=200$ and 60\% missing data for different methods.}
	\label{fig:RNAD62}
\end{figure}

\subsection{Scenario 2: Varying correlation between variables}
To compare the methods with different correlations, we consider the correlation between $X_i $ and $X_j$ in four situations: $p_1$ for $\rho_{ij}=0.2$, $p_2$ for $\rho_{ij}=0.5$, $p_3$ for $\rho_{ij}=0.8$ and $p_4$ for $\rho_{ij}=0.8^{|i-j|}$. Here, we set $\sigma^2$ with $R^2=0.7$. 
All other aspects remain the same as in Scenario 1. For the missing rate, two settings are considered, where

\begin{itemize}
	\item $(a, b, c)=(0.1,-2,-1)$, so that the missing rate is approximately 60\%, 
	\item $(a, b, c)=(0.7,-3.5,-4)$, so that the missing rate is approximately 90\%.
\end{itemize}

\begin{figure}[H]
	\centering
	\subfigure[]{
		\begin{minipage}[t]{0.33\linewidth}
			\centering
			\includegraphics[width=1.9in]{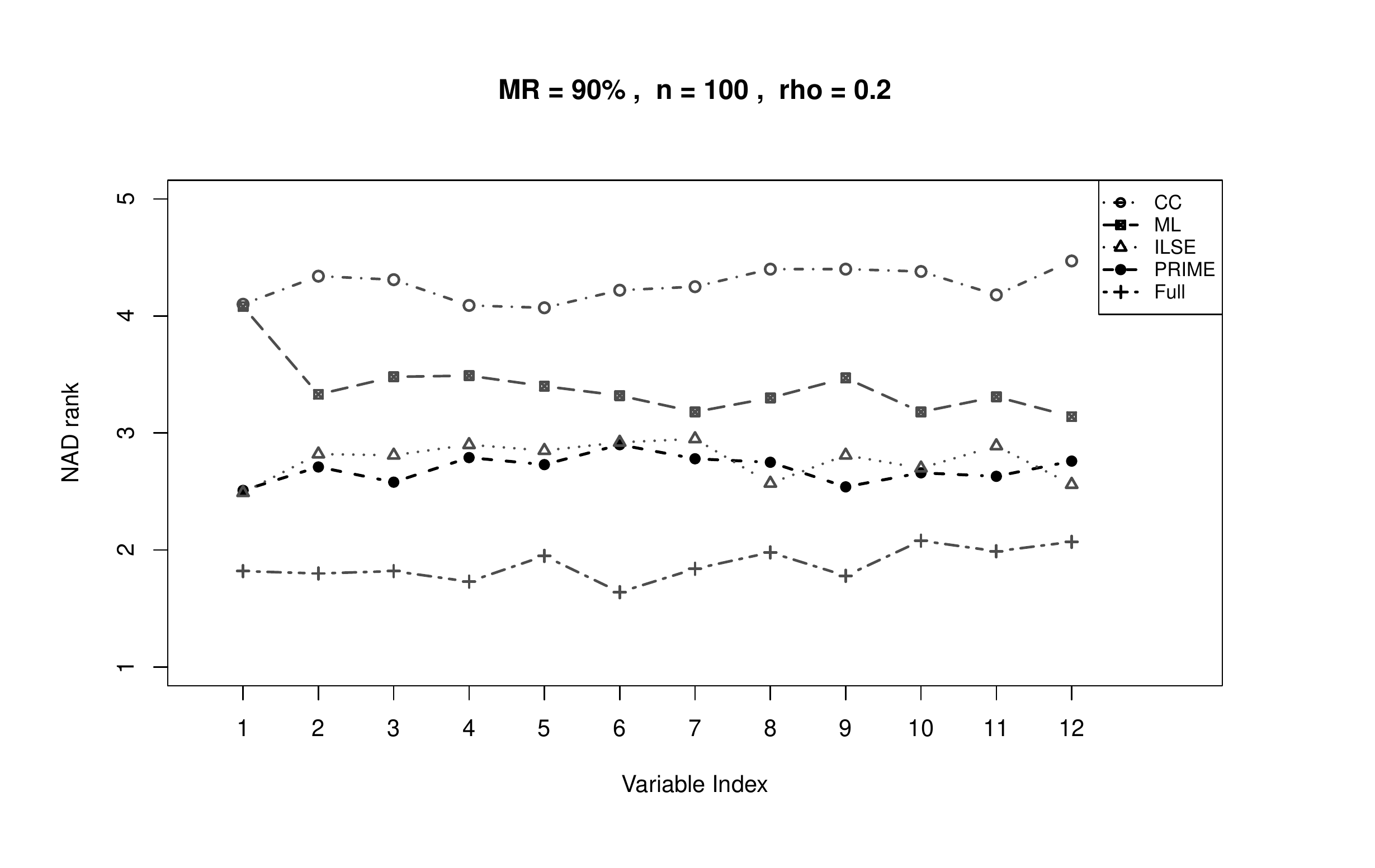}
		\end{minipage}%
	}%
	\subfigure[]{
		\begin{minipage}[t]{0.33\linewidth}
			\centering
			\includegraphics[width=1.9in]{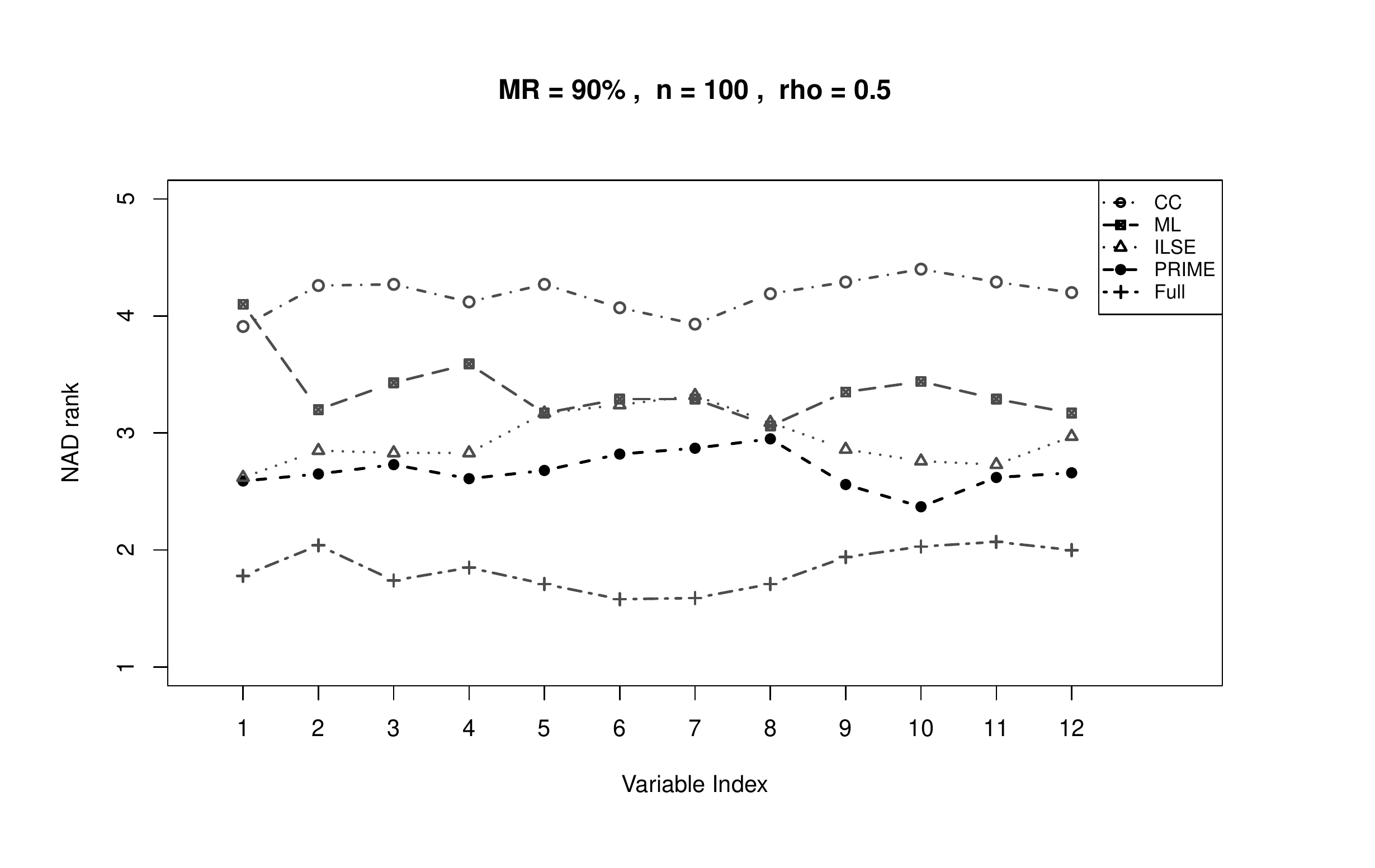}
		\end{minipage}
	}%
	\subfigure[]{
		\begin{minipage}[t]{0.33\linewidth}
			\centering
			\includegraphics[width=1.9in]{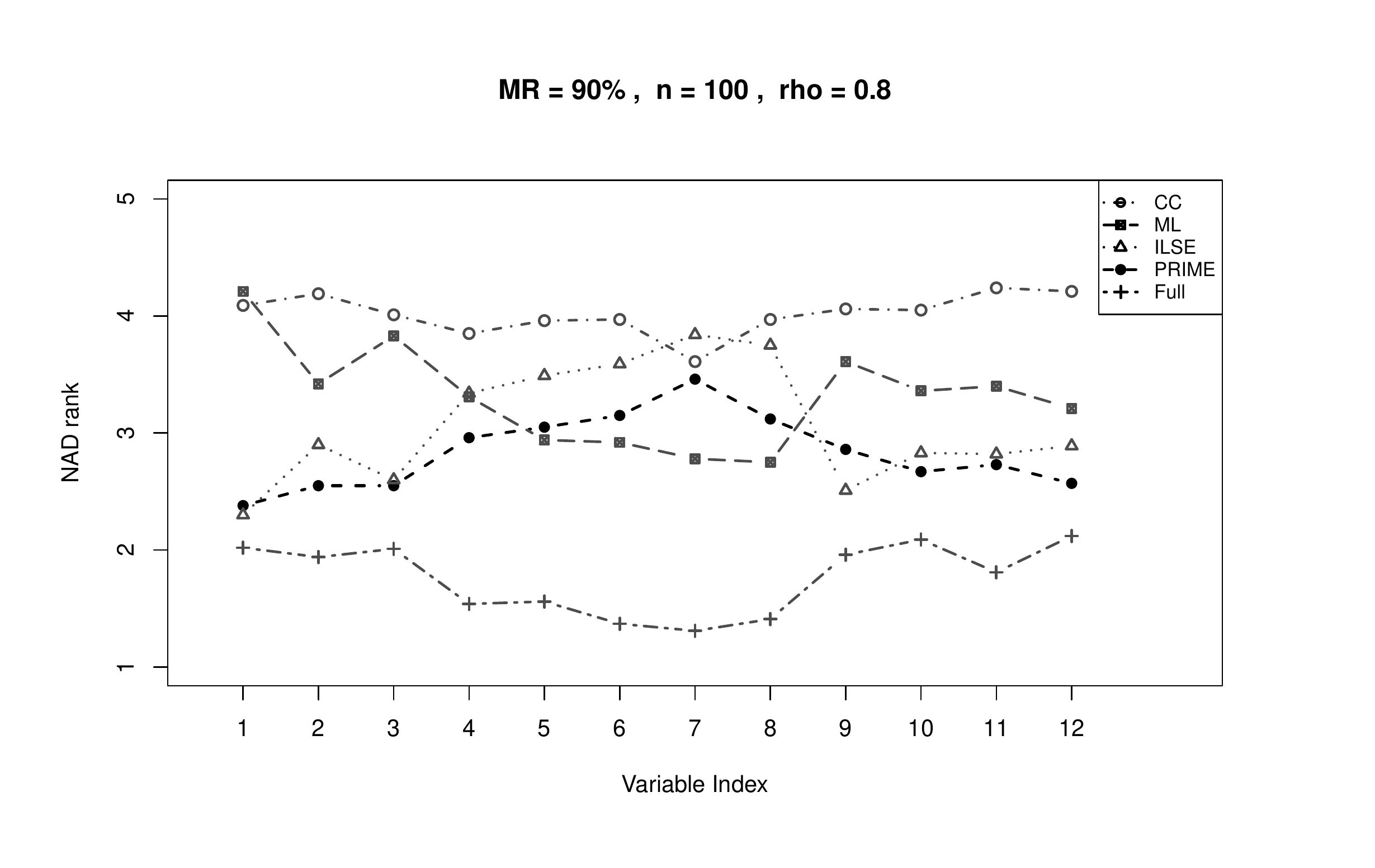}
		\end{minipage}
	}%
	\centering
	\caption{NAD with $n=100$ and 90\% missing data for different methods.}
	\label{fig:RHONAD91}
\end{figure}

\begin{figure}[H]
	\centering
	\subfigure[]{
		\begin{minipage}[t]{0.33\linewidth}
			\centering
			\includegraphics[width=1.9in]{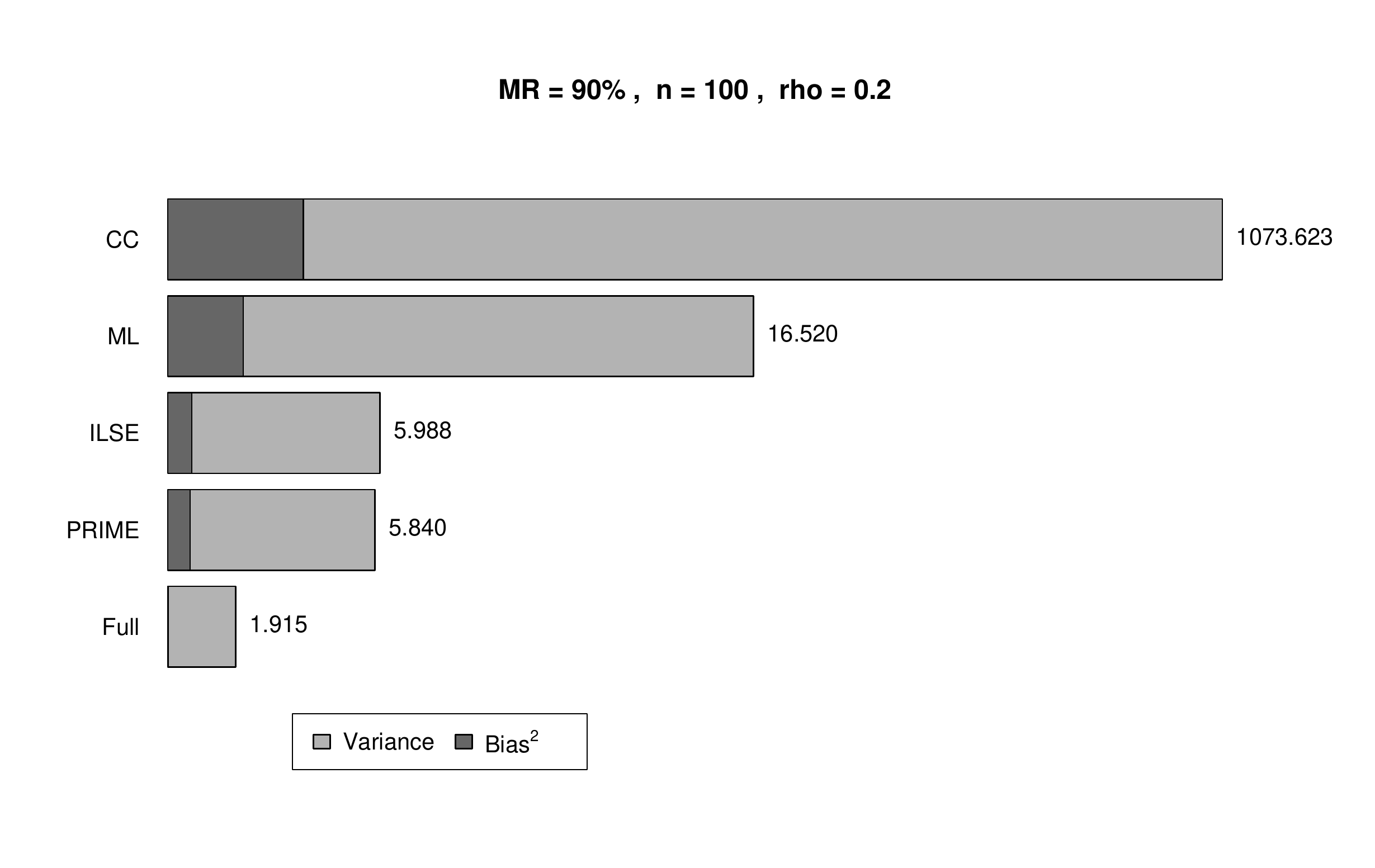}
		\end{minipage}%
	}%
	\subfigure[]{
		\begin{minipage}[t]{0.33\linewidth}
			\centering
			\includegraphics[width=1.9in]{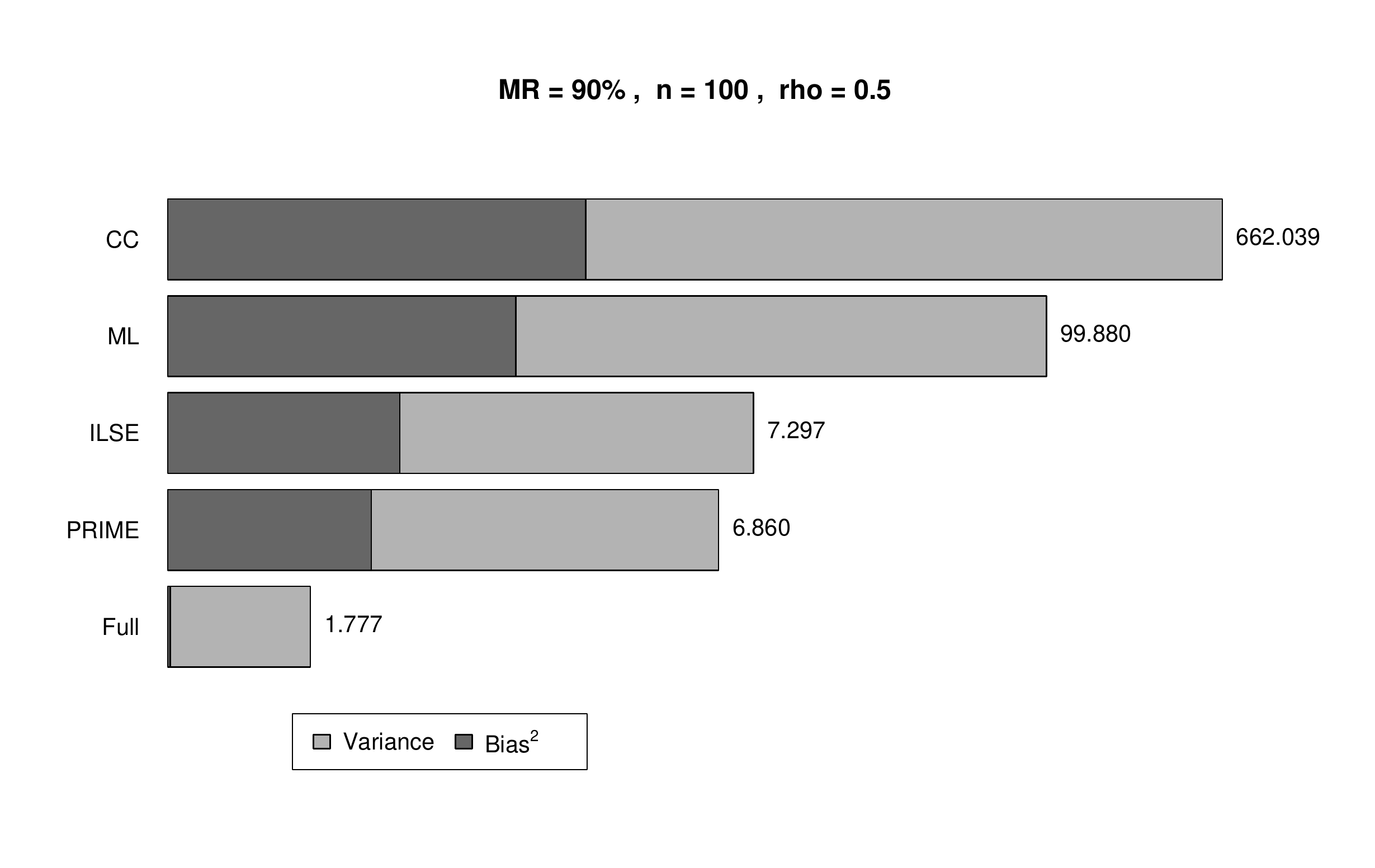}
		\end{minipage}
	}%
	\subfigure[]{
		\begin{minipage}[t]{0.33\linewidth}
			\centering
			\includegraphics[width=1.9in]{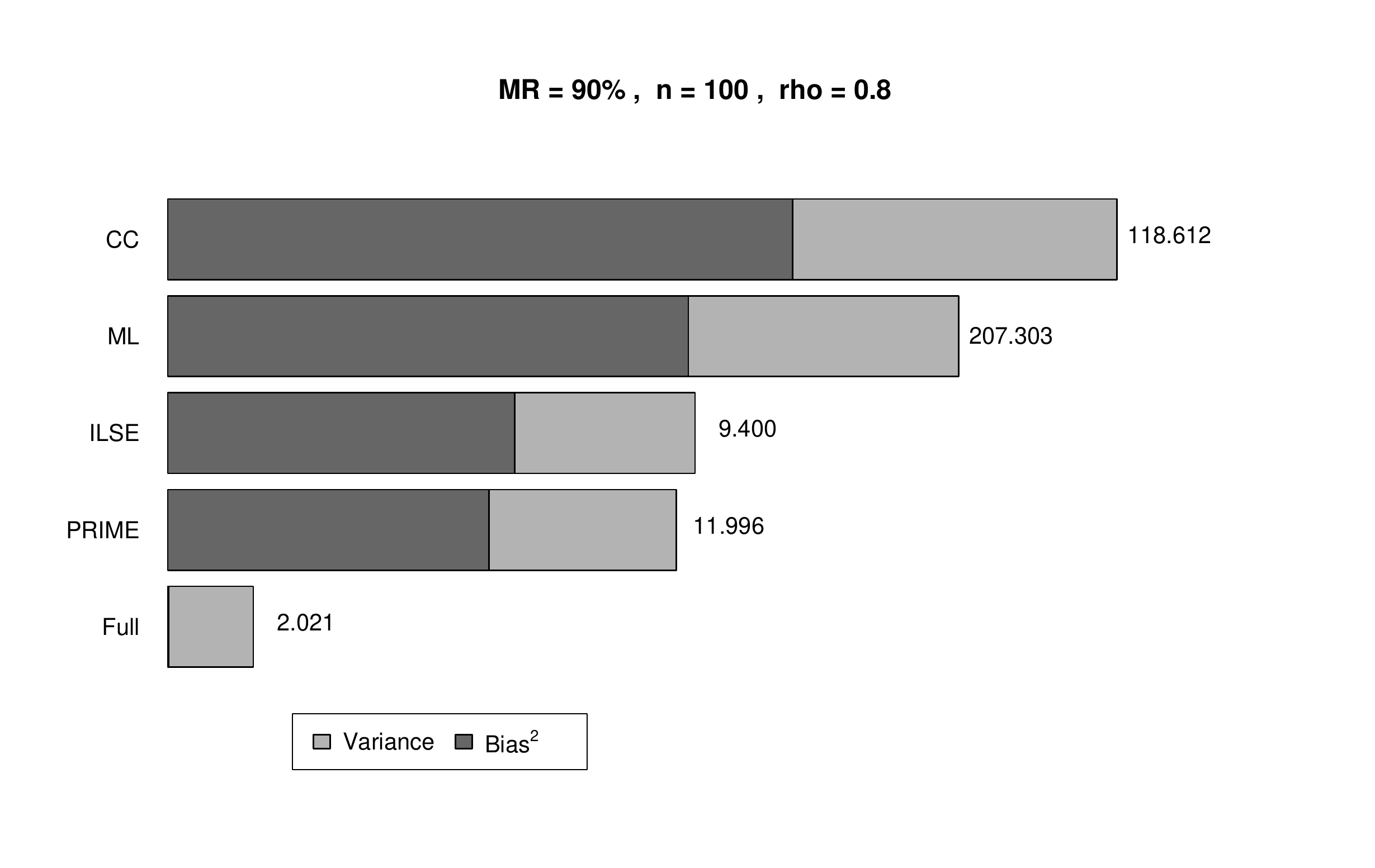}
		\end{minipage}
	}%
	\centering
	\caption{MSE with $n=100$ and 90\% missing data for different methods.}
	\label{fig:RHOMSE91}
\end{figure}

\begin{figure}[H]
	\begin{center}
		\includegraphics[width=10cm,height=7cm]{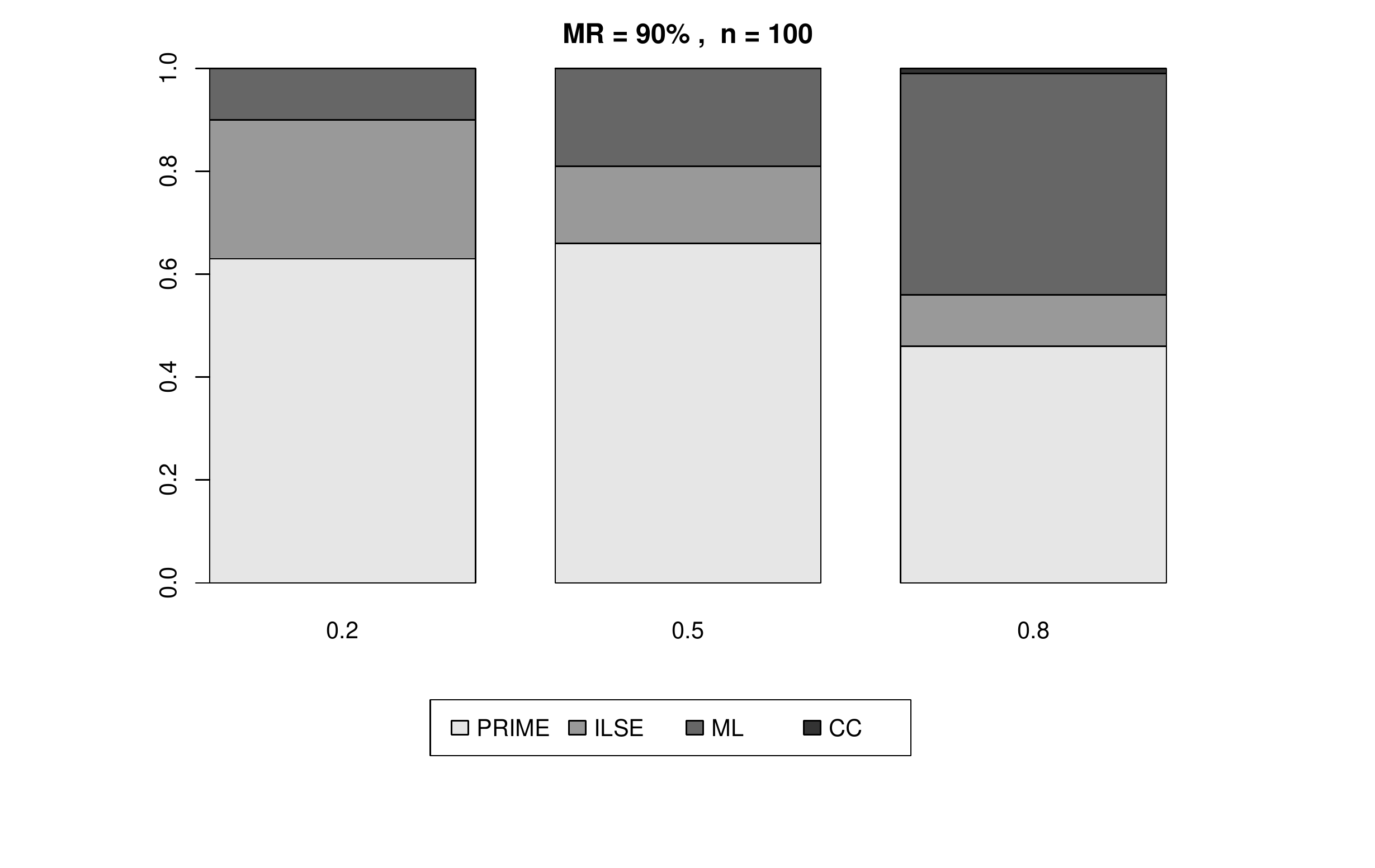}
	\end{center}
	\caption{Optimal rate of MSE with $n=100$ and 90\% missing data for different methods.}
	\label{fig:ORRHO91}
\end{figure}

\begin{figure}[H]
	\centering
	\subfigure[]{
		\begin{minipage}[t]{0.33\linewidth}
			\centering
			\includegraphics[width=1.9in]{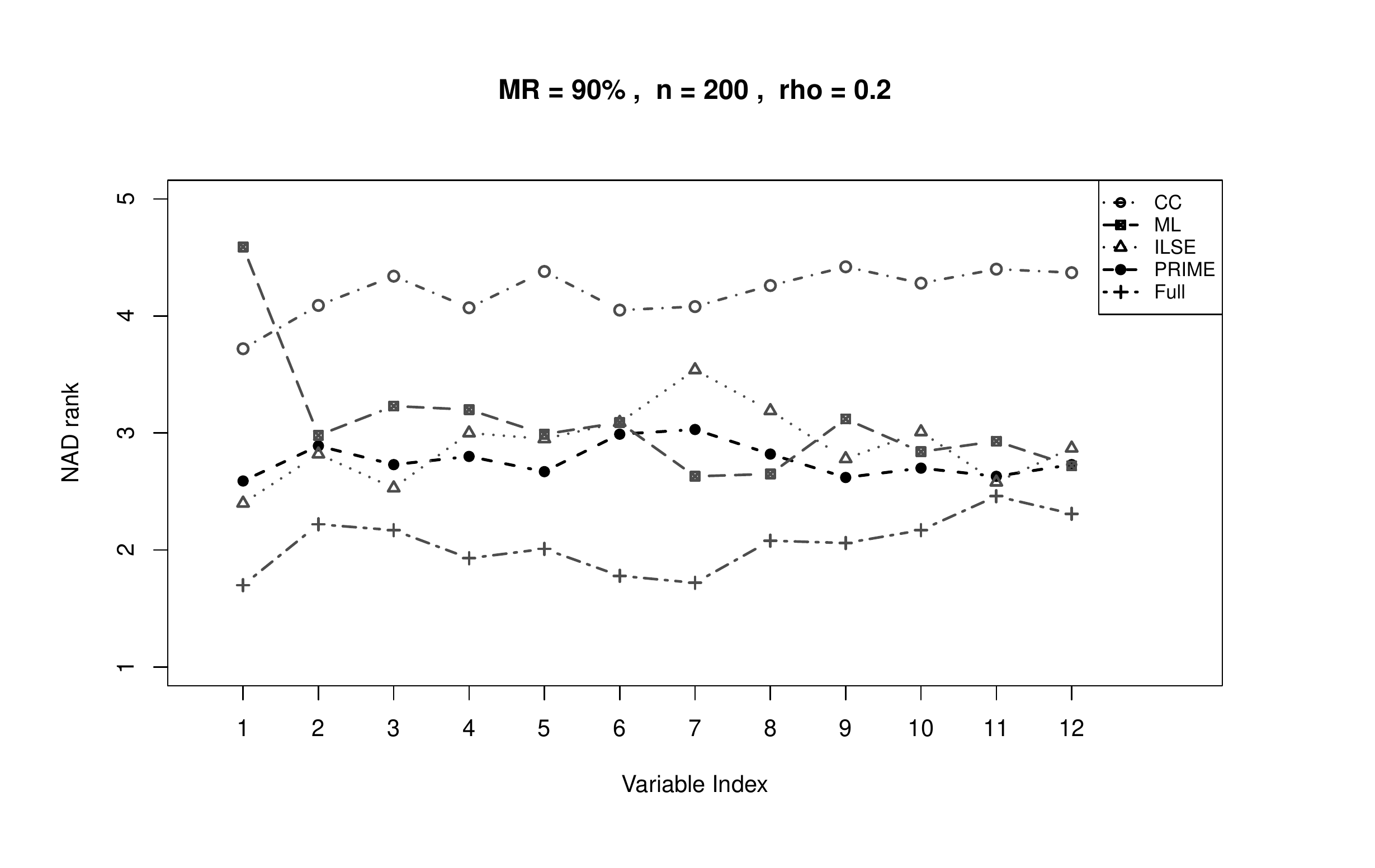}
		\end{minipage}%
	}%
	\subfigure[]{
		\begin{minipage}[t]{0.33\linewidth}
			\centering
			\includegraphics[width=1.9in]{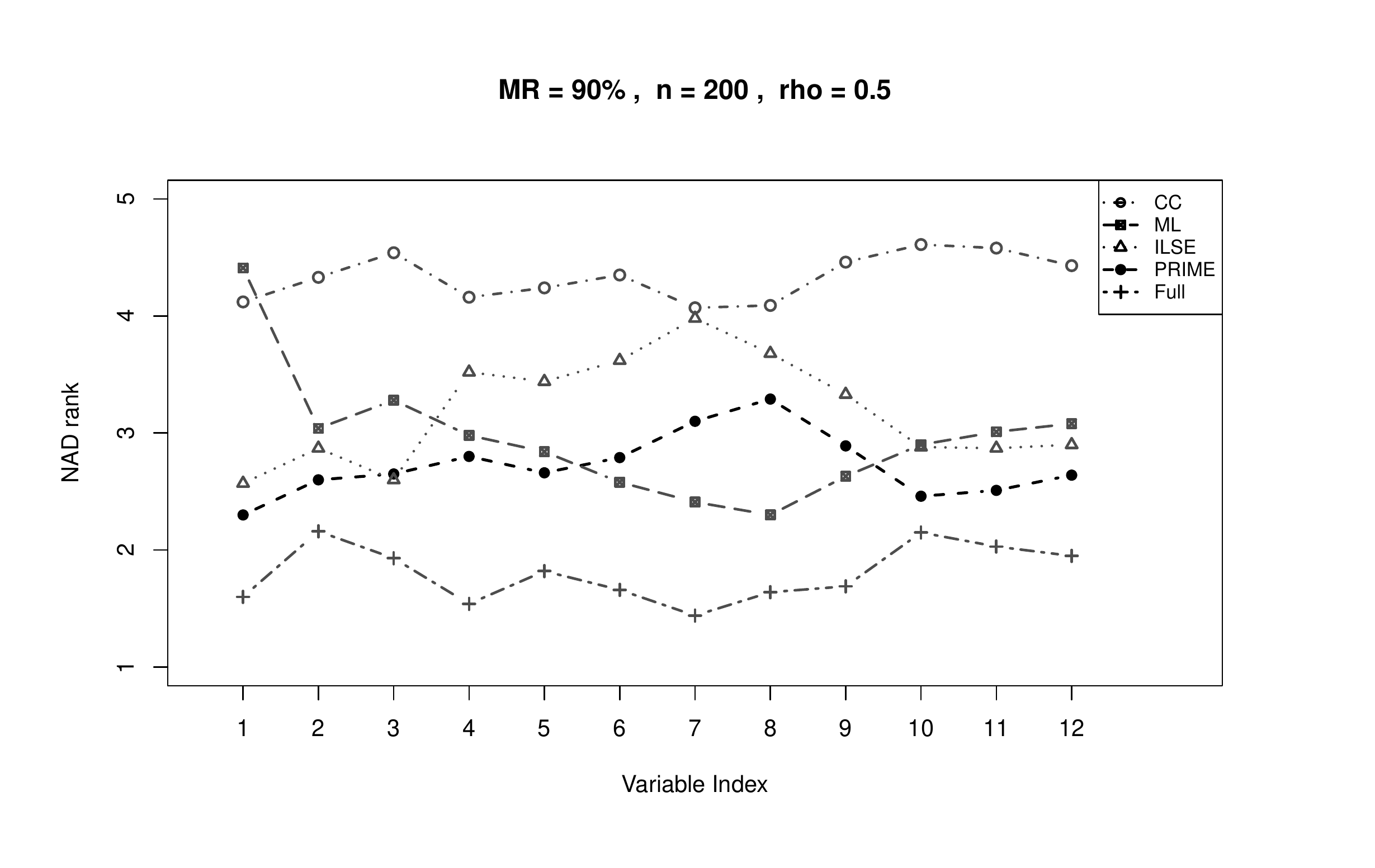}
		\end{minipage}
	}%
	\subfigure[]{
		\begin{minipage}[t]{0.33\linewidth}
			\centering
			\includegraphics[width=1.9in]{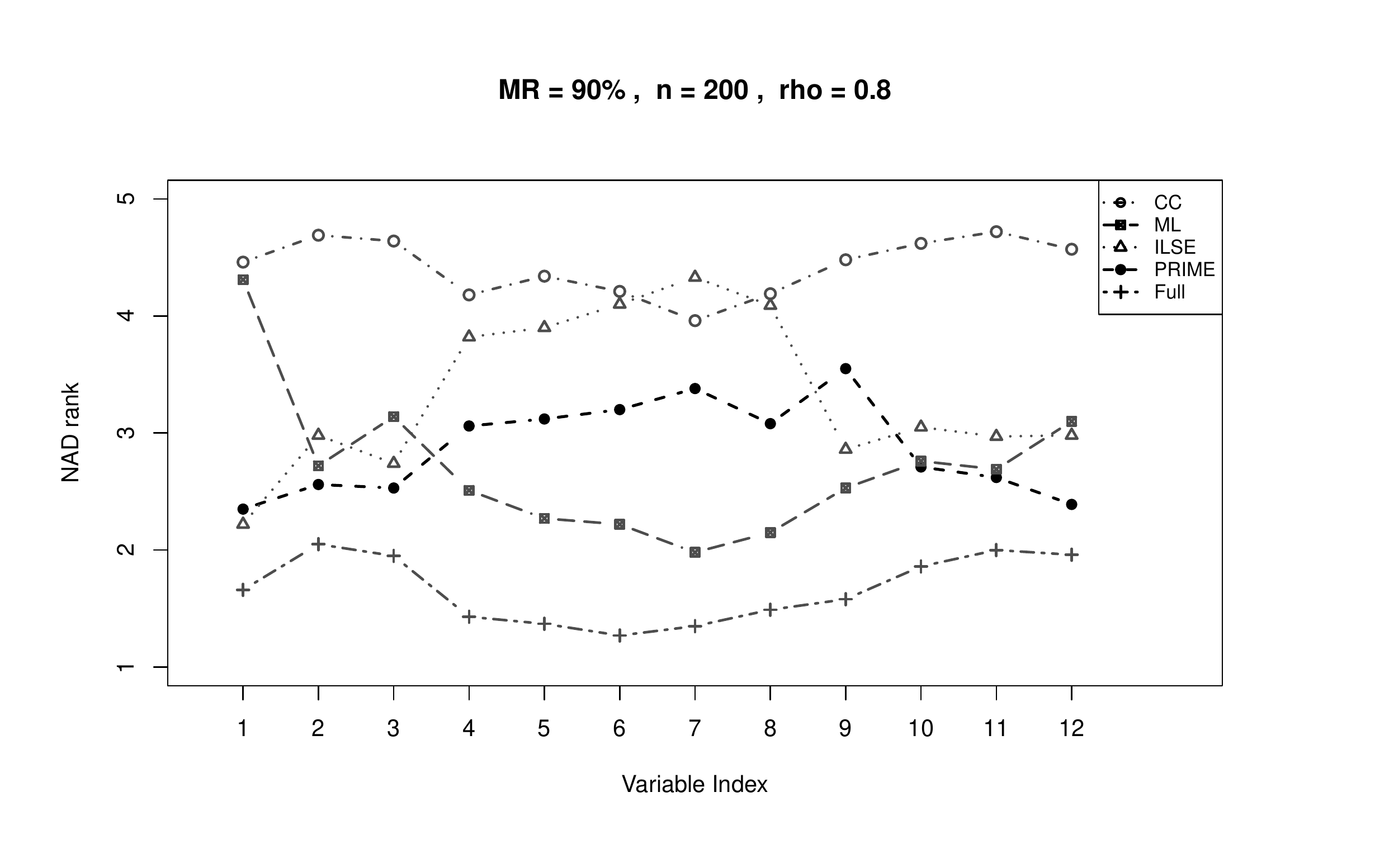}
		\end{minipage}
	}%
	\centering
	\caption{NAD with $n=200$ and 90\% missing data for different methods.}
	\label{fig:RHONAD92}
\end{figure}

\begin{figure}[H]
	\centering
	\subfigure[]{
		\begin{minipage}[t]{0.33\linewidth}
			\centering
			\includegraphics[width=1.9in]{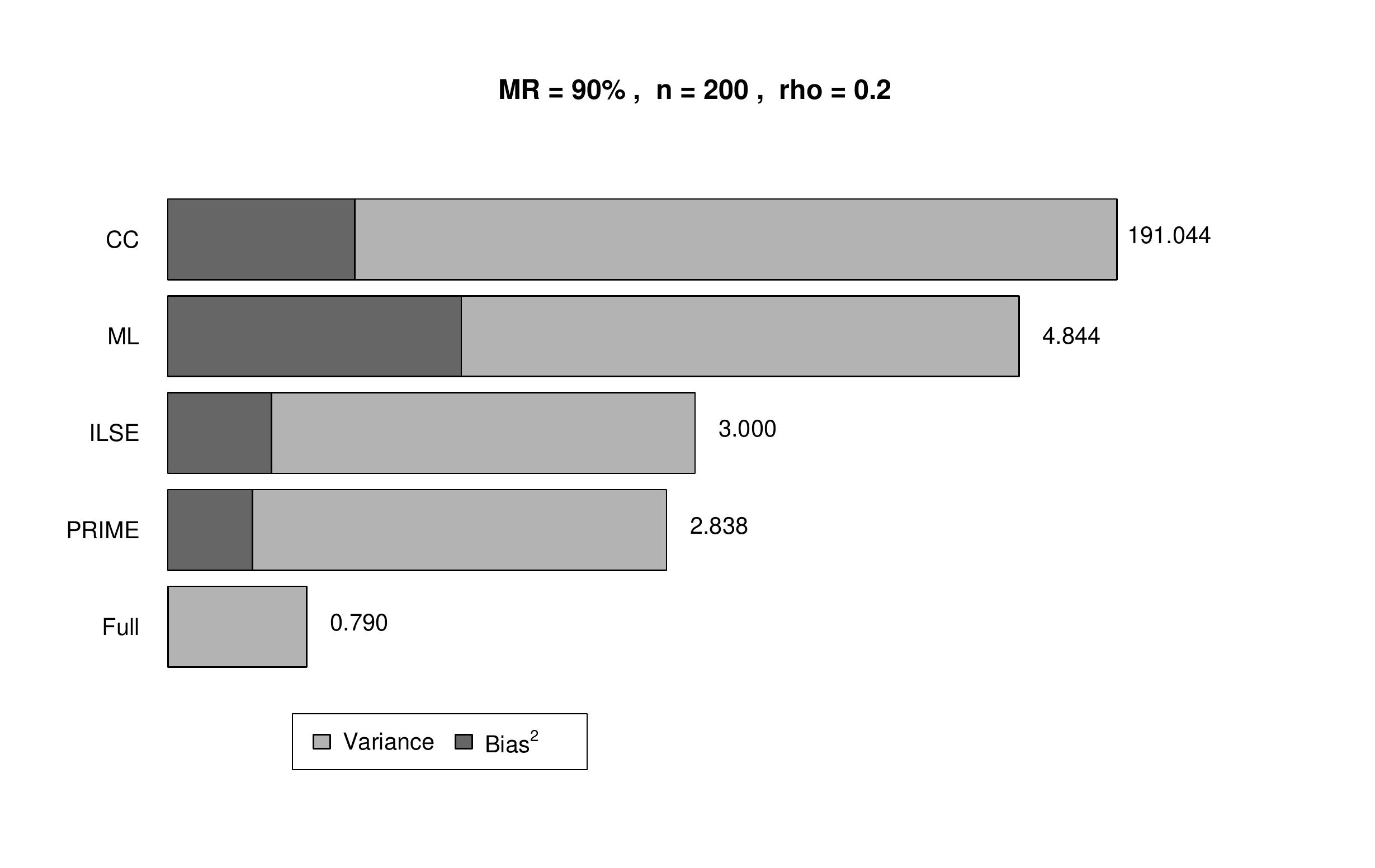}
		\end{minipage}%
	}%
	\subfigure[]{
		\begin{minipage}[t]{0.33\linewidth}
			\centering
			\includegraphics[width=1.9in]{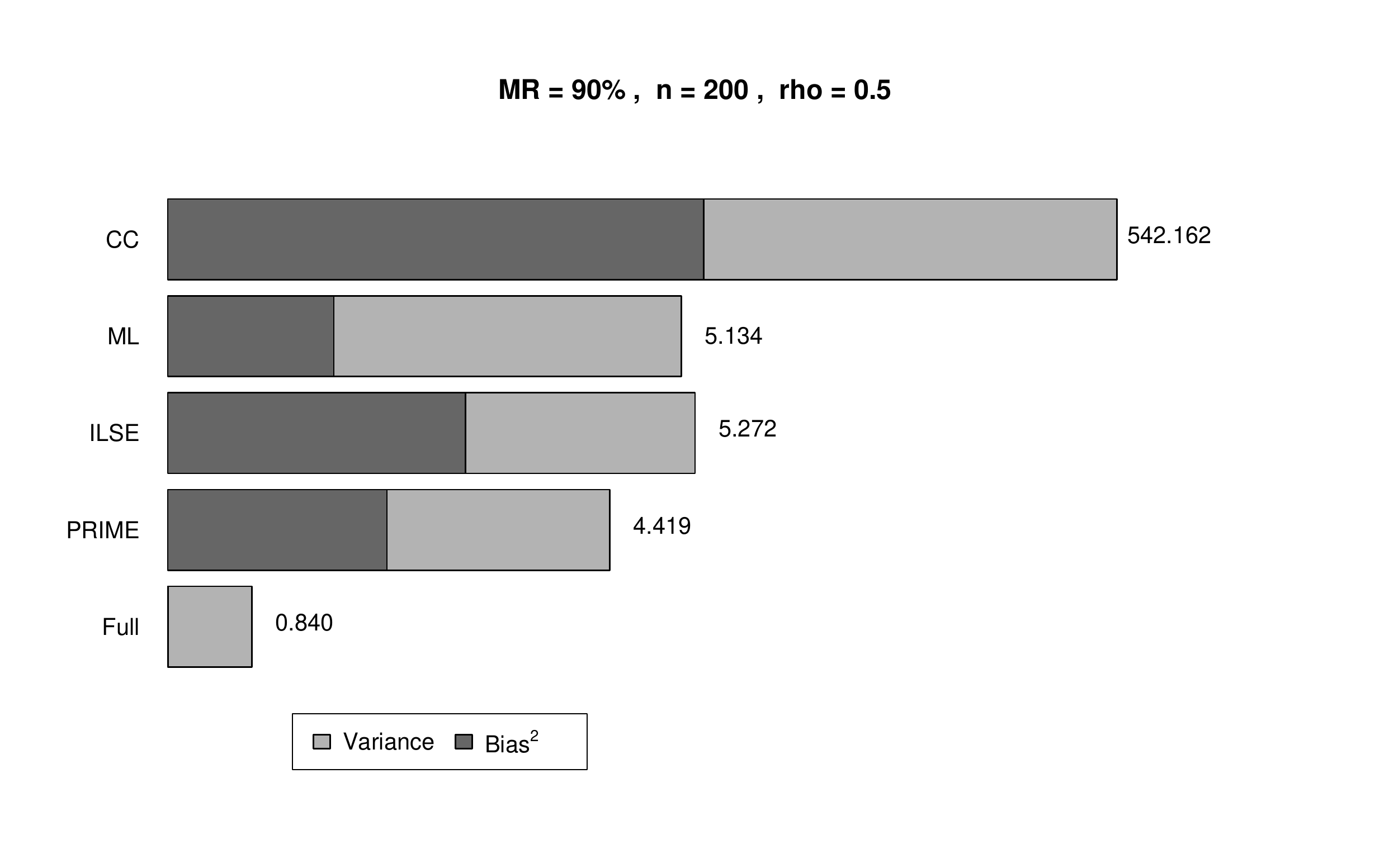}
		\end{minipage}
	}%
	\subfigure[]{
		\begin{minipage}[t]{0.33\linewidth}
			\centering
			\includegraphics[width=1.9in]{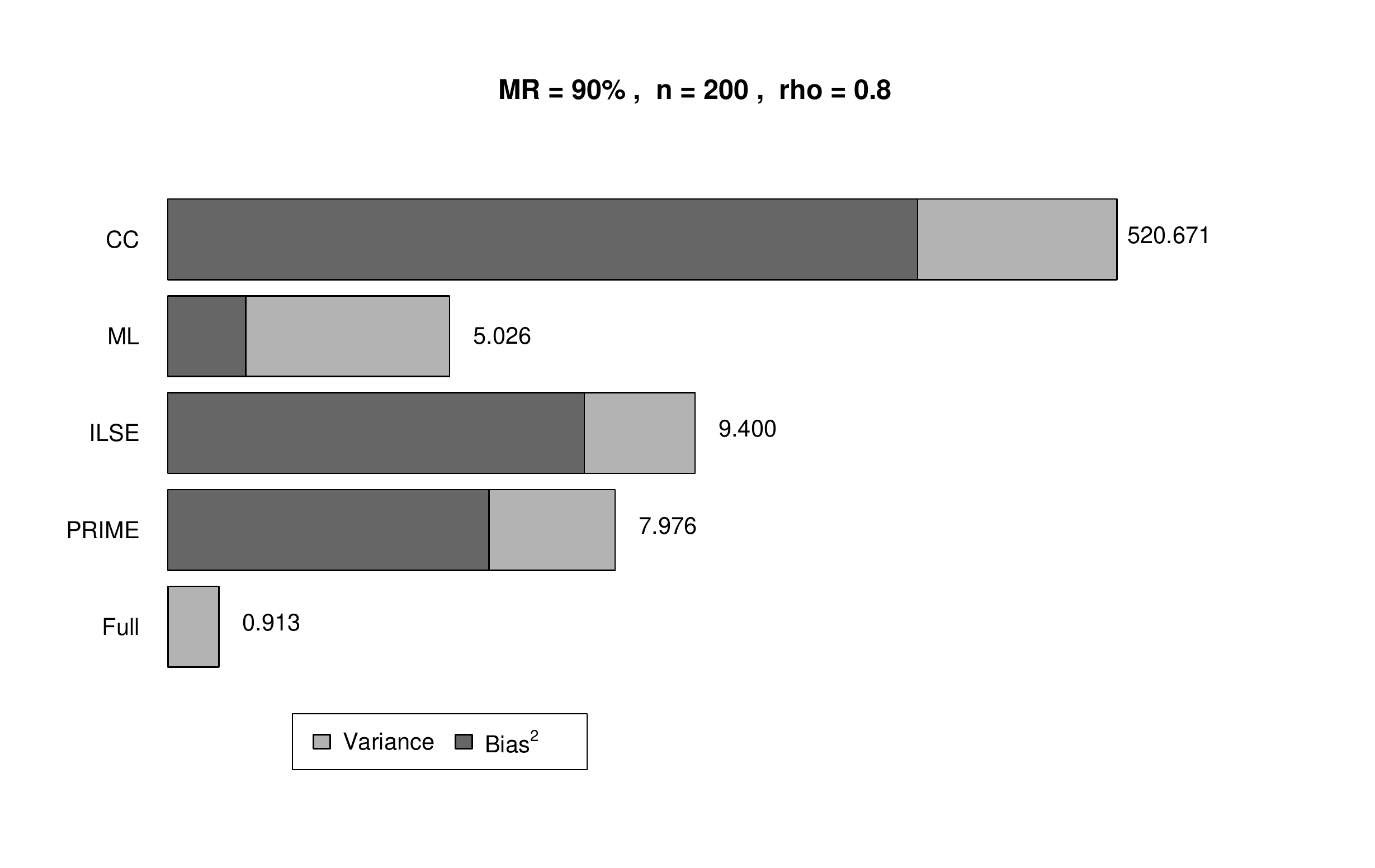}
		\end{minipage}
	}%
	\centering
	\caption{MSE with $n=200$ and 90\% missing data for different methods.}
	\label{fig:RHOMSE92}
\end{figure}

\begin{figure}[H]
	\begin{center}
		\includegraphics[width=12cm,height=7cm]{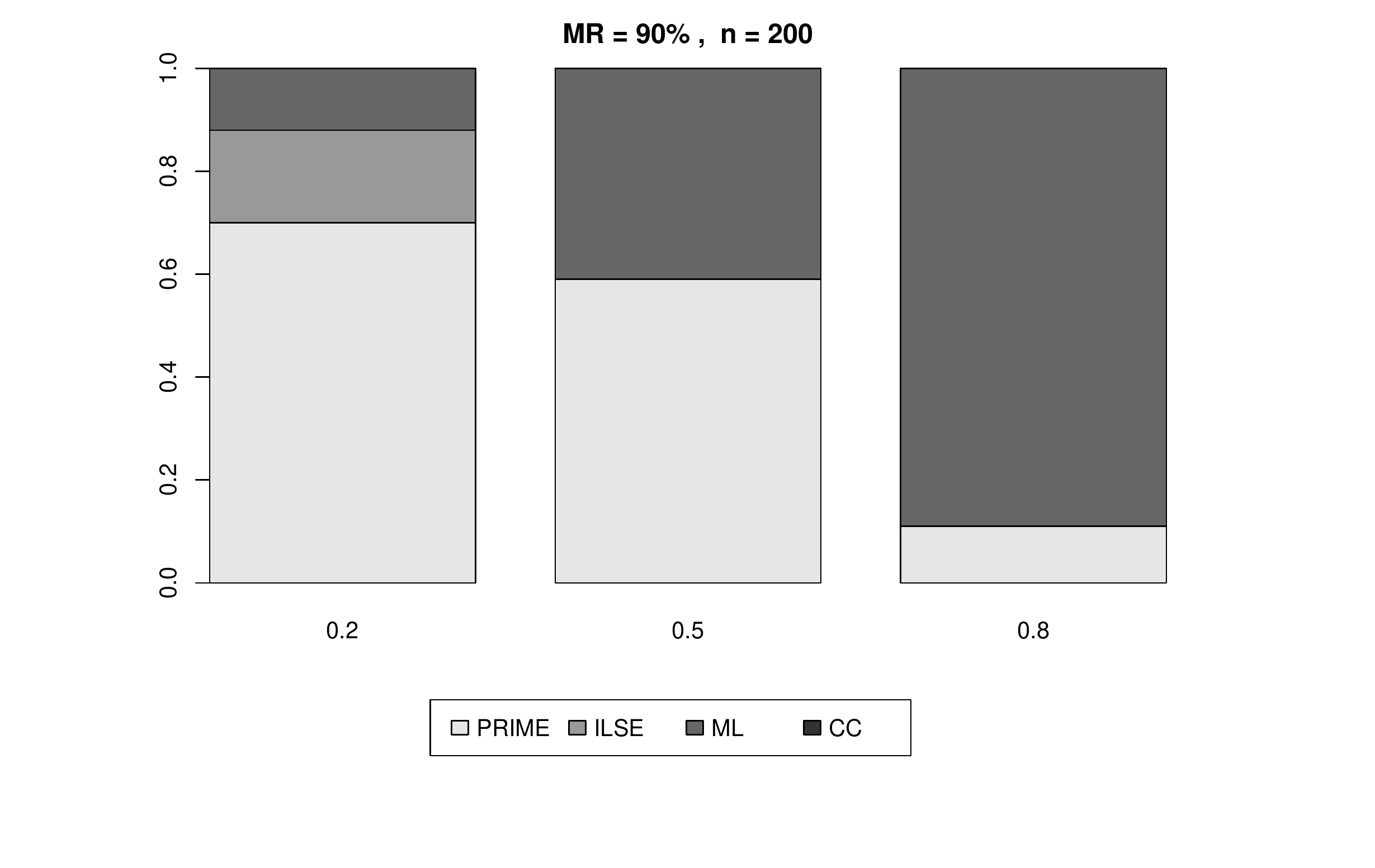}
	\end{center}
	\caption{Optimal rate of MSE with $n=200$ and 90\% missing data for different methods.}
	\label{fig:ORRHO92}
\end{figure}

\begin{figure}[H]
	\centering
	\subfigure[]{
		\begin{minipage}[t]{0.33\linewidth}
			\centering
			\includegraphics[width=1.9in]{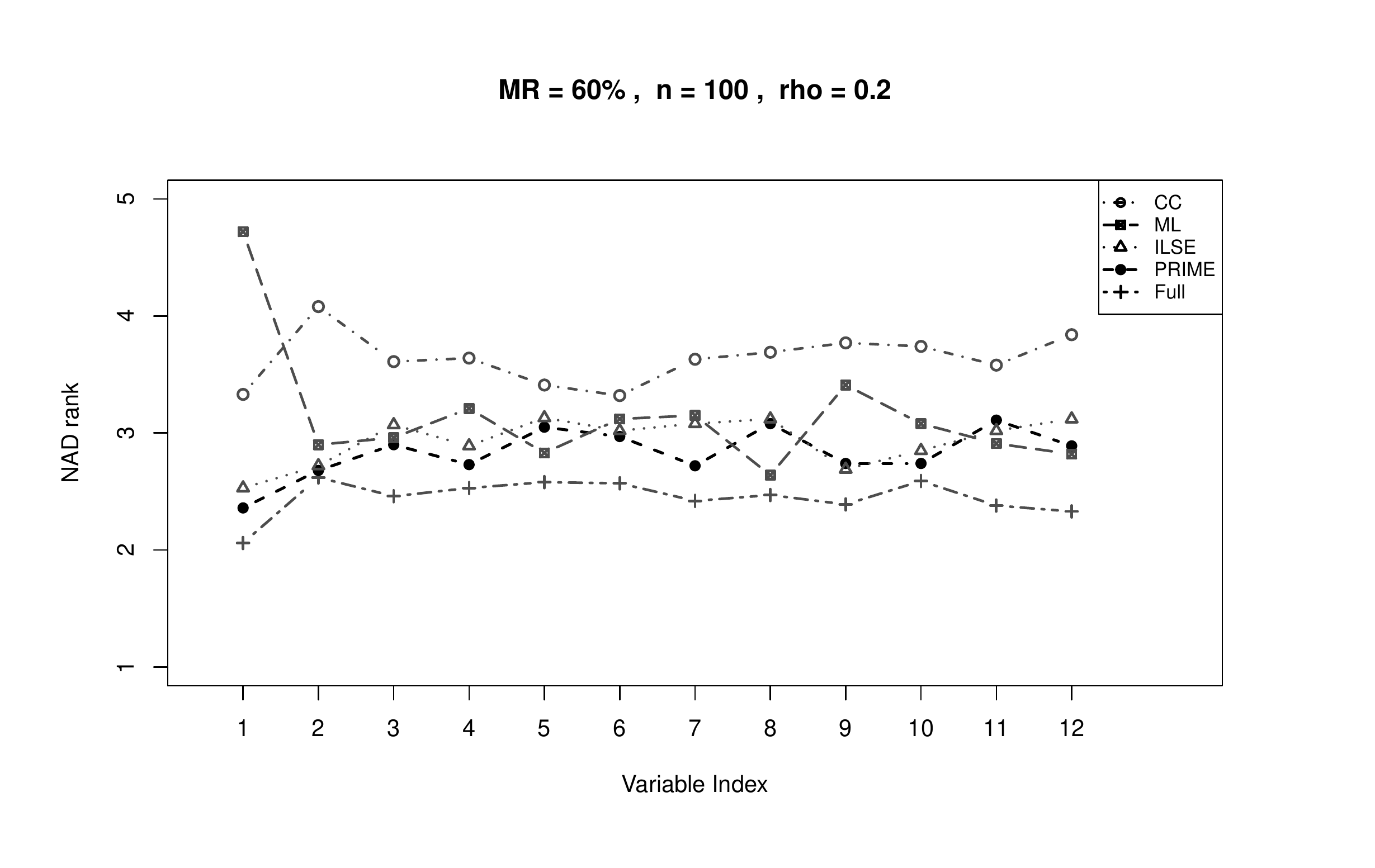}
		\end{minipage}%
	}%
	\subfigure[]{
		\begin{minipage}[t]{0.33\linewidth}
			\centering
			\includegraphics[width=1.9in]{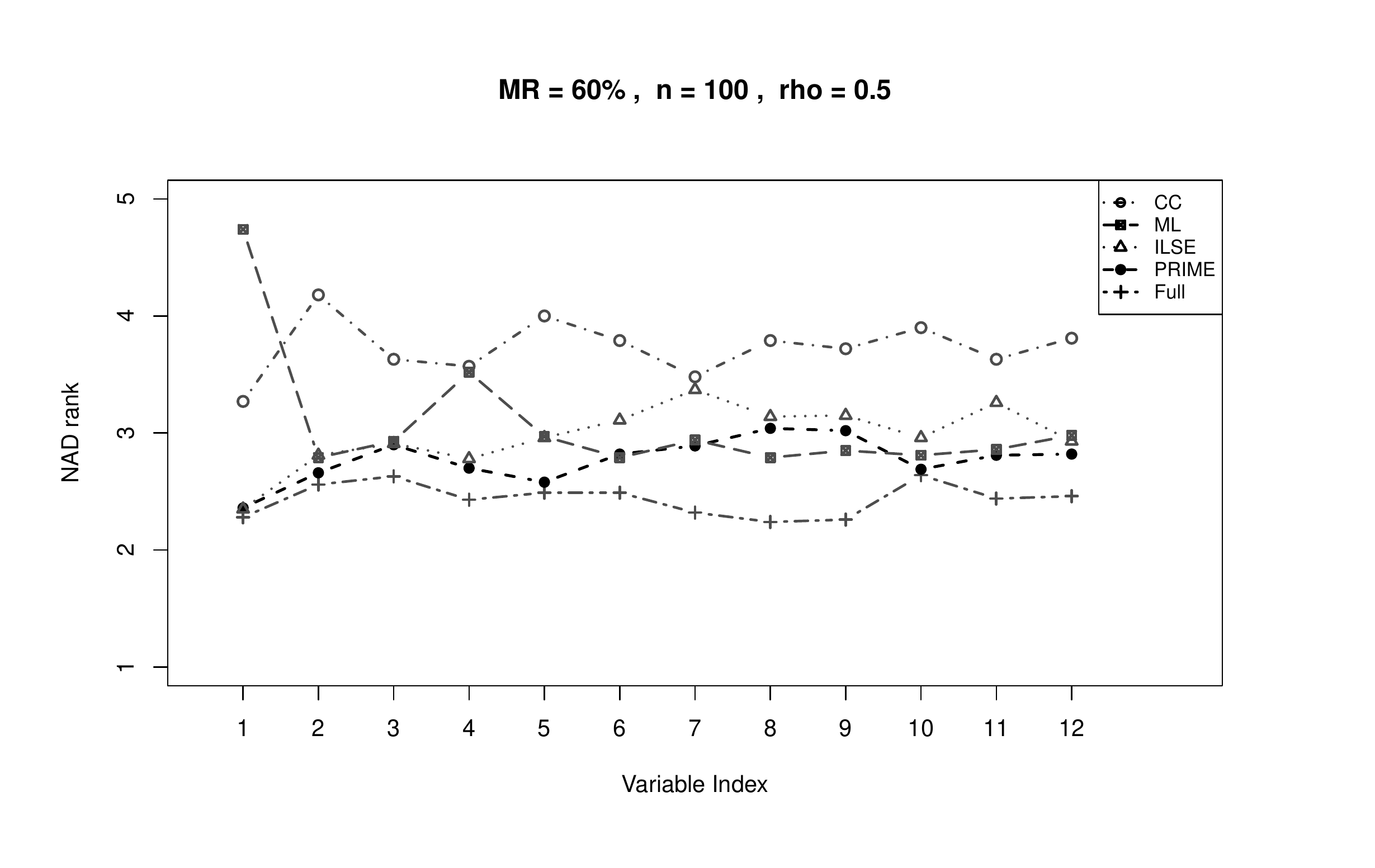}
		\end{minipage}
	}%
	\subfigure[]{
		\begin{minipage}[t]{0.33\linewidth}
			\centering
			\includegraphics[width=1.9in]{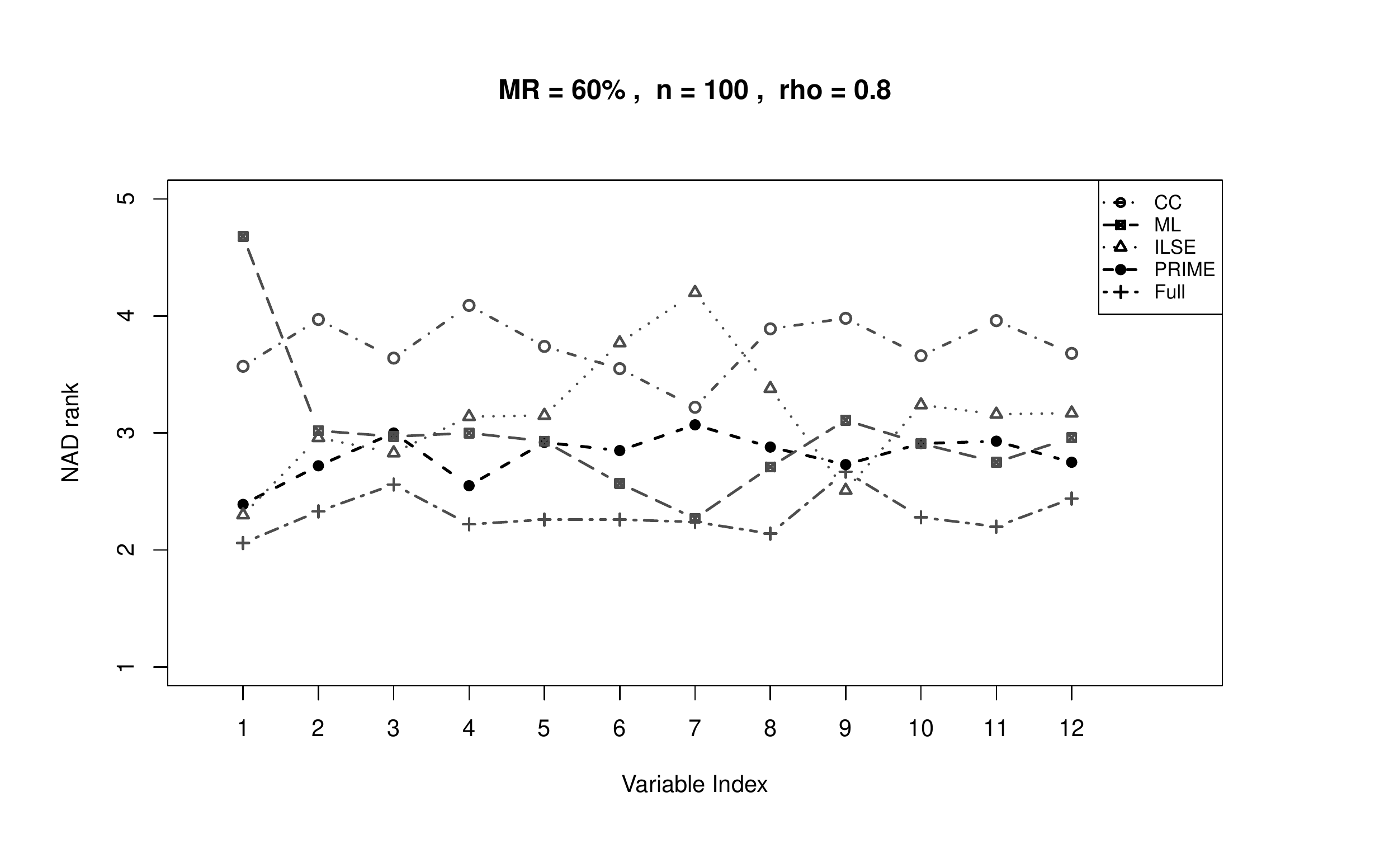}
		\end{minipage}
	}%
	\centering
	\caption{NAD with $n=100$ and 60\% missing data for different methods.}
	\label{fig:RHONAD61}
\end{figure}

\begin{figure}[H]
	\centering
	\subfigure[]{
		\begin{minipage}[t]{0.33\linewidth}
			\centering
			\includegraphics[width=1.9in]{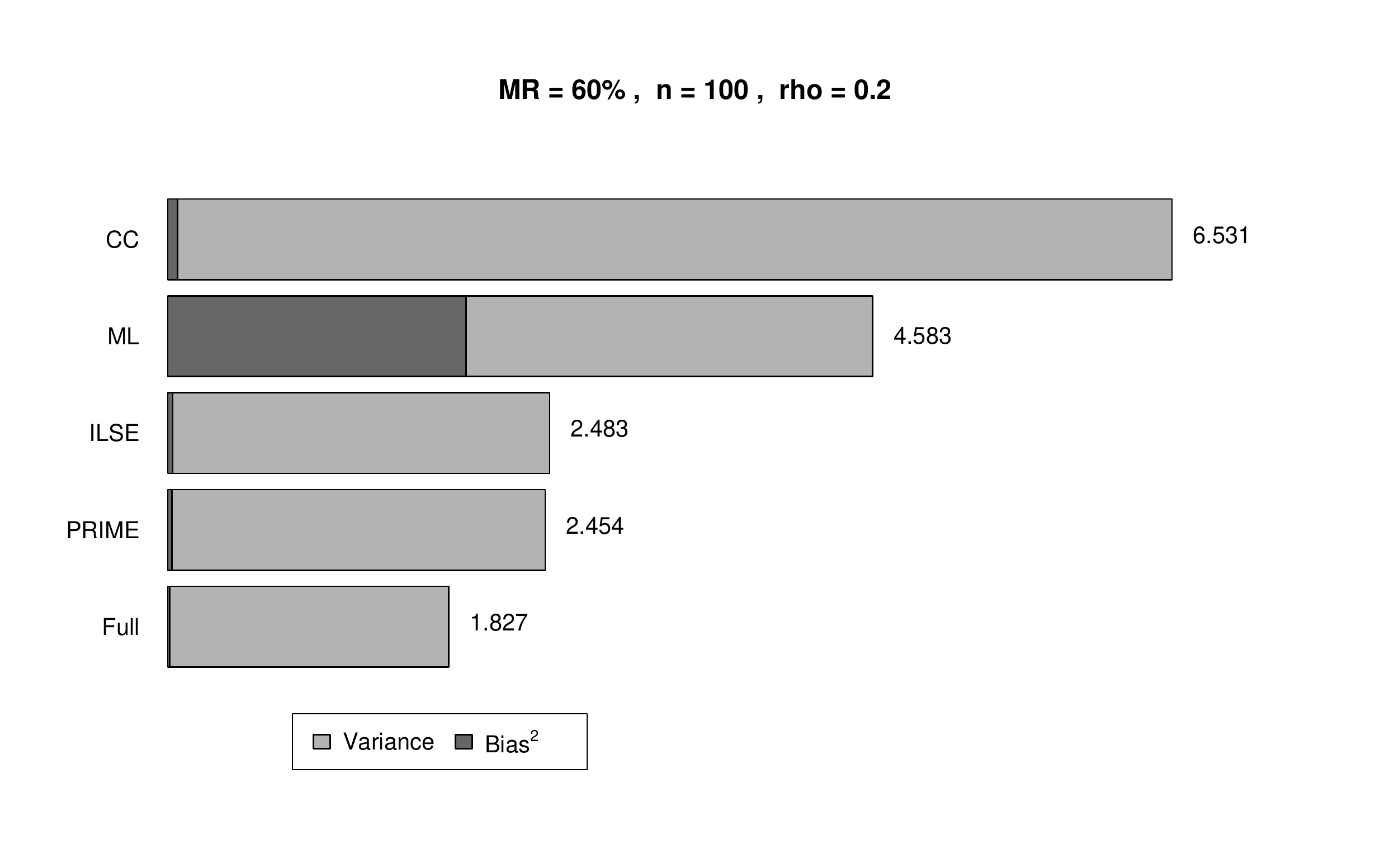}
		\end{minipage}%
	}%
	\subfigure[]{
		\begin{minipage}[t]{0.33\linewidth}
			\centering
			\includegraphics[width=1.9in]{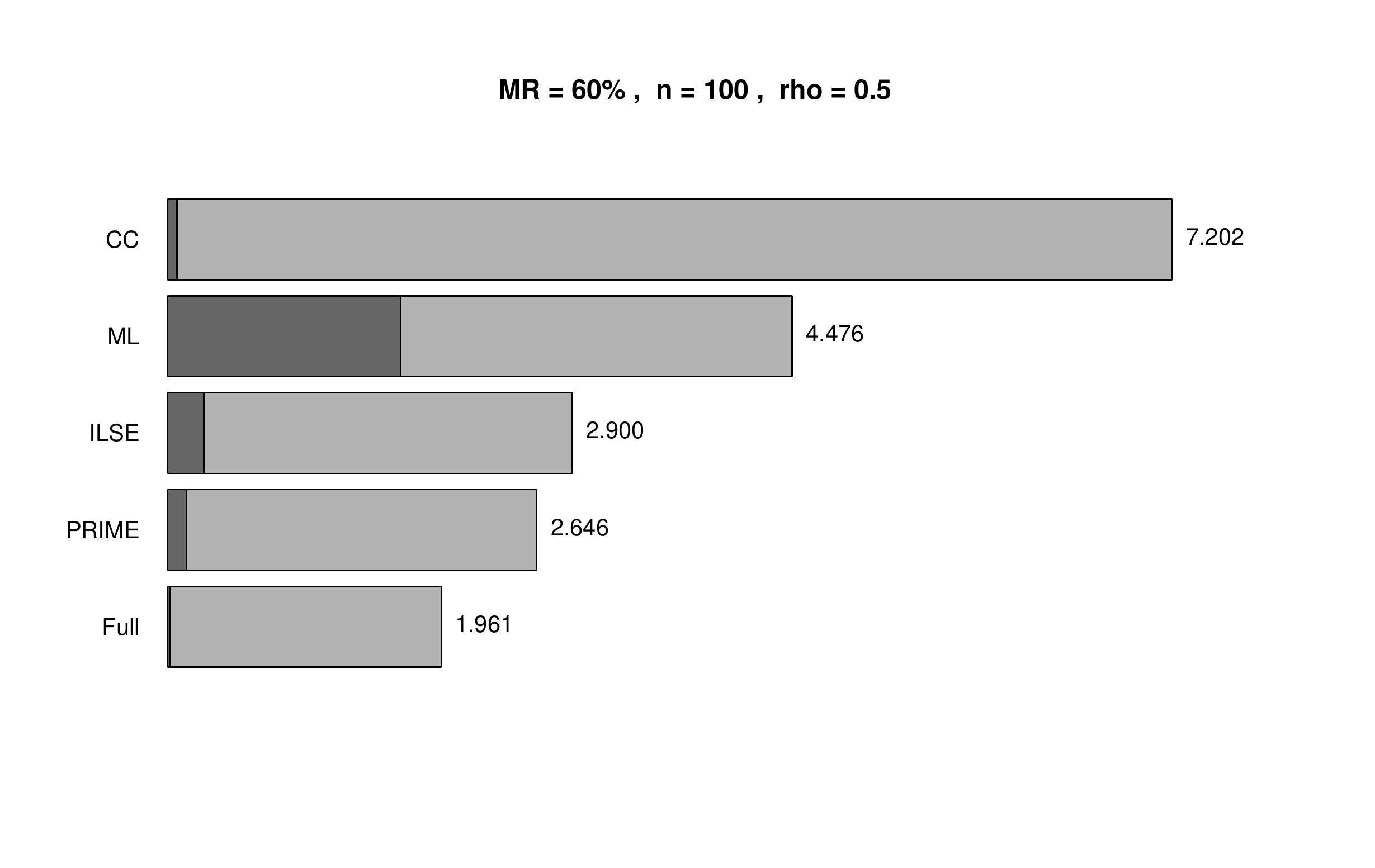}
		\end{minipage}
	}%
	\subfigure[]{
		\begin{minipage}[t]{0.33\linewidth}
			\centering
			\includegraphics[width=1.9in]{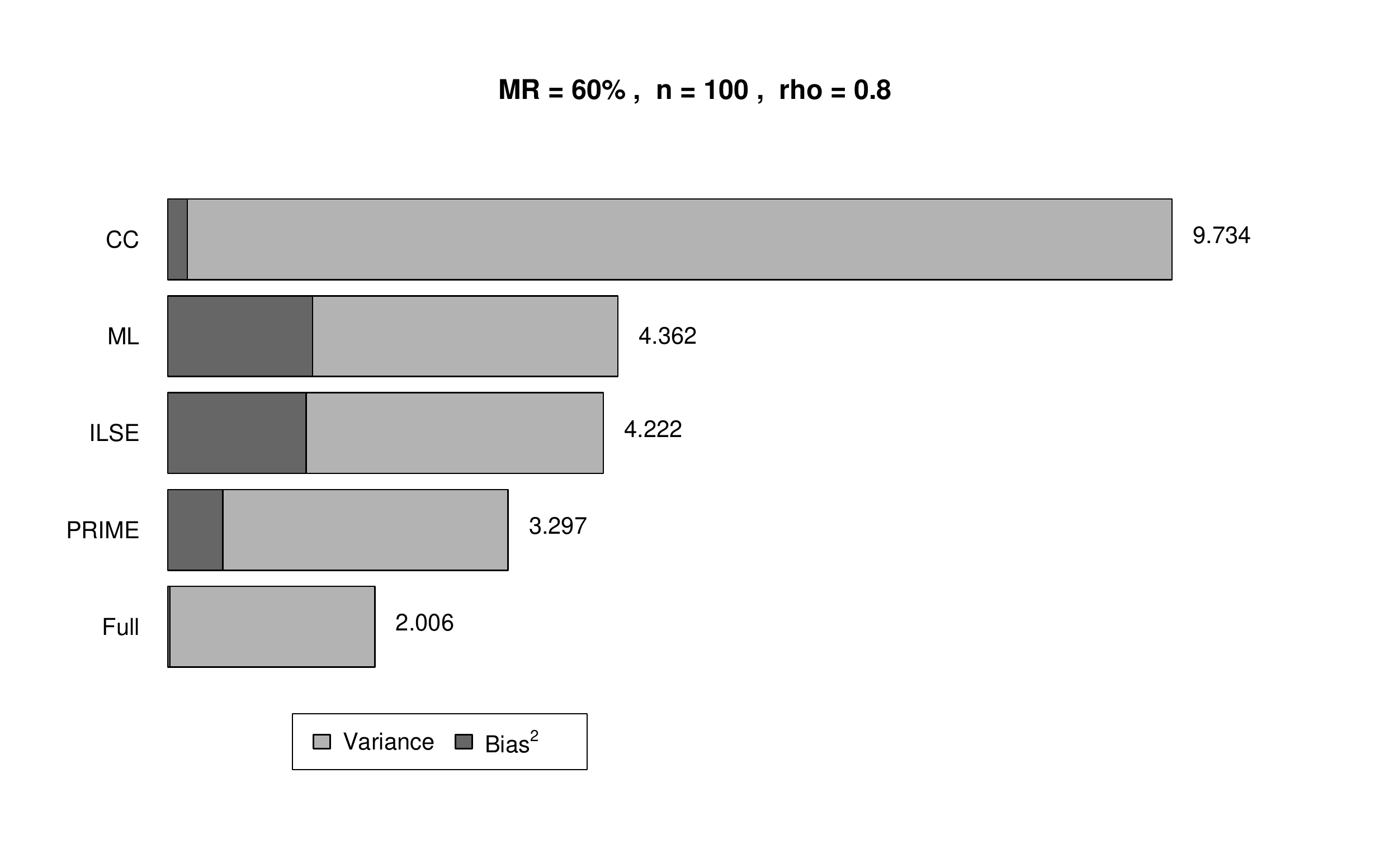}
		\end{minipage}
	}%
	\centering
	\caption{MSE with $n=100$ and 60\% missing data for different methods.}
	\label{fig:RHOMSE61}
\end{figure}

\begin{figure}[H]
	\begin{center}
		\includegraphics[width=10cm,height=7cm]{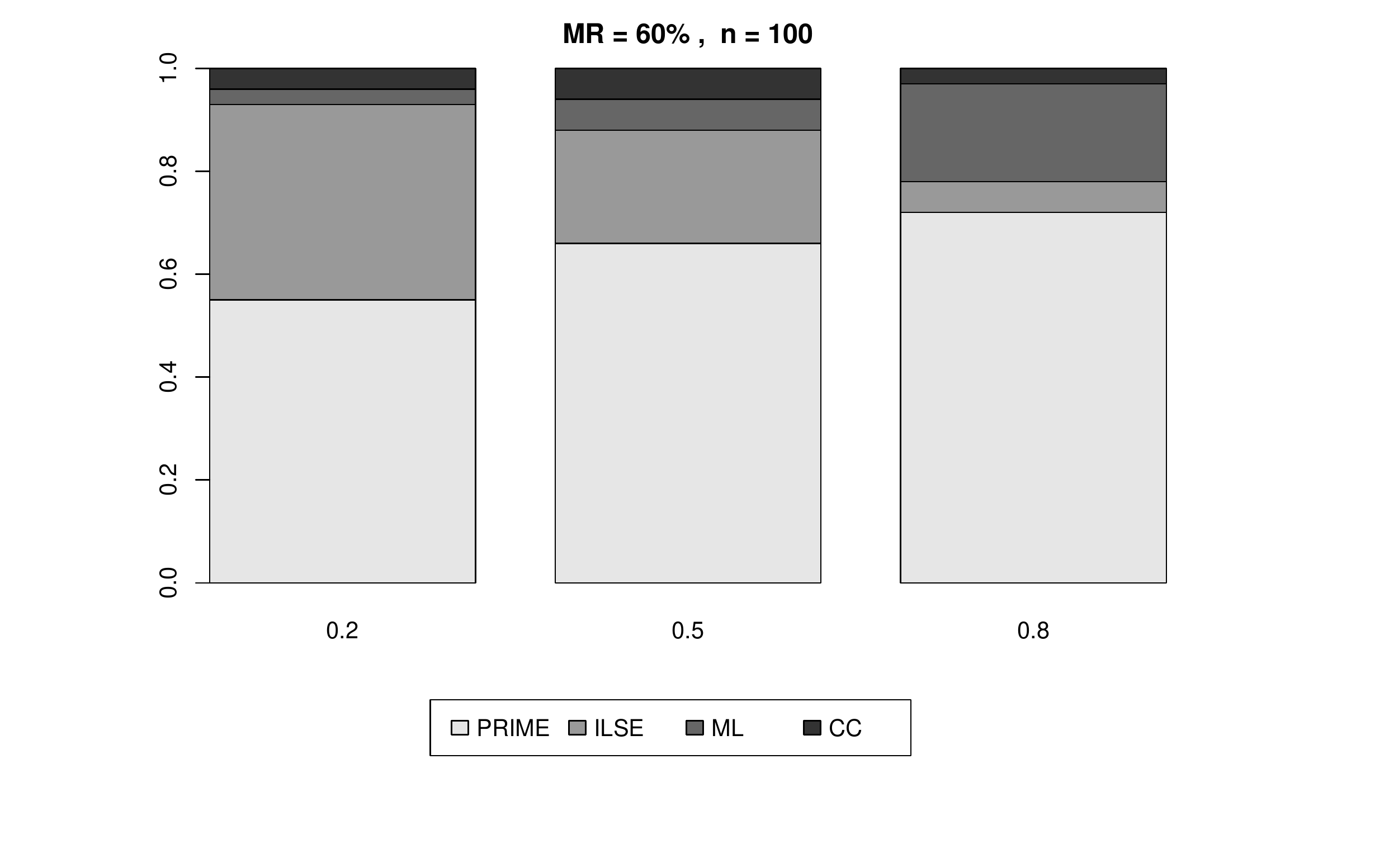}
	\end{center}
	\caption{Optimal rate of MSE with $n=100$ and 60\% missing data for different methods.}
	\label{fig:ORRHO61}
\end{figure}

\begin{figure}[H]
	\centering
	\subfigure[]{
		\begin{minipage}[t]{0.33\linewidth}
			\centering
			\includegraphics[width=1.9in]{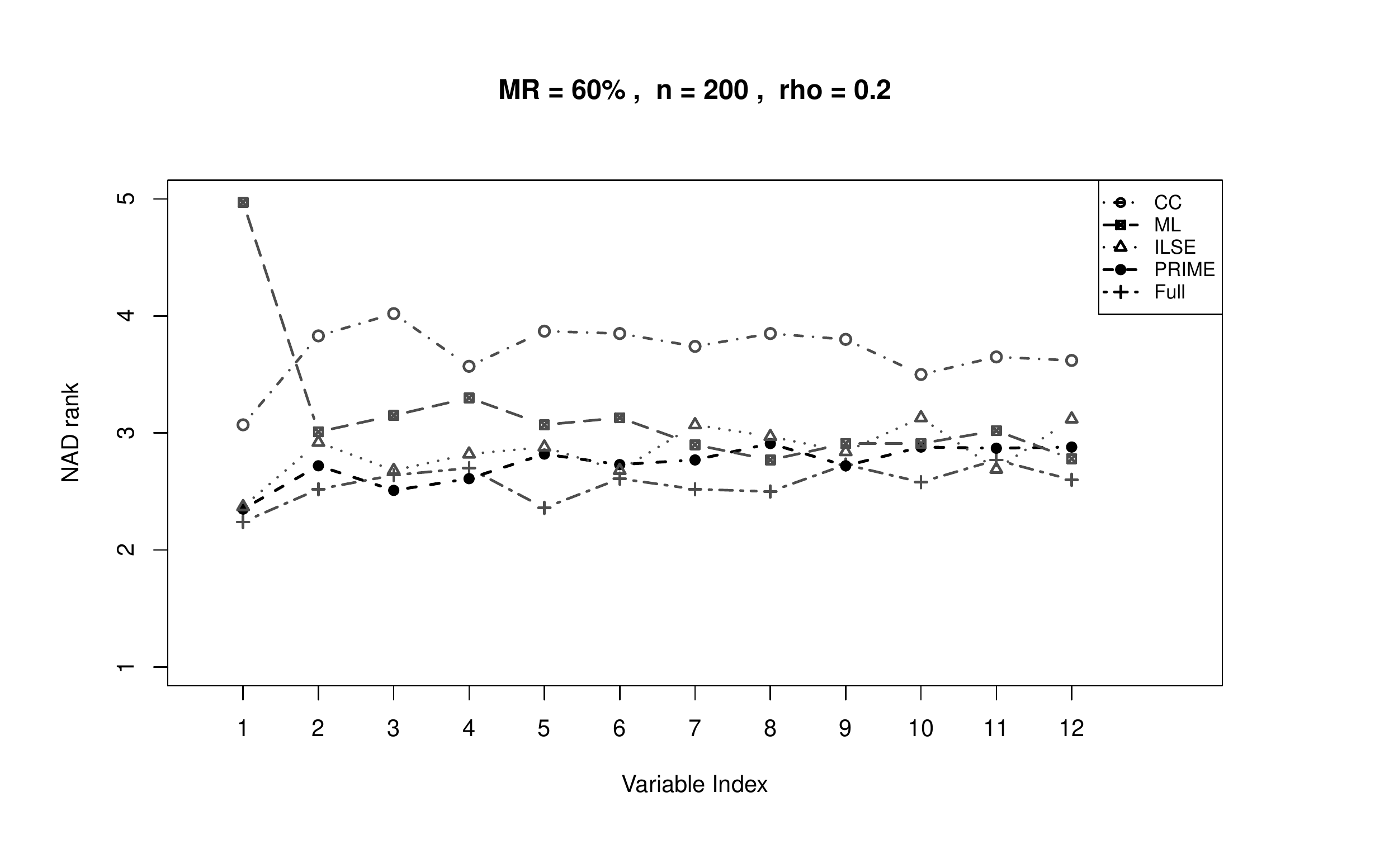}
		\end{minipage}%
	}%
	\subfigure[]{
		\begin{minipage}[t]{0.33\linewidth}
			\centering
			\includegraphics[width=1.9in]{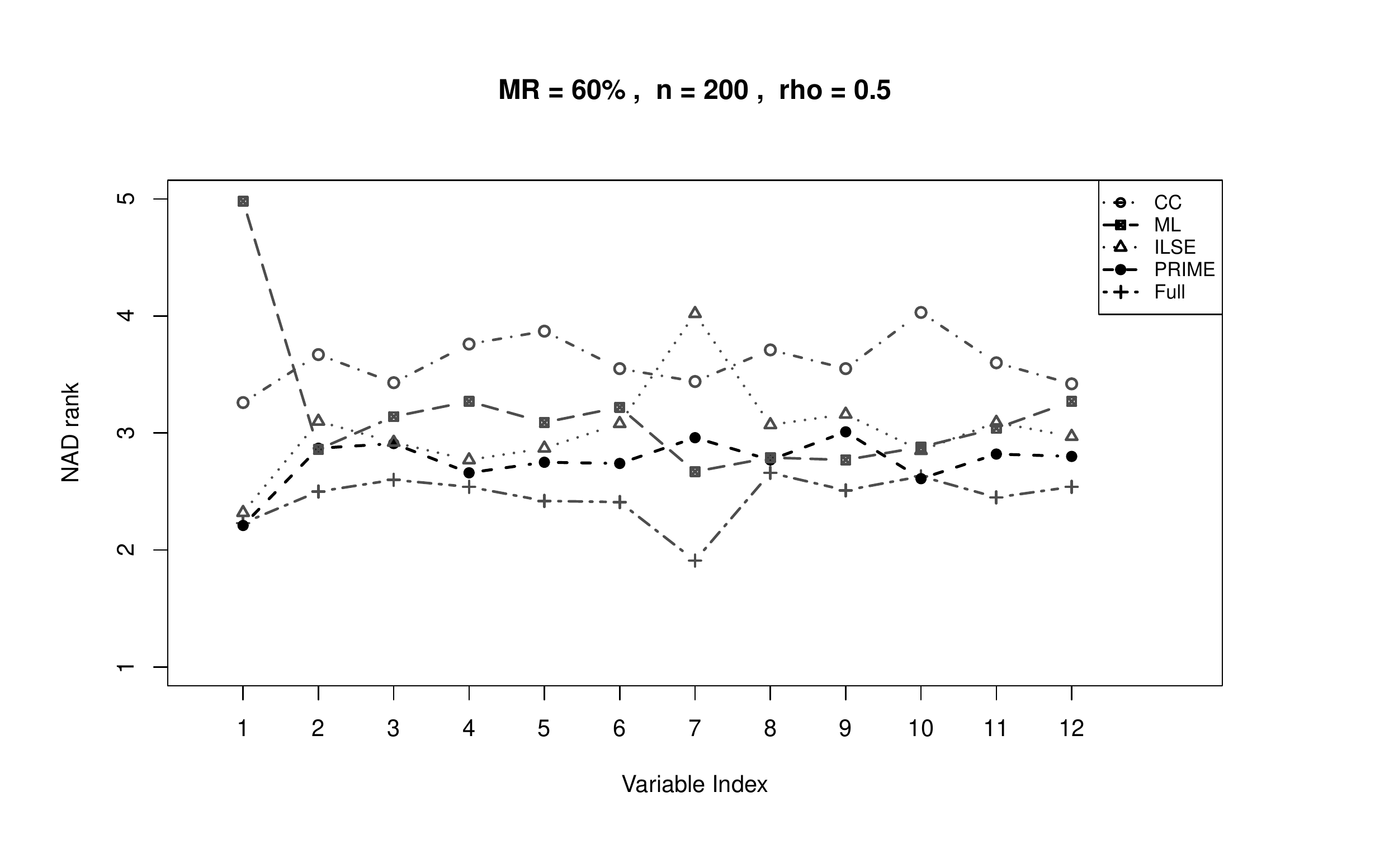}
		\end{minipage}
	}%
	\subfigure[]{
		\begin{minipage}[t]{0.33\linewidth}
			\centering
			\includegraphics[width=1.9in]{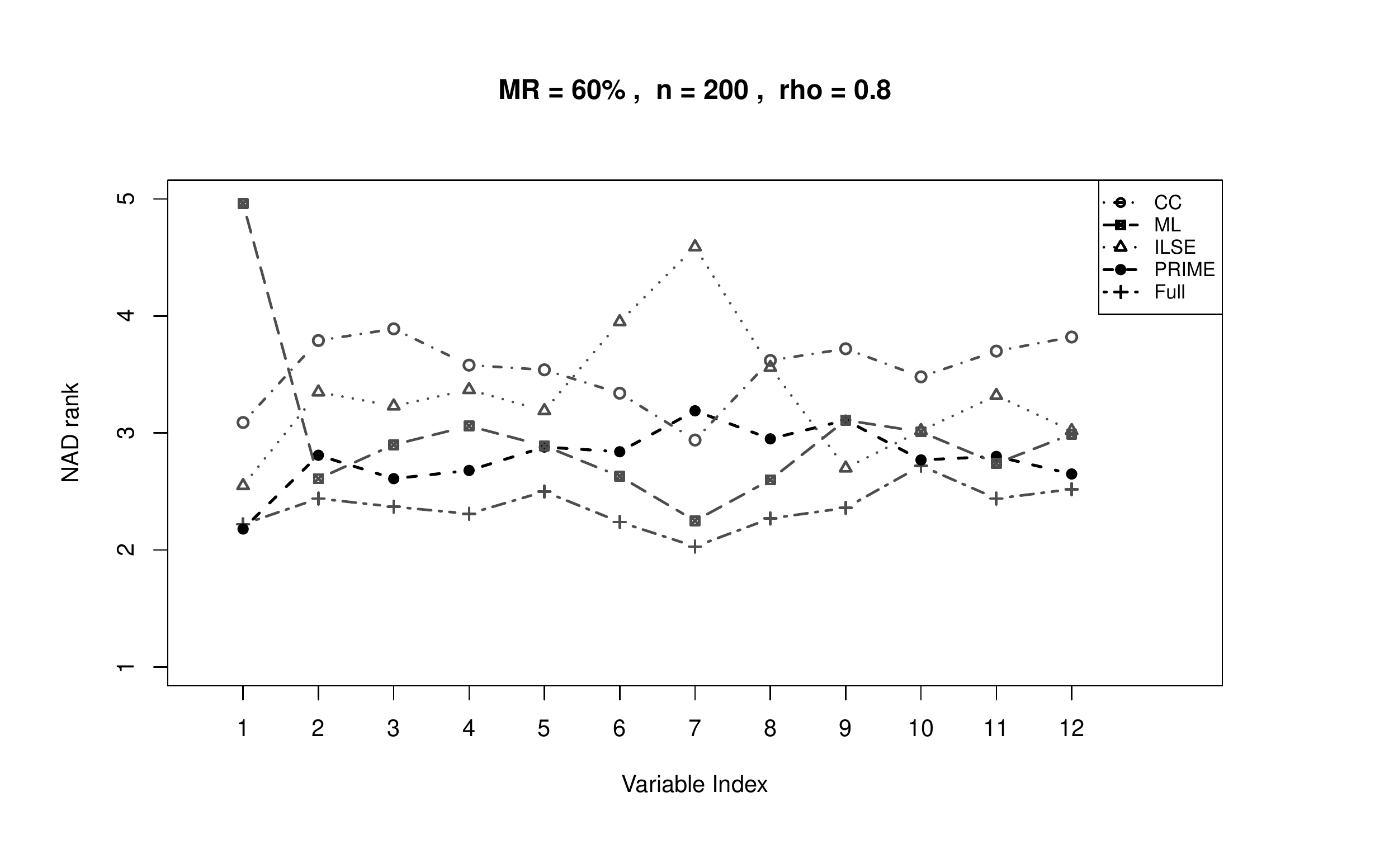}
		\end{minipage}
	}%
	\centering
	\caption{NAD with $n=200$ and 60\% missing data for different methods.}
	\label{fig:RHONAD62}
\end{figure}

\begin{figure}[H]
	\centering
	\subfigure[]{
		\begin{minipage}[t]{0.33\linewidth}
			\centering
			\includegraphics[width=1.9in]{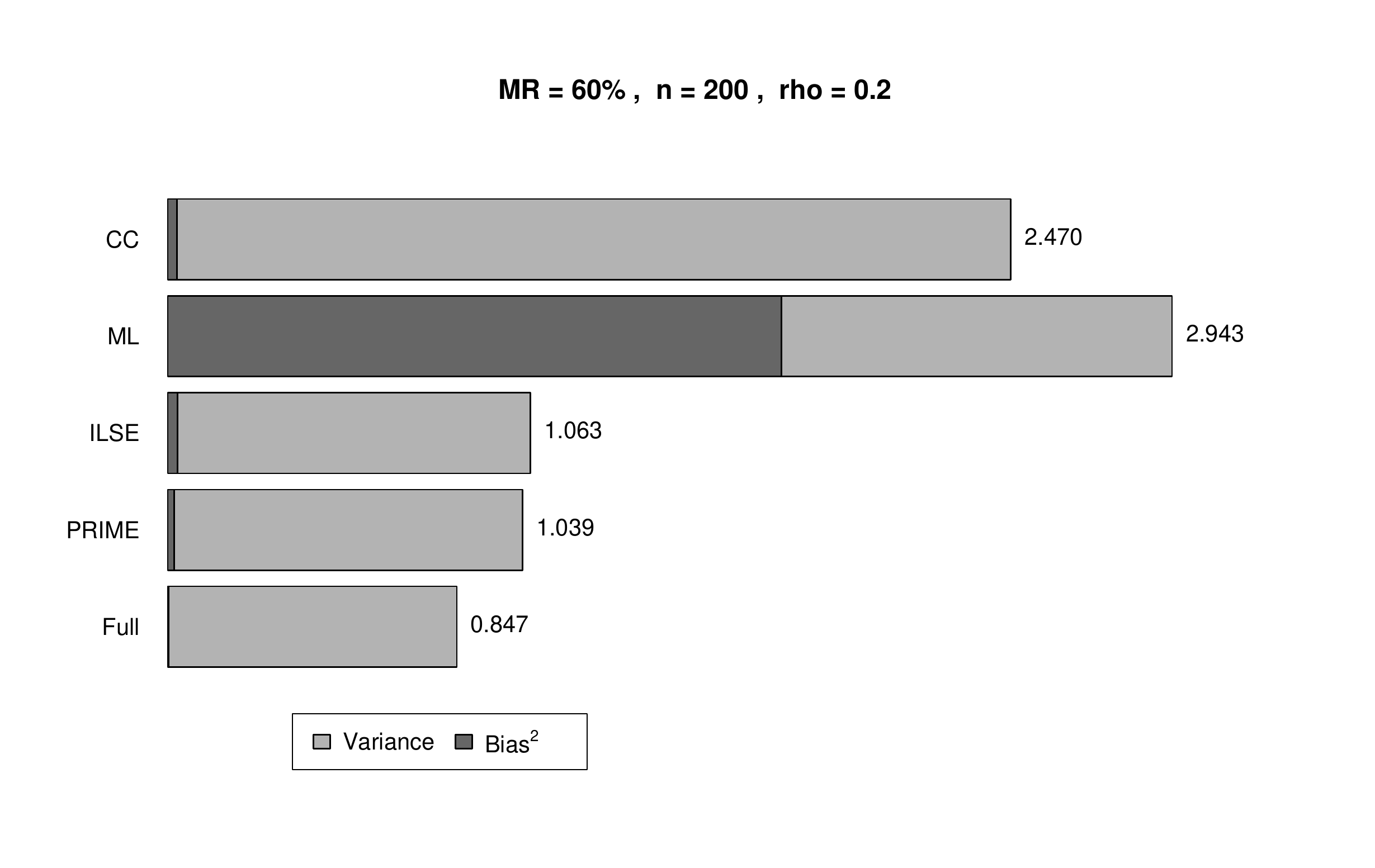}
		\end{minipage}%
	}%
	\subfigure[]{
		\begin{minipage}[t]{0.33\linewidth}
			\centering
			\includegraphics[width=1.9in]{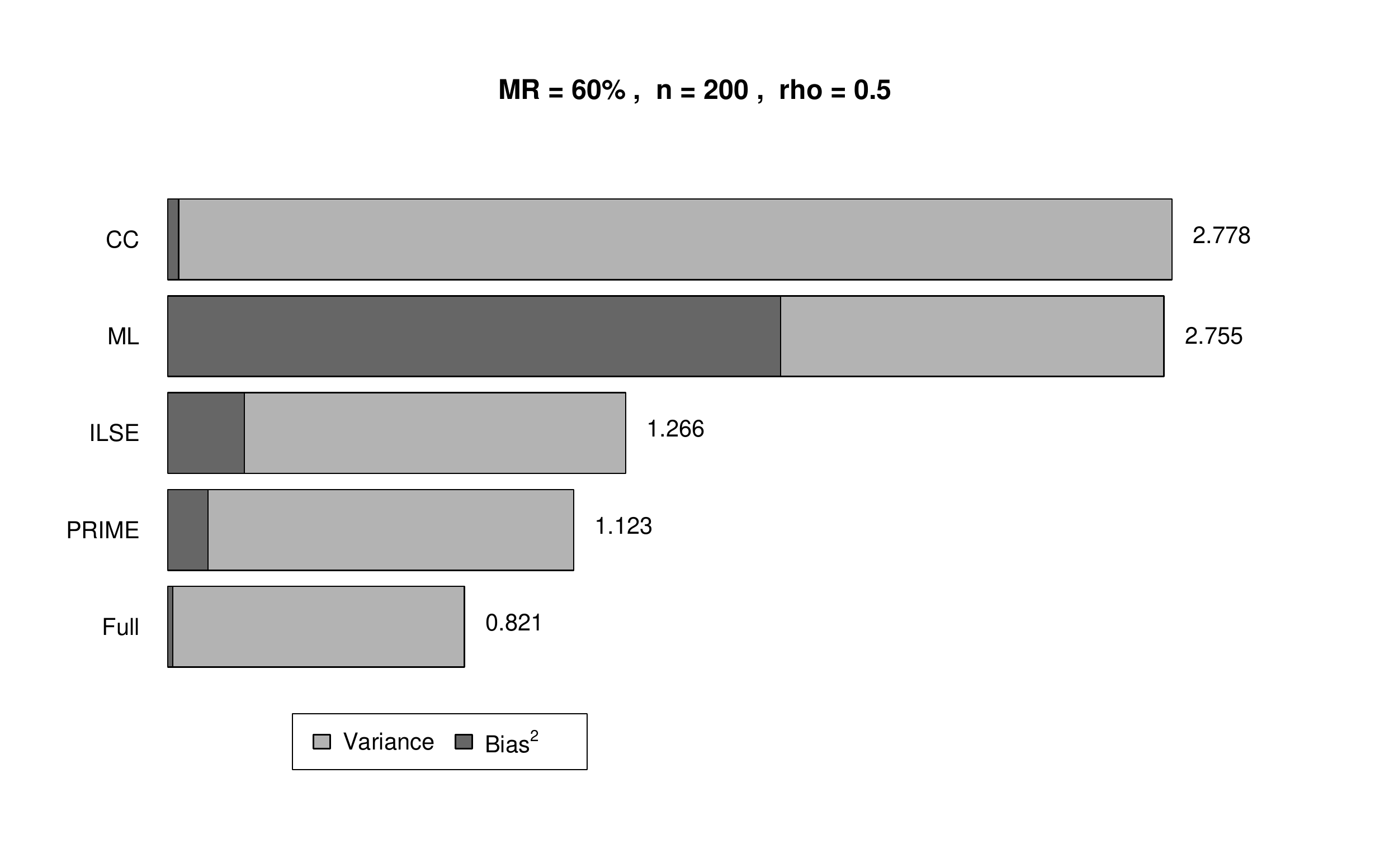}
		\end{minipage}
	}%
	\subfigure[]{
		\begin{minipage}[t]{0.33\linewidth}
			\centering
			\includegraphics[width=1.9in]{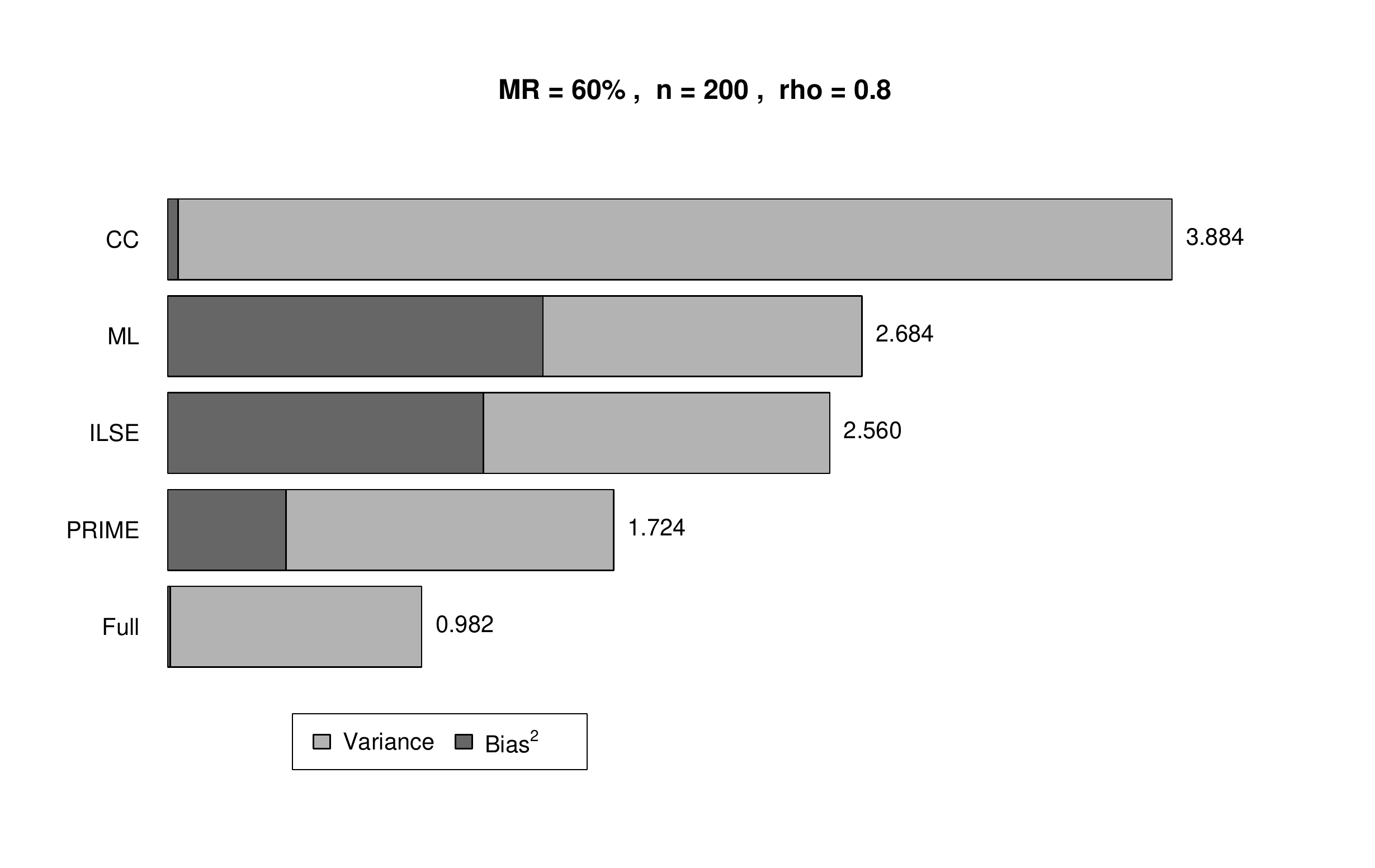}
		\end{minipage}
	}%
	\centering
	\caption{MSE with $n=200$ and 60\% missing data for different methods.}
	\label{fig:RHOMSE62}
\end{figure}

\begin{figure}[H]
	\begin{center}
		\includegraphics[width=10cm,height=7cm]{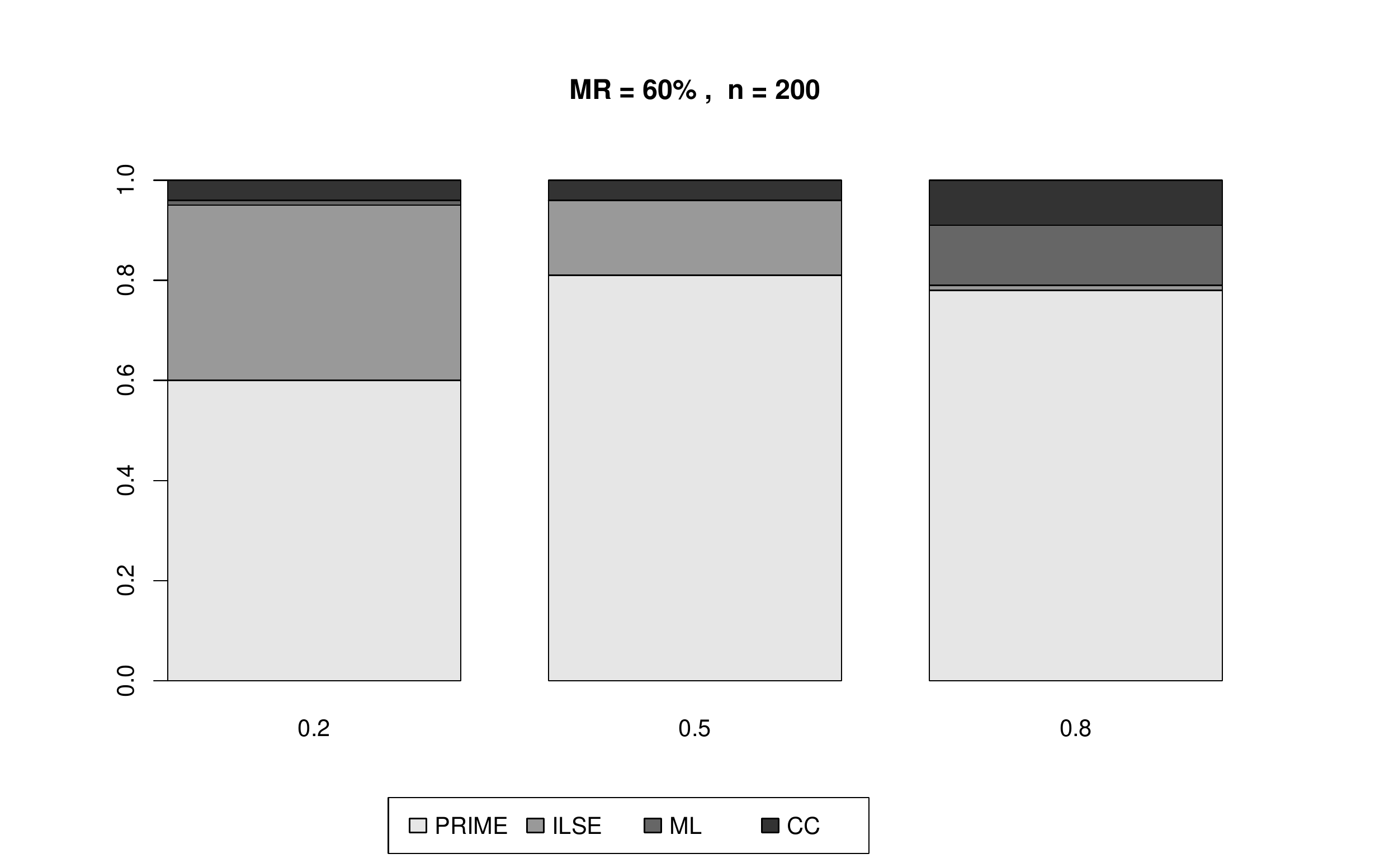}
	\end{center}
	\caption{Optimal rate of MSE with $n=200$ and 60\% missing data for different methods.}
	\label{fig:ORRHO62}
\end{figure}

The NAD and MSE results are shown in Figures \ref{fig:RHONAD91}, \ref{fig:RHOMSE91}, \ref{fig:RHONAD92}, \ref{fig:RHOMSE92}, \ref{fig:RHONAD61}, \ref{fig:RHOMSE61}, \ref{fig:RHONAD62}, and \ref{fig:RHOMSE62}. The optimal MSE rates are shown in Figures \ref{fig:ORRHO91}, \ref{fig:ORRHO92}, \ref{fig:ORRHO61}, and \ref{fig:ORRHO62}. The main conclusions are as follows:\\

\begin{enumerate}[1.]
	\item  Scenario 2 results yield conclusions similar to those of Scenario 1. PRIME produces the smallest NADs and MSEs in almost all cases. As shown in Figures \ref{fig:RHOMSE61} and \ref{fig:RHOMSE62}, although CC has the smallest estimation error, its estimation variance is extremely high, which causes problems in the MSE.
	
	\item The optimal rate results show that PRIME has obvious advantages over other methods because it produces the smallest MSE in almost all situations except when $\text{MR}=90\%, n=200,$ and $\rho_{ij}=0.8$. However, when $\rho_{ij}$ increases, the gap between PRIME and other methods increases. CC still has the worst MSE performance among the four methods. 
	
\end{enumerate}

\subsection{Scenario 3: Taking sparse structure into consideration}
In this scenario, we illustrate the proposed SPRIME by studying the data from simulations. We consider penalized Full (denoted as SFull), complete-case analysis with a penalty (denoted as SCC), ILSE, and ML as the alternatives. ILSE in Lin et al. \cite{Lin2019} and ML in Jiang et al. \cite{Jiang2020} were proposed without considering the sparse assumption; hence, we use them directly instead of using the penalized estimation form. We acknowledge that there are other approaches such as those in Xue and Qu \cite{Xue2020} that can be used to address a high-dimensional missing-data problem. However, the missing-data patterns in these methods are different from the individual-specific case.

The model used to generate data has the linear expression
\begin{eqnarray*}
	Y_i=\sum_{j=1}^p\beta_jX_{ij}+\varepsilon_i,\ i=1,2,\cdots,n,
\end{eqnarray*}
where $n=200$, $p=30$, $(\beta_1,\cdots,\beta_{12})^{\top}=(1, -0.6, 1.5, 1, 1.2,0.4, -1, -0.7, 1.3, 0.5, 1.1, -1.4, 0.9)$, and $\beta_j=0\ (j=13,\cdots,30)$. 
We generate $(X_{i1},\cdots,X_{ip})$ from the multivariate normal distribution $N_{p}(\bf{0},\Sigma)$. We set the non-diagonal element $\rho_{ij}$ of $\Sigma$ equal to $0.5$. For $\varepsilon_i$, we use the error distribution
$N(0,\sigma^2)$, where $\sigma^2$ changes with $R^2=\text{Var}({\bm X}_i^{\top}{\bm \beta})/\{\text{Var}({\bm X}_i^{\top}{\bm \beta})+\sigma^2\}$. We consider only the case in which $R^2=0.7$.

\begin{figure}[H]
	\centering
	\subfigure[]{
		\begin{minipage}[t]{0.5\linewidth}
			\centering
			\includegraphics[width=3in]{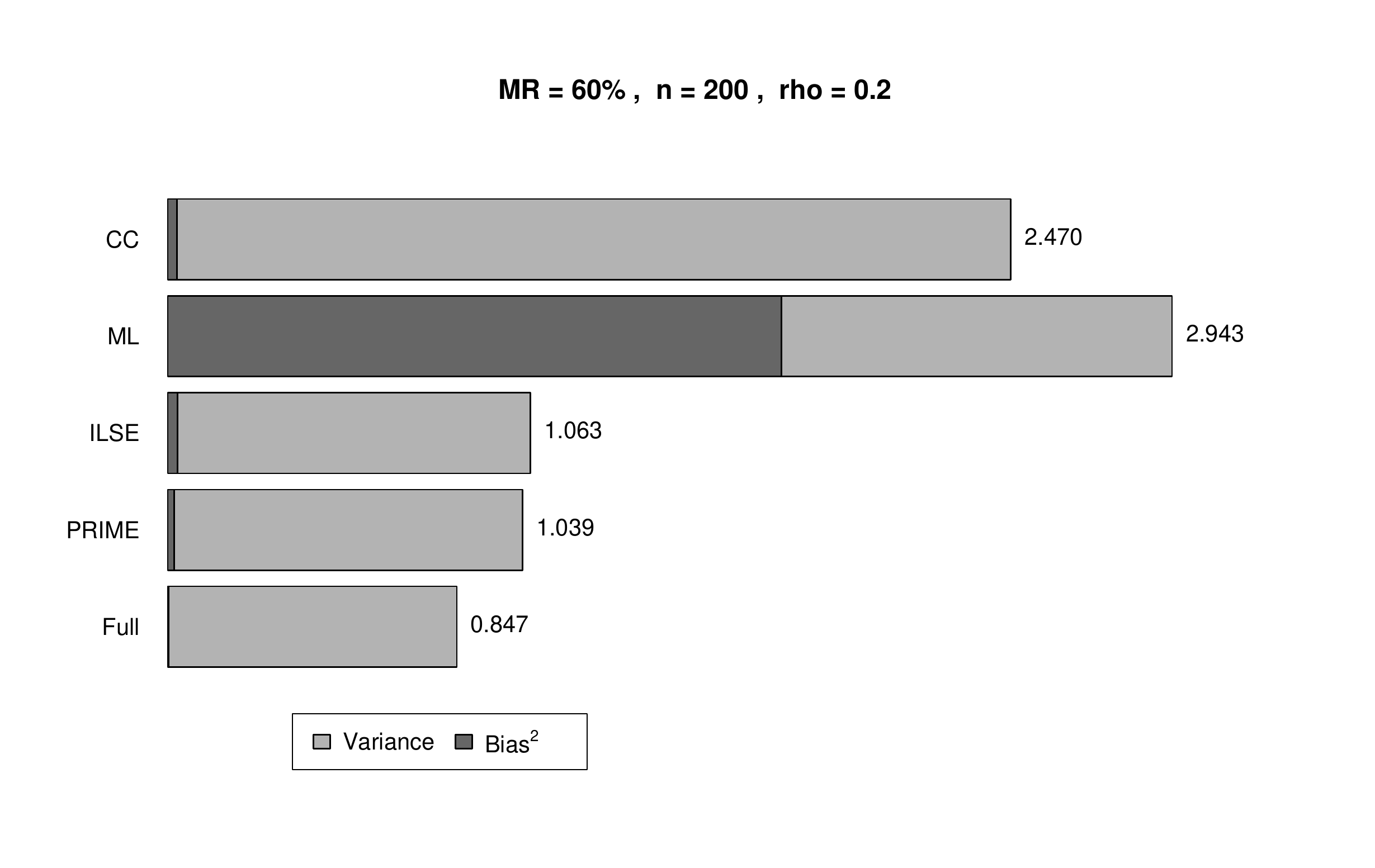}
		\end{minipage}%
	}%
	\subfigure[]{
		\begin{minipage}[t]{0.5\linewidth}
			\centering
			\includegraphics[width=3in]{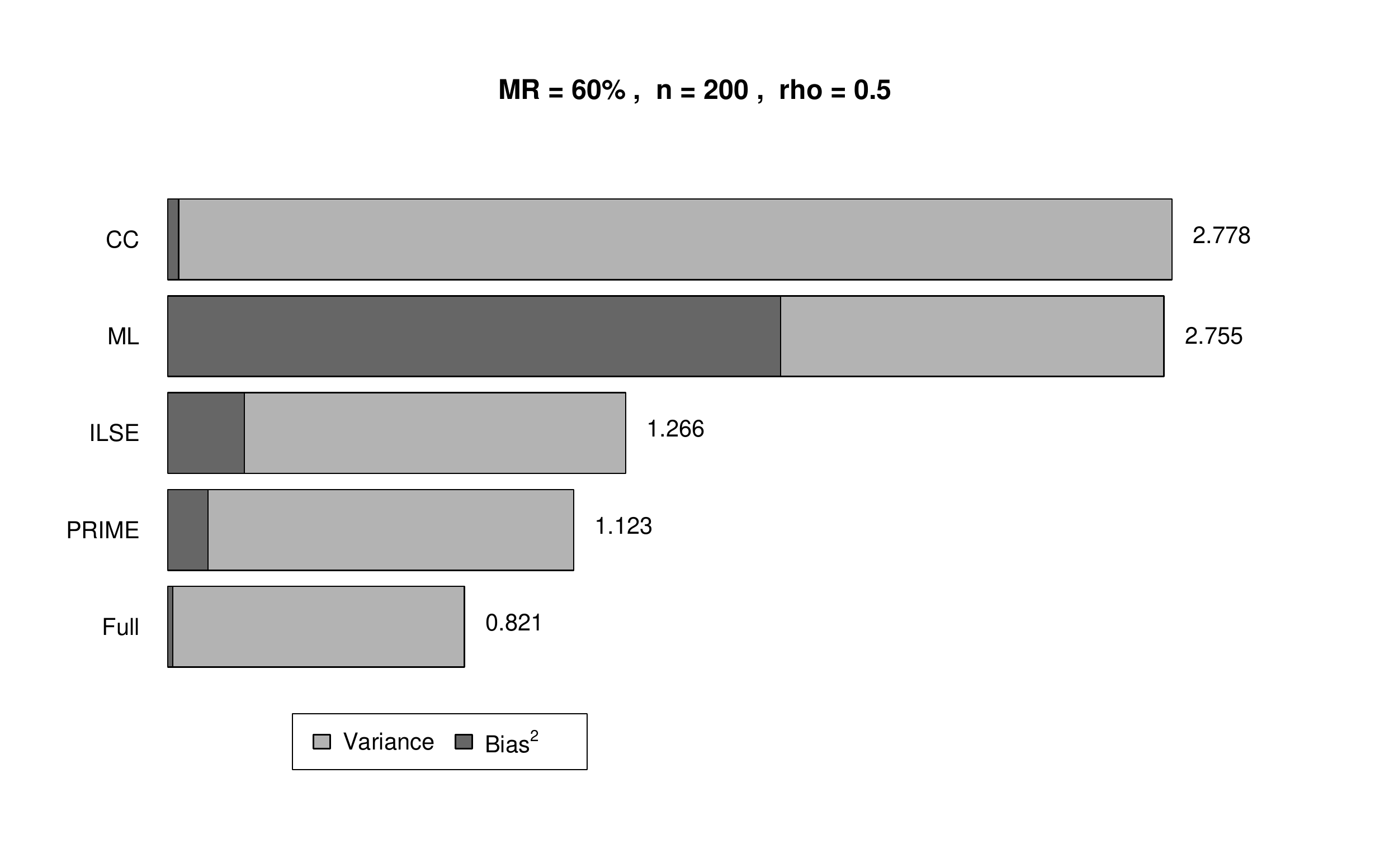}
		\end{minipage}
	}%
	\centering
	\caption{MSE with $n=200$ missing data for different methods.}
	\label{fig:SDMSER5}
\end{figure}

When taking sparse structure into consideration, the coefficients $\beta_j \ (j=13,\cdots,30)$ are equal to 0. Thus, the criterion, \text{NAD} used in Scenario 1 and 2, is no longer meaningful. So we only use MSE and the optimal rate of MSE to assess the performance of different methods. Several conclusions can be drawn from the figures:

\begin{enumerate}[1.]
	\item  Scenario 3 results yield conclusions similar to those of Scenarios 1 and 2. From Figure \ref{fig:SDMSER5} we can see that SPRIME produces the smallest MSEs in almost all cases. 
	
	\item When MR=60\%, the optimal rates of MSE are 0.95, 0.01, 0.00, 0.04 for SPRIME, ILSE, CC and ML, respectively. When MR=90\%, the optimal rates of MSE are 0.94, 0.05, 0.01, 0.00 for SPRIME, ILSE, CC and ML, respectively. Not surprisingly, SPRIME that considers the sparse structure performs better than ILSE and ML without the penalty, which reconfirms the superiority of PRIME-type approach.
\end{enumerate}

\section{Cardiac surgery-associated acute kidney injury study}
\label{section5}
In this example, we illustrate the proposed method by analyzing data regarding Cardiac surgery-associated acute kidney injury (CSA-AKI). CSA-AKI is the second condition for acute kidney injury in the intensive care setting and sometimes causes death \cite{Chen2020}. However, because of a general lack of effective treatment for CSA-AKI, tools or methods for earlier identification are very important
for prevention and management of the syndrome. To find more predictive biomarkers for CSA-AKI, Chen et al. \cite{Chen2020} collected 32 plasma cytokines, including CTACK, FGFa, G-CSF, HGF, interferon-$\alpha$2,
interferon-gamma (IFN-$\gamma$), IL-1$\alpha$, IL-1$\beta$, IL-2, IL-4, IL-6, IL-7, IL-8, IL-9, IL-10, IL-12p70, IL-12p40, IL-16, IL-17$\alpha$, IL-18, IP-10, MCP-1, MCP-3, M-CSF, MIF, MIG, MIP-1$\alpha$ (macrophage inflammatory
protein-1 $\alpha$), MIP-1$\beta$ (macrophage inflammatory protein-1 $\beta$), SCF, SCGF-$\beta$, SDF-1$\alpha$, and tumor necrosis factor-$\alpha$. CSA-AKI severity is evaluated primarily by deltaScore, which is measured by serum creatinine alterations before and after surgery. Serum creatinine concentrations before and after surgery are measured by an identical testing platform in the clinical laboratory of the hospital.

We use the continuous-variable deltaScore as the response. For simplicity, we conduct a standardized
transformation to scale the both response and covariates. Furthermore, we exclude subjects with missing deltaScores because the aforementioned methods (PRIME and SPRIME) apply primarily to the missing covariates. Finally, 321 patients are enrolled for statistical analysis, of which only approximately 60\% have complete covariate information. 

Because the number of related variables is not expected to be large, as in Scenario 3 in the simulation study, we use SPRIME and SCC to simultaneously select and estimate the coefficients of the factors that might shed light on the deltaScore. We also use ILSE and ML directly without considering the sparse assumption. The regression coefficient estimates obtained from the four methods are listed in Table \ref{tab:coef}. Among them, IL-8, IL-10, IFN-$\gamma$, IL-16, and MIP-$\alpha$ are also found to be related to CSA-AKI in Chen et al. \cite{Chen2020}. However, in real-world data, it is difficult to objectively evaluate the performance of candidate methods. Therefore, we delete the subjects with missing covariates and construct missing data manually for the complete-case data of CSA-AKI. For the same reason as before in Scenario 3, we consider only MSE  to evaluate the estimation accuracy. However, because of the unknown true coefficients, we are unable to evaluate MSE as described in the simulation. Hence, we calculate $\text{MSE}_{\text{Full}}$ instead, as follows:

\[
\text{MSE}_{\text{Full}}=\frac{1}{N}\sum_{i=1}^{N}\frac{1}{p}\sum_{j=1}^{p}(\hat{\beta}_{\text{Full},j}-\hat{\beta}_j)^2.
\]
where $\hat{\beta}_{\text{Full},j}\ (j=1,2,\cdots,p)$ are the estimated coefficients obtained using Full method. 

The setting of missing-data patterns is the same as that in the simulation study except for the missing probability function. We randomly assign the missing patterns in ${\bm A}^{(1)}$ to the sample with missing probability $P=a$. Furthermore, we set the missing probability of the $i$th unit for the patterns in ${\bm A}^{(2)}$ as $P=\{1+\exp(b\epsilon_i+c)\}^{-1}$, where $\epsilon_i=Y_i-{\bm X}_i^{\top} \hat{\bm \beta}_{\rm Full}$. Then, we randomly assign the patterns in ${\bm A}^{(2)}$ to the missing samples. For the missing rate, we set $(a,b,c)=(-1.5,-2,0.4)$. Consequently, the missing rate is approximately 90\%. This is repeats $N=100$ times to randomly generate missing data. Here, like in simulation, we also set bandwidth $h = n^{-1/3}$.

\begin{figure}[H]
	\begin{center}
		\includegraphics[width=10cm,height=6cm]{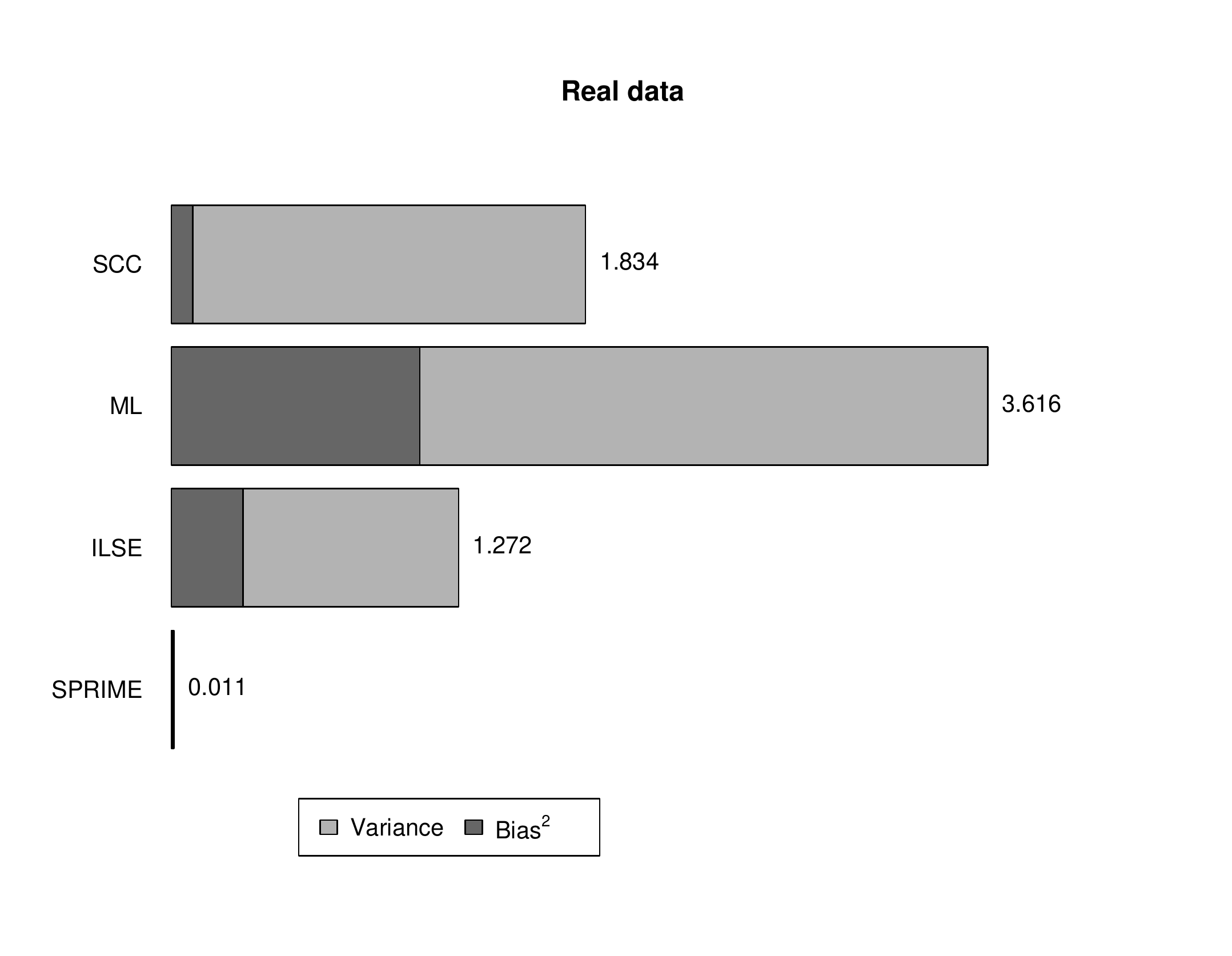}
	\end{center}
	\caption{MSE with real data for different methods.}
	\label{fig:realMSE}
\end{figure}

The $\text{MSE}_{\text{Full}}$ results are shown in Figure \ref{fig:realMSE}. The optimal rates of MSE are 1.00, 0.00, 0.00, 0.00 for SPRIME, ILSE, CC and ML, respectively. The resuls show that SPRIME has advantages over the other methods in estimation accuracy as it produces the smallest $\text{MSE}_{\text{Full}}$. The missing mechanism and the wrong model assumption may give rise to the worse performance of ILSE, CC and ML. 

\begin{table}[ht]
	\caption{Regression coefficients of CPLSD with regression coefficient  estimates (The symbol "-" means that the current variable or predictor has not been selected).}
	\label{tab:coef}
	\centering
	\begin{tabular}{lcccc}
		\hline
		\textbf{Variable} & \textbf{SPRIME} & \textbf{SCC} & \textbf{ILSE} & \textbf{ML} \\ 
		\hline
		age & -0.058 & -0.099& -0.167 & -0.012 \\ 
		BMI & $-$ & $-$ & 0.040 & 0.048 \\ 
		hospitalized time & 0.051 & 0.045 & 0.069 & 0.080 \\ 
		CTACK & $-$ & 0.020 & 0.207 & 0.117 \\ 
		FGFa & $-$ & $-$ & 0.137 & 0.137 \\ 
		G-CSF & $-$ & $-$ & -0.403 & -0.462 \\ 
		HGF & $-$ & $-$ & -0.120 & -0.162 \\ 
		IFN-$\alpha$2 & $-$ & $-$ & 0.233 & 0.393 \\ 
		IFN-$\gamma$ & 0.089 & 0.063 & 0.207 & 0.199 \\ 
		IL-1$\alpha$ & $-$ & $-$ & -0.126 & -0.113 \\ 
		IL-1$\beta$ & $-$ & $-$ & -0.257 & -0.256 \\ 
		IL-2 & $-$ & $-$ & 0.349 & 0.280 \\ 
		IL-4 & $-$ & $-$ & -0.011 & -0.012 \\ 
		IL-6 & $-$ & $-$ & -0.126 & -0.148 \\ 
		IL-7 & $-$ & $-$ & -0.165 & -0.182 \\ 
		IL-8 & 0.149 & 0.155 & 0.527 & 0.624 \\ 
		IL-9 & $-$ & -0.008 & -0.020 & -0.034 \\ 
		IL-10 & 0.040 & 0.023 & 0.043 & 0.064 \\ 
		IL-12p70 & $-$ & $-$ & 0.347 & 0.361 \\ 
		IL-12p40. & $-$ & $-$ & -0.029 & -0.099 \\ 
		IL-16 & 0.100 & 0.109 & 0.143 & 0.118 \\ 
		IL-17$\alpha$ & $-$ & $-$ & -0.322 & -0.303 \\ 
		IL-18 & $-$ & $-$ & -0.071 & -0.080 \\ 
		IP-10 & $-$ & $-$ & -0.057 & -0.170 \\ 
		MCP-1 & $-$ & $-$ & 0.081 & 0.060 \\ 
		MCP-3 & $-$ & $-$ & -0.099 & -0.233 \\ 
		M-CSF & $-$ & 0.080 & 0.024 & 0.042 \\ 
		MIF & $-$ & $-$ & -0.131 & -0.146 \\ 
		MIG & 0.024 & 0.009 & 0.169 & 0.342 \\ 
		MIP-1$\alpha$ & 0.036 & 0.039 & 0.315 & 0.330 \\ 
		MIP-1$\beta$ & $-$ & $-$ & -0.187 & -0.165 \\ 
		SCF & 0.175 & 0.143 & 0.166 & 0.175 \\ 
		SCGF-$\beta$ & 0.093 & 0.0923 & 0.060 & 0.121 \\ 
		SDF-1$\alpha$ & $-$ & $-$ & -0.096 & -0.057 \\ 
		preLVEF & $-$ & $-$ & 0.056 & 0.073 \\ 
		TNF-$\alpha$ & $-$ & $-$ & -0.100 & -0.071 \\ 
		\hline
	\end{tabular}
\end{table}

\section{Conclusions}
\label{section6}
In this study, we propose a projective resampling imputation mean estimation method to estimate the regression coefficients for a high rate of missing-data covariates. Our first set of random features projects data points onto a randomly chosen line and then averages the resulting scalar values to yield a comprehensive result. The random lines are drawn from the standard normal distribution to ensure less loss of information. We experimentally evaluate the performance of the PRIME method, and the results showe that the proposed method is a feasible alternative.

However, the aforementioned work has been developed for a classical setting. Developing PRIME to combine generalized linear models or Cox models with missing data warrants future research. Furthermore, we considered only missing covariates even though it is common to encounter cases where both covariates and responses are missing. Hence, developing methods to address practical issues will be the focus of our future work. 

\clearpage


\bibliographystyle{unsrt} 
\bibliography{ref}

\clearpage
\section*{Appendix}
\noindent{\bf Proof of Theorem \ref{th1}} \\
\begin{proof}
	Following the proof of Lin et al. \cite{Lin2019}, we first define 
	\begin{equation*}
	\widehat{E}_{j, A}({\bm w}_A)=\frac{\sum_{i'=1}^{n} I\left(A_{i'} \supset A \cup j\right) X_{i^{\prime} j} \prod_{b=1}^{B}\left[K_{h}\left(\mathbf{X}_{i',A}^{\top} {\bm v}_{b,A}-{\bm w}^{\top}_A{\bm v}_{b,A}\right)\right]^{\frac{1}{B}}}{\sum_{i'=1}^{n} I\left(A_{i'} \supset A \cup j\right) \prod_{b=1}^{B}\left[K_{h}\left(\mathbf{X}_{i',A}^{T}{\bm v}_{b,A}-{\bm w}^{\top}_A{\bm v}_{b,A}\right)\right]^{\frac{1}{B}}}, \quad\left(j \notin A\right)
	\end{equation*}
	and then 
	\begin{equation*}
	{\bm Z}_i=(X_{ij}I(j\in A_i)+\widehat{E}_{j,A_i}({\bm X}_{i,A_i})I(j \notin A_i: 1\le j\le p),
	\end{equation*}
	
	\begin{equation*}
	U({\bm \beta})=\frac{1}{n}\sum_{i=1}^n{\bm Z}_i\{Y_i-{\bm Z}_i^{\top}{\bm \beta}\}.
	\end{equation*}
	Let
	\begin{equation*}
	\mathbf{z}_{i}=\left(X_{i j} I\left(j \in A_i\right)+e_{j, A_i}\left(\mathbf{X}_{i, A_i}\right) I\left(j \notin A_i\right):\right.1 \leq j \leq p)^{\top}
	\end{equation*}

	\begin{equation*}
	u(\boldsymbol{\beta})=E\left[{\mathbf{z}}_{i}(\boldsymbol{\beta})\left\{\mathbf{X}_{i}^{\mathrm{T}} \boldsymbol{\beta}_{0}-\mathbf{z}_{i}^{\top} \boldsymbol{\beta}\right\}\right].
	\end{equation*}
	Similar to Lin et al. \cite{Lin2019}, we first prove that ${\bm \beta}_0$ is a solution of $u({\bm \beta})$. We have
	\begin{equation*}
	E(Y_i|{\bm X}^{\top}_{i,A_i},A_i)={\bm z}_i^{\top}{\bm \beta}_0.
	\end{equation*}
	and we also have $E(Y_i|{\bm X}_{i,A_i},A_i)=E({\bm X}^{\top}_{i}|{\bm X}_{i,A_i},A_i){\bm \beta}_0$, hence we can get $E\left\{({\bm X}_i-{\bm z}^{\top}_i)|{\bm X}_{i,A_i},A_i\right\}{\bm \beta}_0=0$. Noting that ${\bm z}_i$ is a function of $({\bm X}_{i,A_i},A_i)$, then
	\begin{equation}\label{u1}
	u({\bm \beta}_0)=E\left\{{\bm z}_i\left({\bm X}_i^{\top}{\bm \beta}_0-{\bm z}_i^{\top}{\bm \beta}_0\right)\right\}=0
	\end{equation}
	
	Similar to Lin et al. \cite{Lin2019}, it is easy to show that ${\bm \beta}_0$ is the unique solution of $u({\bm \beta})=0$.
	
	To prove Theorem \ref{th1}, it suffices to show that 
	\begin{equation}\label{uu}
	\sup_{{\bm \beta}\in \mathcal{B}}\|U({\bm \beta})-u({\bm \beta})\|_2=0.
	\end{equation}
	We rewrite 
	\begin{equation*}
	U({\bm \beta})-u({\bm \beta})=U_1-U_2+U_3+U_4, 
	\end{equation*}
	where
	\begin{eqnarray*}
		&&U_1=\frac{1}{n} \sum_{i=1}^{n}\left({\mathbf{Z}}_{i}-{\mathbf{z}}_{i}\right) Y_{i},\\
		&&U_2=\frac{1}{n} \sum_{i=1}^{n}\left({\mathbf{Z}}_{i} \mathbf{Z}_{i}^{T}-{\mathbf{z}}_{i} \mathbf{z}_{i}^{T}\right){\bm \beta},\\
		&&U_3=\frac{1}{n} \sum_{i=1}^{n}\left[{\bm z}_{i}\left\{\mathbf{X}_{i}^{T} \boldsymbol{\beta}_{0}-\mathbf{z}_{i}^{T} \boldsymbol{\beta}\right\}-E\left\{{\bm z}_{i}\left(\mathbf{X}_{i}^{T} \boldsymbol{\beta}_{0}-\mathbf{z}_{i}^{T} \boldsymbol{\beta}\right)\right\}\right],\\
		&&U_4=\frac{1}{n} \sum_{i=1}^{n}{\mathbf{z}}_{i}\varepsilon_{i 0}.
	\end{eqnarray*}
	Then $\sup_{{\bm \beta}\in  \mathcal{B}}\|U_3\|_2=o_p(1), \sup_{{\bm \beta}\in  \mathcal{B}}\|U_4\|_2=o_p(1)$ follows from the weak law of large numbers. Noting that 
	\begin{equation}\label{Ee}
	\hat{E}_{j, A}({\bm w}_A)-e_{j, A}({\bm w}_A)=\frac{\sum_{i^{\prime}=1}^{n} I\left(A_{i'} \supset A \cup j\right)\left\{X_{i^{\prime} j}-e_{j, A}({\bm w}_A)\right\} \left[\prod_{b=1}^{B}K_{h}\left(\mathbf{X}_{i',A}^{\top} {\bm v}_{b,A}-{\bm w}_A^{\top}{\bm v}_{i,A}\right)\right]^{\frac{1}{B}}}{\sum_{i^{\prime}=1}^{n} I\left(A_{i'}\supset A \cup j\right) \left[\prod_{b=1}^{B}K_{h}\left(\mathbf{X}_{i',A}^{\top} {\bm v}_{b,A}-{\bm w}_A^{\top}{\bm v}_{i,A}\right)\right]^{\frac{1}{B}}}
	\end{equation}
	For gaussian kernel generates better empirical performance than do other types of kernels, here we assume $K_h(\cdot)$ is a gaussian kernel function, then $\left[\prod_{b=1}^{B}K_h({\bm X}^{\top}_{i', A_i}{\bm v}_{b,A_i}-{\bm w}^{\top}_{i, A}{\bm v}_{b,A})\right]^{\frac{1}{B}}$ can be written as $\exp\left\{-\left\|\frac{1}{\sqrt{B}}V{\bm X}_{i',A}-\frac{1}{\sqrt{B}}V{\bm w}_{i,A}\right\|^2\right\}$, where $V$ be a $B\times |A|$ random matrix whose entries are chosen independently from either $N(0,1)$ or $U(-1,1)$. Hence, we can show that $\left[\prod_{b=1}^{B}K_h({\bm X}^{\top}_{i', A}{\bm v}_{b,A}-{\bm X}^{\top}_{i, P}{\bm v}_{b,A})\right]^{\frac{1}{B}}=K_h({\bm X}_{i',A}-{\bm X}_{i,A})+o_p(1)$ using Theorem 1 in Arriaga and Vempala \cite{Arriaga2006}. Then, (\ref{Ee}) can be writtern as 
	\begin{eqnarray*}
		&&\hat{E}_{j, A}({\bm w}_A)-e_{j, A}({\bm w}_A)=\frac{\sum_{i^{\prime}=1}^{n} I\left(A_{i'} \supset A \cup j\right)\left\{X_{i^{\prime} j}-e_{j, A}({\bm w}_A)\right\} \left[K_{h}\left(\mathbf{X}_{i',A}-{\bm w}_A\right)+o_p(1)\right]}{\sum_{i^{\prime}=1}^{n} I\left(A_{i'}\supset A \cup j\right) \left[K_{h}\left(\mathbf{X}_{i',A}-{\bm w}_A\right)+o_p(1)\right]}\\
		&&~~~~~~~~~~~~~~~~~~~~~~~~=\frac{\sum_{i^{\prime}=1}^{n} I\left(A_{i'} \supset A \cup j\right)\left\{X_{i^{\prime} j}-e_{j, A}({\bm w}_A)\right\} K_{h}\left(\mathbf{X}_{i',A}-{\bm w}_A\right)}{\sum_{i^{\prime}=1}^{n} I\left(A_{i'} \supset A \cup j\right) \left[K_{h}\left(\mathbf{X}_{i',A}-{\bm w}_A\right)+o_p(1)\right]}\\
		&&~~~~~~~~~~~~~~~~~~~~~~~~+\frac{\sum_{i^{\prime}=1}^{n} I\left(A_{i'} \supset A \cup j\right)\left\{X_{i^{\prime} j}-e_{j, A}({\bm w}_A)\right\} o_p(1)}{\sum_{i^{\prime}=1}^{n} I\left(A_{i'}\supset A \cup j\right) \left[K_{h}\left(\mathbf{X}_{i',A}-{\bm w}_A\right)+o_p(1)\right]}\\
		&&~~~~~~~~~~~~~~~~~~~~~~~~=E_1+E_2.
	\end{eqnarray*}
	Under the conditions given, following the proof of theorem 1 in Lin et al. \cite{Lin2019} and the lemma 4 in Chen et al. \cite{Chen2010}, we know that 
	\begin{equation*}
	\sup_{{\bm w}_A}\|E_1\|_2\le \sup_{{\bm w}_A}\left\|\frac{\sum_{i^{\prime}=1}^{n} I\left(A_{i'} \supset A \cup j\right)\left\{X_{i^{\prime} j}-e_{j, A}({\bm w}_A)\right\} K_{h}\left(\mathbf{X}_{i',A_i}-{\bm w}_A\right)}{\sum_{i^{\prime}=1}^{n} I\left(A_{i'} \supset A \cup j\right) \left[K_{h}\left(\mathbf{X}_{i',A}-{\bm w}_A\right)+o_p(1)\right]}\right\|_2=o_p(1)\\
	\end{equation*}
	Let $R({\bm w}_A;j)=1/n\sum_{i'=1}^{n}I\left(A_{i'} \supset A \cup j\right)K_h\left(\mathbf{X}_{i',A}-{\bm w}_A\right)$, and $\sup_{{\bm w}_A}|R({\bm w}_A;j)-P_r\left(A_{i'}\supset A \cup j\right)f({\bm w}_A)|=O_p\left(\sqrt{\log n}/\sqrt{nh}+h^2\right)$ follows from lemma 4 in Chen et al. \cite{Chen2010}.  Using Assumption (A5), then $\inf_{{\bm w}_A}R({\bm w}_A;j)>C=C_0-O_p\left(\sqrt{\log n}/\sqrt{nh}+h^2\right)$. Hence, we have
	\begin{eqnarray*}
		\sup_{{\bm w}_A}\left\|E_2\right\|\le \sup_{{\bm w}_A}\left\|\frac{1}{n}\sum_{i^{\prime}=1}^{n}\frac{ I\left(A_{i'} \supset A \cup j\right)\left\{X_{i^{\prime} j}-e_{j, A}({\bm w}_A)\right\} K_{h}\left(\mathbf{X}_{i',A_i}-{\bm w}_A\right)}{C+\frac{1}{n}\sum_{i^{\prime}=1}^{n} I\left(A_{i'} \supset A \cup j\right)o_p(1)}o_p(1)\right\|_2. 
	\end{eqnarray*}
	Let $S_{i'}=I\left(A_{i'} \supset A \cup j\right)\left\{X_{i' j}-e_{j, A}({\bm w}_A)\right\}/\{C+\frac{1}{n}\sum_{i^{\prime}=1}^{n} I\left(A_{i'} \supset A \cup j\right)o_p(1)\}$. Clearly, for the condition $A_i\bot {\bm X}_i$
	\begin{eqnarray*}
		&&E(S_{i'})=E\left[\frac{I\left(A_{i'} \supset A \cup j\right)X_{i' j}}{C+\frac{1}{n}\sum_{i^{\prime}=1}^{n} I\left(A_{i'} \supset A \cup j\right)o_p(1)}\right]-E\left[\frac{I\left(A_{i'} \supset A \cup j\right)E(X_{ij}|{\bm X}_{i,A}={\bm w}_A)}{C+\frac{1}{n}\sum_{i^{\prime}=1}^{n} I\left(A_{i'} \supset A \cup j\right)o_p(1)}\right]\\
		&&~~~~~~~~=E\left[\frac{X_{i'j}}{C+o_p(1)}\Bigg|I\left(A_{i'} \supset A \cup j)\right)=1\right]P_r(A_{i'}\supset A\cup j)\\
		&&~~~~~~~~~~~~~~-E\left\{E\left[\frac{X_{ij}}{c+o_p(1)}\Bigg|{\bm X}_{i',A}={\bm w}_A\right]\Bigg|I\left(A_{i'} \supset A \cup j\right)=1\right\}P_r(A_{i'}\supset A\cup j)\\
		&&~~~~~~~~=\frac{1}{C}\left\{E(X_{i'j})P_r(A_{i'}\supset A\cup j)-E\left[E(X_{ij}|{\bm X}_{i,A}={\bm w}_A)\right]P_r(A_{i'}\supset A\cup j)\right\}\\
		&&~~~~~~~~=\frac{1}{C}\left\{E(X_{i'j})P_r(A_{i'}\supset A\cup j)-E(X_{ij})P_r(A_{i'}\supset A\cup j)\right\}=0.
	\end{eqnarray*}
	Then using the weak law of large number, we have
	\begin{eqnarray*}
		\sup_{{\bm w}_A}\left\|E_2\right\|\le o_p(1).    
	\end{eqnarray*}
	Thus, we obtain $\sup_{{\bm \beta}\in  \mathcal{B}}\|U_1\|_2=o_p(1)$ and the same argument can also apply to $\sup_{{\bm \beta}\in  \mathcal{B}}\|U_2\|_2$. The above convergences imply that $\sup_{{\bm \beta}\in  \mathcal{B}}\|U({\bm \beta})-u({\bm \beta})\|_2=0$. Following the technical derivation follow from Lin et al. \cite{Lin2019}, Theorem \ref{th1} holds.
	
\end{proof}

\noindent{\bf Proof of Theorem \ref{th2}} \\
\begin{proof}
	It is obvious that the equations (\ref{u1}) and (\ref{uu}) are always correct no matter the assumption $j \in \mathcal{I}_1, j=1,2,\cdots,p$ is satisfied or not. Then, the proof of Theorem \ref{th2} is easily conducted by using the theorem 1 and 2 in Fu \cite{Fu1998} and thoerem 3 in Fu \cite{Fu2003}. 
\end{proof}

\end{document}